\documentclass[11pt,twoside]{article}
\usepackage[english]{babel}
\usepackage{times,subeqnarray}
\usepackage{url}
\newif\myifpdf
\ifx\pdfoutput\undefined
   \usepackage[dvips]{graphicx}
\else
   \pdfoutput=1        
   \usepackage[pdftex]{graphicx}
\fi
\usepackage{apatitlepages}
\usepackage{psydraft}
\usepackage{apa}       

\newcommand{\wij}{w_{ij}}

\newcommand{\oneo}[1]{\frac{1}{#1}}

\myifpdf
  \DeclareGraphicsExtensions{.pdf,.eps,.png,.jpg,.mps,.tif}
\fi

\def\myheading{ Goal-Driven Cognition }

\newcommand{\PVest}{\widehat{PV}}
\newcommand{\pPV}{PV^+}
\newcommand{\pPVest}{\widehat{PV^+}}
\newcommand{\nPV}{PV^-}

\newcommand{\pLVd}{\delta_{lv+}}
\newcommand{\nLVd}{\delta_{lv-}}

\begin{document}
\bibliographystyle{apa}

\sloppy
\raggedbottom

\def\mytitle{Goal-Driven Cognition in the Brain: A Computational Framework}

\def\myauthor{Randall C. O'Reilly, Thomas E. Hazy, Jessica Mollick, Prescott Mackie \& Seth Herd\\
  Department of Psychology and Neuroscience\\
  University of Colorado Boulder \\
  345 UCB\\
  Boulder, CO 80309\\
  {\small randy.oreilly@colorado.edu}\\}

\def\mynote{arXiv.org q-bio.NC preprint version of submitted manuscript}

\def\myabstract{
Current theoretical and computational models of dopamine-based reinforcement learning are largely rooted in the classical behaviorist tradition, and envision the organism as a purely reactive recipient of rewards and punishments, with resulting behavior that essentially reflects the sum of this reinforcement history.  This framework is missing some fundamental features of the affective nervous system, most importantly, the central role of goals in driving and organizing behavior in a teleological manner.  Even when goal-directed behaviors are considered in current frameworks, they are typically conceived of as arising in reaction to the environment, rather than being in place from the start. We hypothesize that goal-driven cognition is primary, and organized into two discrete phases: goal selection and goal engaged, which each have a substantially different effective value function. This dichotomy can potentially explain a wide range of phenomena, playing a central role in many clinical disorders, such as depression, OCD, ADHD, and PTSD, and providing a sensible account of the detailed biology and function of the dopamine system and larger limbic system, including critical ventral and medial prefrontal cortex.  Computationally, reasoning backward from active goals to action selection is more tractable than projecting alternative action choices forward to compute possible outcomes.  An explicit computational model of these brain areas and their function in this goal-driven framework is described, as are numerous testable predictions from this framework.

Keywords: goals; motivation; ventral medial prefrontal cortex; dopamine; computational model
}

\titlesepage{\mytitle}{\myauthor}{\mynote}{\myabstract}





\pagestyle{myheadings}

\section{Introduction}

There are two countervailing perspectives at work in the minds of those who study human cognition and behavior, which can lead to an apparent contradiction regarding the chain of causal events driving behavior. The first perspective is a simplistic temporal model of causality, adopted by the behaviorists, which tells us that, in the normal flow of events, {\em stimuli} (states of the world) trigger {\em actions}, which then produce {\em outcomes} (new states of the world).  But, our own day-to-day experience tells us that much of the time it is instead our {\em goals} (mental representations of desired future states of the world) that determine our actions, which then lead to new states of the world (which can then be compared with those envisioned in our goals). These two perspectives work against each other, especially in the context of typical experimental paradigms, because we usually try our best to eliminate any contamination from a subject's own set of internal goals, and impose our own instead.  We then carefully analyze the patterns of actions and outcomes that arise in response to the stimuli that we have presented.  This then leads to a kind of {\em reverse fundamental attribution error:} we tend to underestimate the role of goals in driving people's behavior.  This has led to somewhat of a bifurcation across different disciplines: in social, personality, and organizational psychology, where motivation and the role of goals in driving behavior is extensively recognized \cite[e.g.,]{Locke68,LockeLatham02,Bandura77,Bandura01,CarverScheier82,SvebakMurgatroyd85,Kunda90,CarverScheier90,WroschScheierMillerEtAl03,Ajzen91,MasicampoBaumeister11}.  However, in the cognitive neurosciences, which have only more recently begun to address fundamental questions in the motivation of behavior, the critical role played by goals is somewhat less clearly recognized.

In particular, the computational paradigm of reinforcement learning (RL), which has largely followed in the behaviorist tradition by emphasizing stimulus -- response (S-R) learning, has become quite prominent \cite{Sutton88,SuttonBarto90,SuttonBarto98,MontagueDayanSejnowski96,SchultzDayanMontague97,Schultz98,HollermanSchultz98,SchultzDickinson00,WaeltiDickinsonSchultz01}.  In the RL framework, actions (triggered by stimuli or {\em states} in the environment) that lead to reward are reinforced, and performed more frequently under similar states in the future, while those that lead to less reward (or overt punishment) are attenuated (Thorndike's Law of Effect; \nopcite{Thorndike11}).  Recently, a new distinction between {\em model-free} and {\em model-based} forms of RL has become an area of active exploration \cite{DawNivDayan05,Dayan09,RangelHare10}.  While model-free RL is the standard S-R-O form, model-based approaches use an internal model of the environment to anticipate the potential outcomes of a sequence of actions, making action-choices based on these forward projections.  We argue that the typical framing of even this newer model-based framework, a form of {\em what-if} analysis, still has things subtly backward in reference to the vast majority of human experience.  Instead, behavior is fundamentally goal-driven in an explicitly teleological way, where we are selecting actions for the most part because they are likely to bring about a specific state of the world envisioned by a currently active goal.  In addition to being more computationally tractable (the forward projection process becomes intractable surprisingly quickly; \nopcite{NewellSimon76,SolwayBotvinick12}), we are finding that such a {\em goals-first} perspective leads to many important insights at both the behavioral and neural levels.

The central goal of this paper is to develop a biologically informed and constrained computational framework based on the idea that much of cognition {\em starts} with goals, and that the architecture of cognition actually implemented in our brains is fundamentally structured in this {\em goals first} fashion.  Our main hypothesis is that there are two qualitatively different states of mental life: {\em goal selection} vs. {\em goal engaged}, and that they have qualitatively different effective value functions, which can explain many seeming paradoxes as elaborated below.  The system is strongly motivated to have at least one active goal engaged at any given time, but it also very carefully weighs the cost-benefit tradeoffs of possible goals during the goal selection phase.  This is because once a goal becomes engaged, progress toward that goal then dominates the effective value function --- we are strongly motivated to achieve our goals, once they have been selected and engaged.  In addition to these main novel hypotheses, we elaborate a computational framework for how goals drive actions, which shares some features with various other frameworks, most notably that articulated by \incite{PezzuloCastelfranchi09}.  This framework is implemented in a biologically-based computational model that demonstrates the basic sufficiency and operation of these principles, and serves as a platform for further development and testing of the theory.  As we proceed, we attempt to make contact with relevant work from the long history of scholarship in the domain of goal-driven behavior
\cite{Zeigarnik27,Lewin28,Tolman32,Tolman48,MillerGalanterPribram60,Powers73,Klinger75,Gollwitzer93} (see \nopcite{AustinVancouver96} for a thorough review, and the other references mentioned above on active work in areas outside of cognitive neuroscience).

The framework outlined here builds upon widely accepted ideas in the neuroscience of affective behavior and cognitive control.  In particular, ventral and medial regions of the prefrontal cortex (v/mPFC), including the orbital frontal cortex (OFC) and the medial wall including anterior cingulate cortex (ACC), are important for representing affective states of various sorts, including what might be called ``hot'' (affectively meaningful) goals \cite{BalleineDickinson98,FrankClaus06,NoonanWaltonBehrensEtAl10,OngurPrice00,PauliHazyOReilly12,SaddorisGallagherSchoenbaum05,Schoenbaum04,SchoenbaumChibaGallagher00,SchoenbaumRoeschStalnakerEtAl09,SchoenbaumSetlow01,PriceDrevets10,ShenhavBotvinickCohen13}.  By virtue of their affective grounding, the activations of these v/mPFC regions exerts a far-reaching, energizing influence on processing throughout the brain, including the important executive control areas in lateral PFC, particularly dorsolateral PFC (dlPFC).  These ``higher'' cognitive areas can be seen as subserving the selection and active maintenance of what might be called ``cold'' goals (or {\em task set} representations; \nopcite{RogersMonsell95}), which can then provide focused, top-down biasing to influence processing elsewhere in the brain in a cognitively-framed, task-focused manner \cite[e.g.,]{FusterAlexander71,GoldmanRakic87,OReillyBraverCohen99,MillerCohen01}.  Our focus in this paper is much more on the ``hot'' side of goals, as compared with the more cybernetic, control-theory oriented focus of many other computationally-based frameworks \cite{MillerGalanterPribram60,Powers73,WolpertKawato98,Hommel04,PezzuloCastelfranchi09}.

\subsection{Motivation and Principles}

The following questions about observations in daily life (in the first person perspective of the first author) were important for motivating the overall goal-driven framework developed here, and serve as a concrete, if anecdotal, springboard for the set of principles to follow:
\begin{itemize}
\item Why are my kids so agitated when they don't have something specific to do right now, and why can they get so obsessed with building Legos and playing video games?
\item Why is it so important to have a progress meter on exercise equipment, and why can't I stop myself from organizing my kid's Legos, or cleaning the pool?
\item Why do I sometimes feel completely overwhelmed, and other times completely in control, when facing the same abyss of an inbox filled with hundreds of possibly important things requiring my attention?
\item Why do I reliably cross the street at different points on the way to a given destination, compared to the way back, crossing at the earliest point in each trajectory?
\item Why does our field persist in exhibiting extreme confirmation bias (and other similar such biases), when everyone knows this is a major problem?
\item Why doesn't my computational model know when it has achieved a subgoal, and can then move on to a new step, and give itself a little shot of intermediate reward?
\end{itemize}

After introducing the following principles that constitute the core of our theoretical framework, we apply these principles to the above questions, and then to a broad set of neural and behavioral data.

{\bf Principle 1: Active goals are primary, and there must be at least one engaged at all times.}  These active goals must be concrete, with a clear criterion for accomplishment (a clear end), with a perceptual component that supports the ability to track progress toward the goal \cite{Locke68,LockeLatham02}, and at least a general action plan for how to proceed.  The time horizon for accomplishment can be variable, but probably no longer than hours to perhaps a few days, with this horizon likely dependent on species, age and various personality factors.  Being in a state without such a goal active is aversive, and the system works assiduously to remedy this situation.  Thus, we can parse our cognition into two discrete states: the {\em goal engaged} state, and the {\em goal selection} state, and as we'll see below, these states have strongly dissociable properties.  Goals can be nested and stacked, and multiple can potentially be engaged at a time, but more distal, latent goals must translate into more proximal active goals that have a more concrete near-time horizon, which then tend to dominate the immediate sphere of cognition (in proportion to some dimension of intensity).  \incite{Klinger75} referred to these active goals as one's {\em current concerns}, and articulated many of the same implications of these engaged goals as we highlight below (this represents an independent replication, as we only became aware of Klinger's paper late in the writing process of this paper).

We use the term {\em goal} to denote a distributed representation that links three key components together: 1) an affectively meaningful outcome; 2) stimulus representations that enable progress toward goal achievement to be estimated; and 3) action plans directed toward achieving the outcome.  Because of this critical link with affectively meaningful outcomes, these goals can be considered ``hot'' goals in contrast to just the ``cold'' plans and strategies thought to be encoded in the dlPFC (although the degree of ``hotness'' certainly varies --- everyday goals may not seem all that hot).  Operationally, a goal is something that, when achieved leads to {\em happiness} or {\em satisfaction}, when obstructed leads to {\em frustration} \cite{Amsel62} or {\em anger}, and when abandoned leads to {\em disappointment} or {\em sadness} --- these are the signature emotions surrounding the goal activation system.  Furthermore, goals do not need to be explicit, consciously considered things (and perhaps subconscious goals are much more prevalent, which is why we tend to take them for granted) --- goals can exist at many different levels and under many different influences, spanning everything from basic drives to seek food to very high-level cognitive goals.  Goals differ from affective states (e.g., being hungry, happy, or sad) in that they drive behavior toward a specific, perceptually-defined end state --- being hungry can lead to the selection of a goal to eat, but this requires a specific goal selection process, and is not automatic (depending on the circumstances, hunger can be ignored while pursuing other goals).  See Figure~\ref{fig.pv_list} below for specific examples that can help to clarify these relationships.

Despite any residual definitional ambiguities surrounding this term, the strong claim is that the notion of an active goal is completely unambiguous {\em in the brain}.  There are special areas of the brain (e.g., v/mPFC, ventral striatum, amygdala, and various brainstem nuclei) that are endowed with the power to select, represent, maintain, and monitor goal states, and have the appropriate connections for these goal states to drive lower-level reward systems, and actions.  All of these issues are discussed in greater detail in what follows, and rendered more fully explicit in our computational modeling framework.  We are mindful of the tautological trap lurking in any discussion of goals (e.g., where anything and everything that drives behavior is labeled as a goal), and we think the constraints of explicit computational models are particularly important in avoiding this problem. 

{\bf Principle 2: Once activated, progress toward goals drives incremental dopamine release, and reinforces actions taken (and subgoals) toward these goal states.}  There are multiple overlapping dimensions of progress, including estimated time to achievement, degree of uncertainty of achievement, and basic perceptual distance from the goal state.  The best kind of progress is gradual and steady, which leads to steady delivery of dopamine over time --- it is nearly impossible to stop oneself from working toward goals with a near time horizon (order minutes) and steady indicators of progress, with small challenges to overcome at each step.  This is the recipe for all things that lead to seemingly inexplicable obsessions, from video games to obsessive cleaning and organizing, etc.  Solitaire is a particularly devilish combination of organizing and incremental progress --- it is striking how many people on airplane flights are playing this evil game.  Computer programming, for certain people, also fits this bill quite nicely, and the hours just melt away.  Stories (in movies and books) also tap into this same essential dynamic: a good story sets up a goal in the form of a mystery or unresolved question (the {\em hook}), and keeps doling out incremental progress toward that goal along the way.  Indeed, is there anything (aside from more basic drives such as hunger, sex, etc) that we derive high levels of pleasure from, that does {\em not} fit this general description?  Even with the more basic level goals, it is important to be mindful of this graded progress dynamic: sitting down with a large quantity of tasty food (e.g., a large bag of popcorn, a full pint of ice cream -- you know the drill) is an invitation to overdose, as each step of ``progress'' toward the ``goal'' of ``cleaning out that container'' drives you ever onward. This idea is largely consistent with Bouton's \cite{Bouton11} theory of the importance of context, where the context acts as stimulus associated with the reinforcement of the next response, making it more difficult to stop appetitive behaviors, but we reframe it so that the goal state itself, arising from a joint function of contextual cues and internal states, provides the essential motivation for actions achieving the current goal.   More broadly, the notion that cognition can be dominated by engaged goals is consistent with the {\em motivated reasoning} framework \cite{Kunda90,WestenBlagovHarenskiEtAl06}, and with the value added by strength of engagement \cite{Higgins06}.

{\bf Principle 3: It is relatively hard to activate new goals --- the brain is at its most perfectionistic in weighing costs and benefits at the goal selection phase.}  Whereas we seemingly don't care a whit about blowing hours and hours on meaningless activities such as solitaire once we've started down that path, we do seem to care a great deal at the point of indecision when selecting the next goal to activate.  It can feel like a real chore to weigh the costs and benefits of all the options --- which game should I play?  Which of the hundreds of pressing issues should I turn my attention to next?  The inability to select that next goal can be debilitating.  Indeed, as we discuss in more detail later, it is certainly a major symptom, if not the major cause, of depression, manifest as fatigue and {\em motivational anhedonia} (as contrasted with a basic inability to experience pleasure) \cite{DemyttenaereDeFruytStahl05,SalamoneCorreaFarrarEtAl07,TreadwayZald11,SalamoneCorrea12}.  This dissociation between the value system during the {\em goal engaged} and {\em goal selection} phases can be quite striking: even a strongly depressed person can overcome their sadness if they manage to get engaged in a goal-driven activity, but at the next goal selection opportunity, the abyss emerges once again.  When you see such a strong dissociation, that is a strong clue that something important is going on.  It means that the goal selection process is extremely important, and is where the value system is at its most balanced and critical in trying to optimize goal selection to the greatest extent possible.  As we elaborate below, perceptual inputs and other current state variables also influence the goal selection process in important ways, and can make the selection process much easier by limiting the space of options and giving a strong bias to one option over others.  In any case, once the value system becomes dominated by the active goal state, it can apparently become essentially oblivious to the overall optimization of cost/benefit tradeoffs.  Any progress toward the currently-active goal delivers those dopamine bursts, even if you opted for an objectively worthless activity such as sorting pixels representing cards into stacks on a computer screen.  This dissociation makes good functional sense: once you've decided upon a course of action, it doesn't make sense to be constantly second-guessing things --- you have to give it a fair shake, and then reevaluate at the next choice point. 

{\bf Principle 4: Despite the positive bias evident during the goal engagement stage, a more balanced estimation of value does accumulate.}  Even though it may not have felt especially effortful (once the goal was engaged) to write a paper or pack for a trip, the next time that alternative is considered, a more veridical sense of effort will weigh against engaging that goal, leading to inevitable procrastination.  Indeed, procrastination can be seen as the highly critical functioning of the goal selection process: if there are other options that lead to higher expected utility (reward versus effort) then those are strongly favored, as, in some objective sense, they should be.  It is actually somewhat of a miracle that people can manage to select higher-effort goals at all, and as we discuss later, there are important tricks that people can use to {\em game} the goal selection process to their (longer-term) advantage.  On the other side of the equation, even though a trashy novel or bad movie can keep you engaged, at the end there is a clearer reckoning of the overall reward value.  How many times have you actually managed to walk out (or turn off the TV) while watching a bad movie, compared to how carefully you evaluate the options prior to watching?  Overall, this points to another important dissociation, where latent knowledge is accumulated without influencing behavior at the time.

{\bf Principle 5: Not all goals are achievable, and the system is poised to detect when to give up on a given goal.}  Again despite the positive bias from goal engagement, a persistent lack of progress will lead to various stages of strategy and goal reevaluation, which is affectively aversive \cite{CarverScheier90}.  This is why the steady, gradual progress toward a goal is so important and so strongly rewarding --- its opposite is highly aversive, and indicates the potential need to abandon a goal.  The balance between perseverance in the face of challenge and flexibility in seeking new goals is an important dynamic with many tradeoffs, and it is unclear if there is any obvious optimal set point.  Certainly the population exhibits significant preserved variability on this dimension, despite whatever selection pressures might be at work.  In any case, it seems clear that we are much more likely to try alternative plans or strategies (i.e., the cold dlPFC side of the equation), rather than abandon an affective goal.  We have greater attachment to our goals than our plans, and giving up on a goal is a much bigger step than a change of plans --- people experience disappointment or get depressed when they give up on their goals, not when they have to adopt new strategies to achieve their sustained goals.


{\bf Principle 6: {\em Aversive goal} is an oxymoron --- all goals are fundamentally appetitive.}  When faced with aversive situations, our goals then focus on ways to avoid or overcome adversity, and steady progress toward doing so is just as motivating (if not more so) than progress toward a purely appetitive goal.  It is important to carefully dissociate the goal from the actions required to achieve the goal, and the emotional state experienced while undertaking the goal.  Often we have to execute unpleasant actions to achieve important goals, and we experience overriding negative affect, but the goal itself has been selected through this careful weighing of costs and benefits, and represents a perceived necessary step toward subsequent happiness.  Taking out the trash, cleaning the house, and numerous other onerous tasks are executed not because of their intrinsic pleasurable value, but because they have been evaluated as better to do than the alternative (and of course there is always Sarah Cynthia Sylvia Stout, who refused to take the garbage out, but that didn't end very well).  Surely some people find pleasure in causing harm to others, and some wallow in their misery, but these ultimately are subjectively appetitive.  In the latter case, wallowing and ``misery loves company'' are adaptive coping strategies for distancing oneself from aversive emotions (by putting it ``out there'' instead of keeping it inside), and by obtaining appetitive social support from others.  Indeed, many of our goals derive ultimately from social drives, and these can lead to apparent contradictions to the importance of appetitive goal activation, as we discuss more later.

{\bf Principle 7: Goal activation is graded, and not everything is goal-driven.}  As a function of individual differences, and various contextual and state factors (e.g., resting, sleeping, conserving energy), the degree of activation on goal states (and arousal level more generally) can be graded.  Sometimes you're just chillin', and do not have any pressing needs to drive goal engagement (referred to as the {\em paratelic} state by \nopcite{SvebakMurgatroyd85}, as contrasted with the {\em telic} or goal-engaged state).  Other times you've got multiple urgent deadlines, or are facing a life-and-death emergency.  Furthermore, some actions are {\em habitual} and not under goal-driven control --- the current goal state does not completely dominate all of cognition.  As long as the cognitive system has at least one sufficient goal state activated, plenty of other actions and distractions can be managed.  While engaged in writing a paper, emails pop up and some of them require responses.  Food must be prepared and eaten.  Various other obligations and responsibilities must be tended to.  Many elements of this behavior can potentially be characterized according to the classic habit learning system described by model-free reinforcement learning systems.  But if these things add up to a significant delay in obtaining those steady perceptions of progress toward the most salient engaged goal of writing the paper, then a sense of frustration can emerge, leading to various strategies for mitigating these distractions.  As noted above, this sense of frustration is a strong indicator of the presence of a goal --- you don't typically feel frustration in the context of habitual behavior.  Furthermore, when the engaged goals are strong and strongly reinforced by graded progress, many other basic needs can be severely neglected, reportedly even to the extreme where some people have died from multiple days of continuous video game play.

In summary, the present framework posits a strong dichotomy between two different phases of cognition, {\em goal selection} and {\em goal engaged}, with goal selection being {\em the} {\em prime directive} of the organism.  The rest of cognition can only proceed once goal selection has occurred, and is largely (but not exclusively) subservient to the activated goal(s).  In short, we live life in intervals defined by the reign of these goal states, with the potential for a crisis of succession at every juncture.  In practice, there can be long periods of essentially continuous goal engagement with subgoals driven by larger distal goals cascading and nesting, to carry one along (perhaps a kind of ``flow'' state).  But at some point, a point of reckoning comes where the way forward is less clear, and it becomes difficult to settle on that next concrete goal, and the goal selection process becomes more salient and onerous.  And some times, nothing goes well at all, and all the gradients seem to be negative, and you have difficulty even activating the goal to get out of bed.  We discuss these dynamics and individual differences therein later.  But for now, with this toolkit in hand, we can address those pressing questions raised above:

\begin{itemize}
\item {\em Why are my kids so agitated when they don't have something specific to do right now, and why can they get so obsessed with building Legos and playing video games?}  Because having a proximal appetitive goal is essential.  As in many things, children can provide a clearer window into underlying cognitive dynamics, unobfuscated by all the habits and complexity of adult value systems --- it is truly remarkable the rapid transition in mood and behavior that can occur when a goal has been engaged, vs. the agitated state just prior.  Indeed, it is generally thought that the tantrum state  reflects a profound state of frustration and/or disappointment due to the inability to achieve or engage a desired goal state.  You can safely reject the notion that a tantrum is a cold calculated attempt at manipulation (besides, the energetics just don't make sense): what you are witnessing is the raw power of goal-driven cognition, and the kind of force that is unleashed when it is thwarted.  As for the obsession, these activities, once engaged, deliver steady gradual progress toward concrete achievable goals.

\item {\em Why is it so important to have a progress meter on exercise equipment, and why can't I stop myself from organizing my kid's Legos, or cleaning the pool?}  You need to see progress toward a goal (or progress toward cessation of an aversive state), and even a fruitless task such as organizing your kid's Legos (which will be undone in short order) delivers those inexorable rewards as you make incremental progress.

\item {\em Why do I sometimes feel completely overwhelmed, and other times completely in control, when facing the same abyss of an inbox filled with hundreds of possibly important things requiring my attention?}  It all depends on whether you're in a goal-engaged or a goal-selection (indecision) state.  When engaged, other pressing deadlines and onerous tasks just magically recede into the background.  But upon popping back out, they come right back with familiar feelings of dread and guilt.  This can be rather fleeting if there is an obvious next-most-urgent task, but if there are many seemingly undifferentiable and unappealing obligations, the feeling of indecision can be debilitating.  One trick is to apply a weak general method (a meta-goal) to the indecision 
state by creating a subgoal of prioritizing the todo items, and then following that prioritized list instead of just staring into the abyss or doing something mindlessly rewarding instead (see organizing kid's Legos above).  But it can be hard to get over threshold to even do this prioritization step.

\item {\em Why do I reliably cross the street at different points on the way to a given destination, compared to the way back, crossing at the earliest point in each trajectory?}  It is all about making the most salient progress toward the goal as soon as possible --- there is a lot of uncertainty in timing etc for crossing the street, so you want to get that out of the way, so the remainder of the goal achievement is more certain and more subjectively proximal.

\item {\em Why does our field persist in exhibiting extreme confirmation bias (and other similar such biases), when everyone knows this is a major problem?}  It is hard to give up on a goal that you've worked on for some amount of time (e.g., years).  It is the same as the Zeigarnik effect \cite{Zeigarnik27}, motivated reasoning \cite{WestenBlagovHarenskiEtAl06}, the the gambler's fallacy --- we have a very hard time giving up on goals because these goals are primary, and we face the difficulty of goal selection and indecision and a broader sense of lack of self-efficacy and control when we give up on our goals \cite{Bandura77,Bandura01,MaierSeligman76}.  These are the at-work forces.  Mere facts and principles are no match.  At a larger scale, the Kuhnian structure of paradigm shifts followed by within-paradigm scientific thinking follows this same general structure: the field clings to paradigms as long as possible because they provide a structure and metric of progress and order.  Having to cast about for new paradigms is structurally aversive, and is only undertaken when the old ones are clearly broken.

\item {\em Why doesn't my computational model know when it has achieved a subgoal, and can then move on to a new step, and give itself a shot of intermediate reward?}  Because these (earlier) models did not have a special role for goals, and a constant measuring of distance to the active goal, so they just aimlessly execute actions according to previous patterns of reinforcement, without knowing when anything in particular has been achieved (or not).  These models never get frustrated or disappointed.  Until computational models ``feel'' those core emotions, they will remain perpetually passive and lacking the motivational foundation of cognition.
\end{itemize}

\subsection{ Overview of Proposed Computational Framework }

In the remainder of the paper, we attempt to distill the above insights into a biologically-constrained computational algorithm, leveraging known features of the relevant brain areas and mechanisms.  An initial implementation of this computational framework is then presented, and tested on a simple foraging task.  We hope that this kind of biologically-informed, neural network-based perspective will help crystalize some of the key challenges inherent in a goal-centric perspective on cognition and make clear what it may add to extant theorizing.  While many unanswered questions remain, we think this framework significantly realigns some of the principal dimensions of this problem domain in ways that may make the eventual solution easier.  

Some of the central questions to be addressed include: 1) How do goals gain control of the central neuromodulatory brain areas that project widely to the rest of the brain, and control learning and other dynamics; 2) How do these neuromodulatory signals, along with direct activation dynamics, reinforce goal-driven action plans, in the context of having a particular goal engaged?  3) How are new goals activated during the goal selection process?  4) How is a suitable map of available goal representations learned, to support flexible, adaptive behavior?  In what follows we try to provide some initial answers to these and other relevant questions.

The central features of the framework to be sketched out here build on our existing work with the PVLV (primary value, learned value) model of phasic dopamine firing \cite{OReillyFrankHazyEtAl07,HazyFrankOReilly10}, which drives learning in the striatum of the basal ganglia \cite{Frank05,OReillyFrank06}; which in turn drives selection of actions (at many different levels of abstraction) in the frontal cortex; which in turn drives cognitive control over processing elsewhere in the brain \cite{OReillyBraverCohen99,MillerCohen01,OReilly06}; which in turn feeds back to modulate processing at all those lower levels, resulting in a highly-interactive system capable of complex emergent behavior.  In this paper we provide a preliminary description of how we are beginning to integrate a goals-centric perspective into this existing framework.  We have found that this process is already helping to clarify and deepen our understanding of these foundational mechanisms in important ways, extending its explanatory power.  While our current framework remains dopamine-centered, we recognize the importance of other neuromodulatory systems and have plans to incorporate these into our framework over time.

\begin{figure}
  \centering\includegraphics[width=4in]{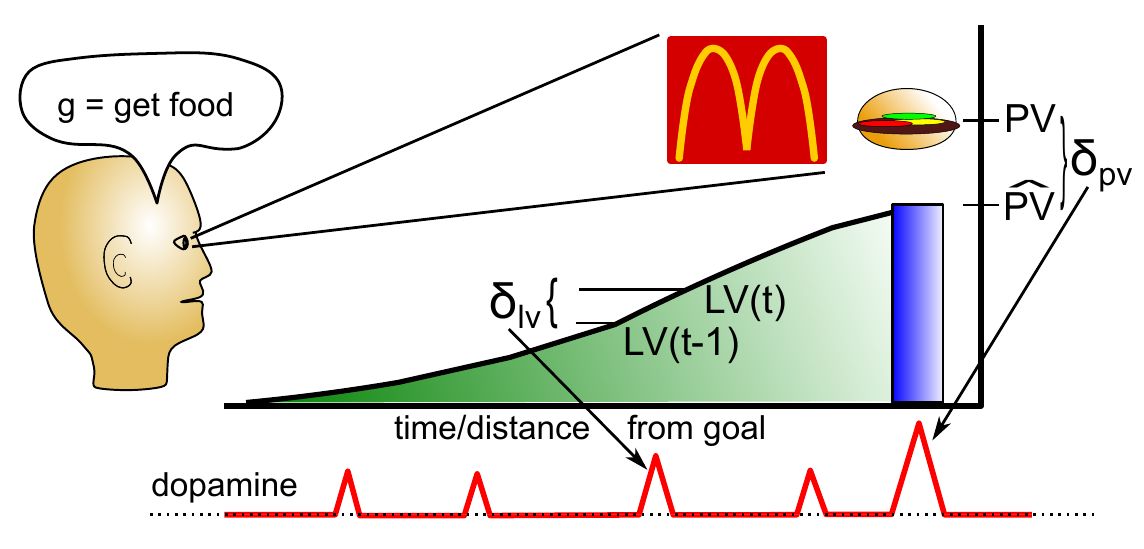}
  \caption{\small Schematic of the PV (primary value) and LV (learned value) components of the goal-driven version of the PVLV algorithm (gdPVLV).  The PV system compares primary outcomes (e.g., food) with expectations thereof ($\widehat{PV}$), and any discrepancy drives phasic dopamine bursts (shown, because the expectation was below the outcome in this case -- a particularly tasty burger) or dips if the expectation is higher than the outcome (especially important if e.g., the restaurant is closed and you don't get anything!).  The LV system learns at the time of PV rewards about sensory cues that can be used to anticipate subsequent PV outcomes, e.g., the logo of a restaurant --- once learned, these can drive phasic dopamine bursts (or dips, not shown) as a function of the perceived progress toward the goal.  In this way the dopamine system reinforces actions that progress toward the goal --- e.g., the actions that bring you closer to the restaurant.  These actions and other factors may result in a more discrete step-function (and trigger the timing of phasic dopamine firing), but the continuous curve shown captures the notion that perception defines a continuous metric for evaluating goal progress.}
  \label{fig.pvlv_food}
\end{figure}

\begin{figure}
  \centering\includegraphics[width=6in]{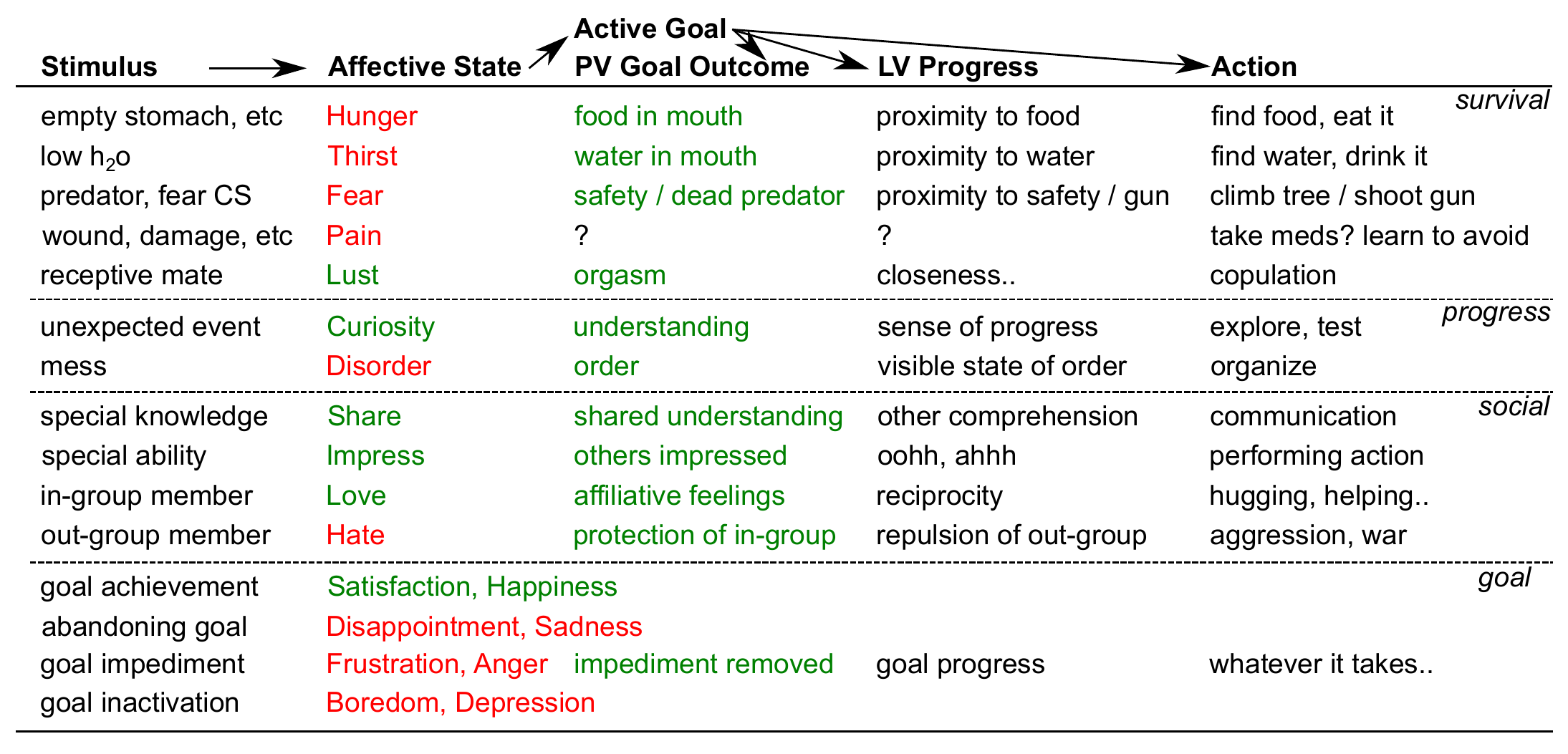}
  \caption{\small Relationship between affective states and PV outcomes for a set of core cases, ranging from basic survival to more ``progressive'' values to a sample of core social values.  The typical causal chain is that stimuli (internal or external) drive activation of affective states, which then influence (depending on their intensity) the goal selection process (but by no means completely determine it --- many other factors can be taken into account as well).  Once a goal is activated, it has an associated PV outcome (always positively valenced (green)), an LV monitoring of goal progress, and actions taken to advance the goal.  Note that the motivating affective states can be either positive or negative (red).  The last three affective states are those associated with the goal process itself, which can be recognized as completing a reasonable set of emotional or affective states (see text later for more possible such states, along with a discussion of the Pain state).}
  \label{fig.pv_list}
\end{figure}

Here is a brief overview of the key features of this framework (relevant literature references are omitted here for brevity, but are provided in the subsequent sections detailing each of these elements):

{\bf Core PVLV learning mechanisms:} The goal processing system comprises an extensive set of specialized brain areas that are continually engaged in processing and being driven by primary affectively-relevant outcomes (i.e., {\em primary values}, or PV in the PVLV framework).  Some of these areas can learn to associate these grounded primary values with other perceptual inputs (i.e., {\em learned values} or LV in the PVLV framework), thereby alerting the system to the presence of specific opportunities (and threats) in its immediate environment, and enabling the computation of a running estimate of the probability that particular primary values will be experienced in the near future.  In the context of a currently-engaged goal state, LV-signals can thus indicate that the probability of goal-realization has changed.  In the case of positive changes, the phasic dopamine bursts so triggered can be interpreted as indicators of progress toward achieving the relevant goal (Figure~\ref{fig.pvlv_food}).  

{\bf The relationship between goals and PVs:} It stands to reason that all goals must be ultimately grounded in the hard-wired values that have been selected through evolution.  Figure~\ref{fig.pv_list} provides some core examples of how goal-associated positive PV outcomes can be motivated by a variety of affective states.  While obviously tightly related, the goal is not exactly isomorphic with these primary values --- these values provide the power to goals by driving the dopamine neuromodulatory learning system, and driving overall affective states (e.g., hunger, thirst) that then provide activation for associated goals, so they are more likely to be selected (but this is far from a guarantee --- many such affective states can be ignored in favor of other stronger goals).  Note that appetitive PVs such as the positive reward associated with eating food can often have a negative affective state driving them (e.g., hunger), so eating can actually be seen as harm avoidance (avoiding starvation) rather than as a purely appetitive state.  One potential connection between PVs and goals comes from neurons in the ventral striatum that learn to expect primary value outcomes --- these neurons may also play a role in the gating of goal representations in v/mPFC (OFC and ACC).  This creates a potential linkage mechanism for PV outcomes to trigger goal deactivation after success (which is just as important as initial goal selection), and LV signals associated with PVs can drive goal selection.

{\bf Open-ended PVs: social rewards, understanding, order:} Despite the firm grounding in primary values, human cognition can be remarkably complex and flexible because our hard-wired reinforcers have come to include (perhaps even be dominated by) a major component of social reinforcement.  This in turn both supports, and is supported by, the powerful machinery of culture, which can then further shape our hard-wired value structure in open-ended ways.  In this way, culture itself has become a unit of selection with some sets of cultural reinforcements ending up more adaptive than others.  Through culture, we initially solve many of the difficult problems of action selection through direct instruction and imitation, which then become internalized and elaborated over time.  Furthermore, our hard-wired reinforcers have come to include abstract things such as new-found  {\em understanding}, often manifest as the motivation we experience as curiosity.  Similarly, we can experience a feeling of {\em order} as rewarding.  These can fuel important goal systems as well (again see Lego organization above).  Thus, it has been possible for evolution to go well beyond primitive reinforcers, but it is important to keep in mind that even the most abstract goals, such as completing a paper about goals, must ultimately be rooted in one or more (evolutionarily-selected) hard-wired reinforcers in some way.  An important challenge in fleshing out a goal-centric framework going forward will be how to properly capture the cocktail of hard-wired reinforcers in a concrete, mechanistically-explicit computational model, 

{\bf The development of a systematic goal space distributed across OFC, ACC, and dlPFC, interconnected with associated subcortical areas:} Another key challenge is to capture the way in which the development, shaped by cultural and other learning experiences, enables us to acquire a rich semantic map of a very high-dimensional {\em goal space} distributed across ventral / medial PFC (OFC and ACC) and other areas of cortex.  This goal space includes not only the universe of possible PV outcome states to be pursued, but also links into the state-action control systems in lateral frontal areas and elsewhere to determine how best to go about pursuing each.  We believe that among the roles of the OFC is a specialization for encoding running estimates of progress toward  desired goal outcomes (primary values, PVs).  Similarly, one of the roles of the ACC is a specialization for processing the costs associated with specific ways of pursuing different possible goals, computing a real-time representation of the net utility expected from each goal-state/action-plan pair.   Once acquired, such a goal space can support flexible goal selection in response to environmental and other challenges, but clearly its acquisition involves a highly complicated interplay between nature and nurture, including many years of hard work during the developmental learning process (even though this hard work is often cleverly disguised as playtime).

{\bf A goal processing pyramid in the brain:} There is a roughly hierarchical ``pyramid'' of goal processing areas in the brain, which are likely to have developed progressively over evolution, with higher areas elaborating and expanding upon the function of existing lower-level systems (the pyramid is inverted, with more neural tissue at the higher end).  This leads to considerable redundancy and complexity overall, and as a practical matter we are thus forced to be selective and to focus (at least initially) on only the most central computationally-relevant parts of the system, as present in mammals and especially primates.  At the highest level of the goal system, we focus on the v/mPFC areas (OFC, ACC) as the primary locus of goal representations, under the influence of basal ganglia gating neurons in the ventral striatum, which are also a key part of the PVLV system.  It is notable that the majority of the rodent prefrontal cortex (prelimbic and infralimbic cortex) consists of analogs of the limbic v/mPFC areas in primates, with just a relatively small portion of the dorsal aspect of the prelimbic cortex thought to be homologous to the vastly expanded lateral PFC areas in primates \cite{NarayananHorstLaubach06}.  Thus, the ``hot'' goal representation areas in v/mPFC are phylogenetically older and also tend to be motivationally dominant over the "colder" processing carried out in lateral PFC areas.

{\bf Active goal maintenance in v/mPFC:} The fact that the ACC and OFC share the same robust mechanisms for active maintenance and adaptive gating so important for lateral frontal areas, combined with their relatively privileged access to other ``limbic'' areas, makes them particularly powerful components of the goal processing system.  Thus, active maintenance can sustain a goal representation for the duration of its engagement.  Furthermore, LV-related goal-attainment probability tracking could be supported by a specialized activation-based temporal integration process in the ACC and/or OFC areas.  With incremental updating of new information into existing active representations, graded overall values such as the running average rate of reward or effort can be computed.  We describe preliminary models of our temporal integration approach below.  Particularly important, we believe, will be the temporal derivatives of these running averages, which could provide both phasic and tonic signals, e.g., whether you are in your {\em zone of proximal development}, or not --- too much success can often be boring, while too little tends to be frustrating \cite{Vygotsky78,CarverScheier90}.

{\bf Goal selection dynamics:}  Many potential factors shape the goal selection process, including internal state of the organism (e.g., level of hunger, thirst, energy), the state of the environment (opportunities and threats), existing goal state representations (higher-order goals, overall priorities, future deadlines, and other latent goal-like representations), and recent history of experience (which can serve to both activate and inhibit recently-activated goals --- you could want to repeat a recent positive experience, or become bored with something and seek to explore new options).  All of these factors interact with learned utility representations to select and engage goals to be actively pursued for the immediate time horizon.  Computationally, our models of this process build upon existing work that captures fundamental aspects of simultaneous bottom-up and top-down constraint satisfaction \cite{OReillyWyatteHerdEtAl13,OReillyMunakata00,OReillyMunakataFrankEtAl12}, along with fundamental aspects of higher-level executive function such as the ability to flexibly recombine existing representations to solve novel problems \cite{OReilly01,RougierNoelleBraverEtAl05,KrieteNoelleCohenEtAl13}.

{\bf Action selection under an engaged goal:} Once a particular goal state has been actively engaged in a single-minded way, the job at hand becomes how best to proceed so as to achieve that goal through some set of often sequential actions.  This can be variously straightforward or quite challenging (see \nopcite{PezzuloCastelfranchi09} for an extensive discussion).  In the straightforward case, there may already be a previously-learned set of instrumental sub-goal states and related action plans that, with input from the engaged overarching goal (including its perceptual associations) and the bottom-up perceptual input, results in a basal-ganglia mediated selection of the action most likely to move you closest to achieving that goal.  Critically, as that plan is carried out and instrumental sub-goals achieved, progress toward the goal can drive phasic dopamine bursts (via the LV, or learned value component of PVLV) that reinforce the activation (and pursuit) of those sub-goals for next time in similar circumstances (with appropriate abstract generalization built in by virtue of a parietal cortex that encodes actions in appropriately abstract and systematic ways that includes anticipated sensory outcomes).  Overall, the goal-first paradigm motivates the use of perceptual and motor representations that operate at a sufficient level of abstraction and compositionality that they can be flexibly recombined to produce novel action plans, formulated in advance through consideration of the current situation \cite{PezzuloCastelfranchi09,RizzolatiCamardaFogassiEtAl88}.  For example, the use of visual indexes may be one important part of this \cite{Pylyshyn00}.
Next, we present a computational model that embodies our principles, within the context of basic rat foraging behavior.  After describing this model, we elaborate on some of the core components of the overall framework, followed by a discussion of some outstanding issues and broader application of the framework to a number of different domains.

\section{Computational Model of Rat Foraging Behavior}

\begin{figure}
  \centering
  \includegraphics[width=6in]{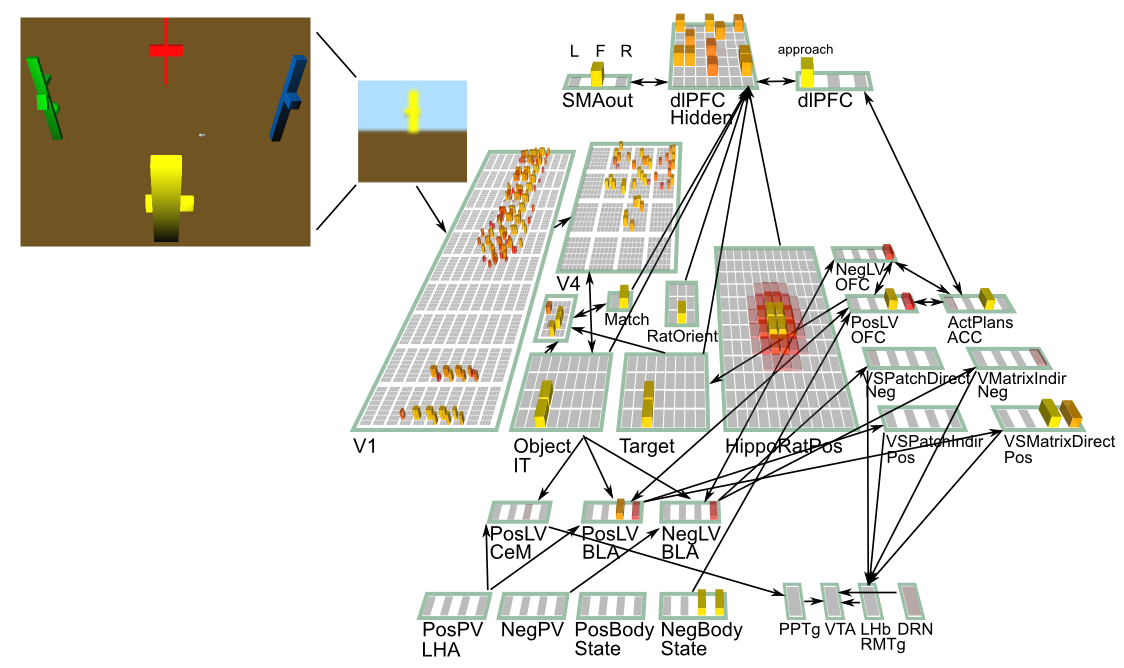}
  \caption{\small The rat (named ``emery'') foraging model and its plus-maze like environment, with four different possible primary value (PV) locations (red = meat, blue = water, yellow = sugar, green = veggies).  In addition to these positive PV outcomes, each location can also have negative outcomes: e.g., veggies can be bitter, and meat can be rotten.  For each run, emery experiences 2 states of negative body state depravation (e.g., thirst and need for vitamins from veggies), which drive goal selection.  Emery processes a bitmap first-person camera view into the environment, to extract an invariant IT Object representation.  Motor control learns to re-orient until this Object representation matches the Target, and then it approaches.  The ACC ActPlans units specify the 4 different targets to approach, coordinating the OFC PosLV representations that learn the specific Target features associated with different PV outcomes, with the specific dlPFC motor plan required (just approach target in this case).  The NegLV OFC representation encodes negative outcomes associated with different targets, and helps constrain the overall ACC action plan selection.  Once a target goal is selected, then those OFC and ACC states are gated in (via simulated Ventral Striatum (VS) gating), and maintained during the goal engaged period.  These top-down signals bias LV signaling via CeM and BLA to drive phasic dopamine in the VTA, which reflects increases in proximity to the desired PV outcome.}
  \label{fig.emery_net}
\end{figure}

To provide a comprehensive platform for current and future investigation of goal-driven behavior, we developed a relatively large-scale model that includes the ability to interact directly with a (virtual) environment through camera-level visual inputs, along with the full set of brain areas that we think are relevant for goal selection and goal-driven behavior in the goal engaged state (Figure~\ref{fig.emery_net}; see Appendix for full details).  We call the simulated rat in this model ``emery'', in reference to the {\em emergent} simulator we use to model it (\verb\http://grey.colorado.edu/emergent\).  Our initial tests presented here involve a simple plus maze environment, with four different locations (each of which is marked with a distinctive visual landmark) where primary value (PV) outcomes can be obtained.  The red landmark signals a location where meat pellets can be obtained, which satisfy emery's need for protein; blue signals water, satisfying thirst; yellow signals a sugar pellet, satisfying a need for calories; and green signals vegetables, satisfying a need for vitamins.  Each location can also have negative outcomes, such as an 80\% chance of experiencing bitterness associated with eating the veggies; a 70\% chance of the sugar pellet also having a sour taste; and 50\% chances of the meat pellet either having a sulferous or rotten taste.  These negative outcomes are intended to be simple proxies for all manner of possible negative costs associated with a given goal choice, such as increased effort, possibility of physical harm, etc.  The overall testing scenario is for emery to explore this environment, and learn to make choices in response to states of multiple depravation (e.g., need for vitamins and thirst) that optimize the cost-benefit tradeoff or {\em utility}.

After an initial brief period of purely random exploration, which enables the development of initial associations between different visual landmarks and PV outcomes (in various brain areas, including the CeM, BLA, and OFC), we begin a more systematic test of goal selection, which is then followed immediately by goal-engaged pursuit of the selected target outcome.  For example, if the negative body state of the model (which is intended to reflect basic brainstem level, likely hypothalamic, mechanisms) is driven by lack of vitamins and water (thirst), then these projections will activate corresponding positive outcome representations in the OFC (PosLV layer), which represent the PV's that satisfy these needs (e.g., veggies and water in this case).  These OFC representations (single localist neurons in this simplified model) have learned interconnections with corresponding ACC action plan value representations, encoding essentially the net utility associated with approaching the associated landmark (with the action plan itself encoded in the dlPFC layer that the ACC interconnects with).  This utility is also shaped by interconnections with the OFC negative outcome neurons (NegLV layer, which may correspond more accurately with the sub-genual ACC, area 25, as discussed later), which has also learned through exploration about the negative outcomes associated with these different actions.  Thus, the ACC action plan representations serve to coordinate multiple different OFC value representations, organized according to the relevant outcomes associated with different actions.  Critically, the negative outcomes have a net negative impact on the associated ACC representations, which enables them to reflect a net overall utility (positive excitation from PosLV minus inhibition from NegLV).

\begin{figure}
  \centering
  \includegraphics[width=4in]{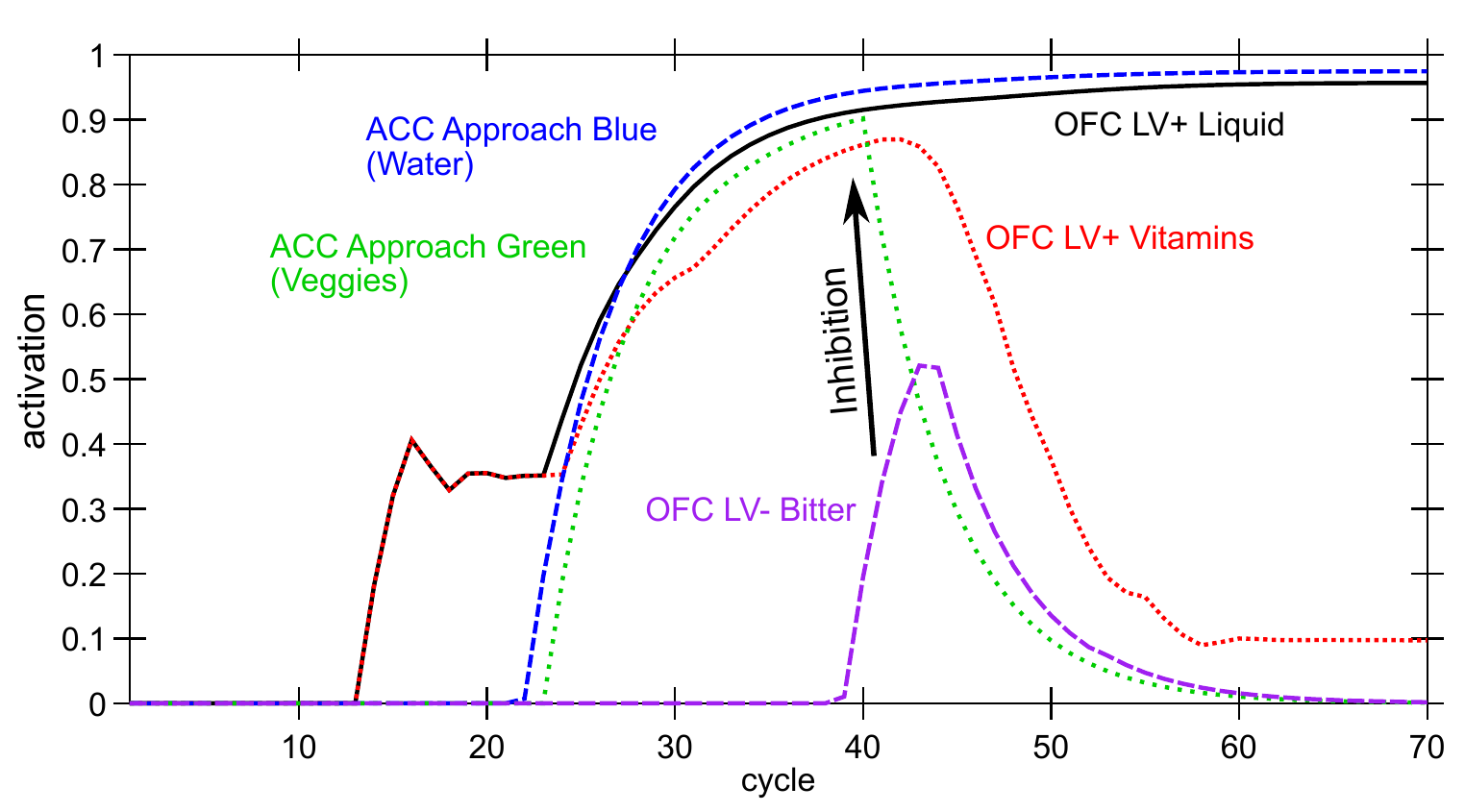}
  \caption{\small Activation dynamics in the OFC and ACC during the goal selection constraint satisfaction settling process, for choosing between veggies and water.  Initially, the OFC positive LV (LV+) representations are activated by the bottom-up negative body state drivers (thirst and need for vitamins), and this spreads to the corresponding ACC action plan value representations.  Once the ACC plan to approach green is activated, it drives the associated negative outcome in OFC LV- layer: bitterness (which is a proxy for all the other possible negative outcomes).  This negative value then feeds back inhibition to the corresponding ACC action plan, leading to its sharp decrease in overall comparative utility relative to the approach water plan, which has no such negative association.  Because neurons in each of these layers are competing with each other, even small overall differences can be magnified, to drive selection of the best overall option.  Once this process settles sufficiently, the winning patterns are gated into a robust active maintenance state, resulting in the transition to the goal engaged state.}
  \label{fig.emery_liq_vs_vit}
\end{figure}

Overall, the resulting dynamic is one of multiple constraint satisfaction among all the different value representations, and actions being considered (Figure~\ref{fig.emery_liq_vs_vit}).  Constraint satisfaction relies on bidirectional excitatory and inhibitory dynamics, and is a fundamental dynamic of cortical processing \cite{OReillyWyatteHerdEtAl13,OReillyMunakata00,OReillyMunakataFrankEtAl12}.  The role of frontal areas using constraint satisfaction processing to select among options has been explored in related computational and empirical work \cite{SnyderHutchisonNyhusEtAl10,SnyderBanichMunakata11}.  This process may result in a paring down of options to a few contenders that satisfy the most constraints.


As shown in the Figure, the constraint satisfaction dynamic between veggies and water is strongly influenced by the negative outcome of bitterness associated with the veggie action plan --- this serves as the decisive tie breaker between two otherwise roughly equally positive outcomes.  In general, competitive dynamics within the ACC and OFC layers can help drive selection of the highest utility option.  But this dynamic is also highly sensitive to the nature of the alternatives being considered --- this strong dependence of decisions on the set of alternatives is well studied in the decision making literature \cite[e.g.,]{SimonsonTversky92}, and more elaborated versions of this model could potentially be used to account for detailed data in that domain.  In any case, once the constraint satisfaction process settles out, the winning pattern of activation across the OFC, ACC, and dlPFC layers constitutes the selected goal representation, and it should then be gated in by the ventral striatum (VS) direct pathway neurons, that should then engage robust active maintenance in these areas \cite{OReillyFrank06,HazyFrankOReilly06,HazyFrankOReilly07}.  In the current model, we simplify by just programmatically engaging a form of robust active maintenance in these layers, but future versions will explore the dopamine-mediated learning dynamics that may govern this gating process, as detailed in our PBWM (prefrontal cortex, basal ganglia working memory) models \cite{OReillyFrank06,HazyFrankOReilly06,HazyFrankOReilly07}.

Once the selected goal becomes engaged, it has several important impacts on subsequent behavior, including driving appropriate motor actions to achieve the goal, and biasing phasic dopamine signals to reflect increments in progress toward this goal.  The motor action system incorporates a basic but powerful form of sensory-driven motor control, where the overall action plan is ``approach a target'', with the specific target being specified in another pathway.  This critically allows the very same motor program to automatically generalize to a very wide range of possible situations, which is of central importance when considering the needs of goal-driven action.  In particular, you need to be able to specify the action plan {\em in advance} of undertaking a given goal, so that you can properly evaluate the cost/benefit tradeoffs associated with that action plan.  Thus, these action plans must exist already, and having a small set of highly generalizable such plans provides an efficient solution to this problem \cite{PezzuloCastelfranchi09,RizzolatiCamardaFogassiEtAl88,Pylyshyn00}.  In our model, the specific target that you want to approach for a given goal is learned during the random exploration period, in the LV (learned value) portions of the network (from the CeM to the BLA and OFC).  Thus, the PosLV OFC representations for ``liquid'' are now capable of activating the corresponding sensory target representation for the blue landmark that was associated with the liquid PV outcome during initial exploration.

The dlPFC approach plan uses inputs from the LV target representation, together with bottom-up processing of the current visual input, which extracts a spatially-invariant object representation, to determine whether the object matches the target (in which case the rat should move forward) or alternatively that it should look around more until it does find a match to the target.  We implemented a scaled-down, simplified version of our recently-published object recognition system to produce this invariant Object representation, going through a primary (V1) and secondary (V4) visual processing pathway \cite{OReillyWyatteHerdEtAl13} (see Appendix for more details).  We also trained an explicit Match representation to compute whether the bottom-up stimulus Object layer matched the Target layer --- this could presumably be learned in the course of developing this generalized motor program, but we wanted to ensure that this information was readily available to drive more robust performance without too much additional training.  Both the visual object recognition and the approach target motor plan were trained in advance of the specific goal selection task, reflecting background developmental learning processes.


\begin{figure}
  \centering
  \includegraphics[width=3in]{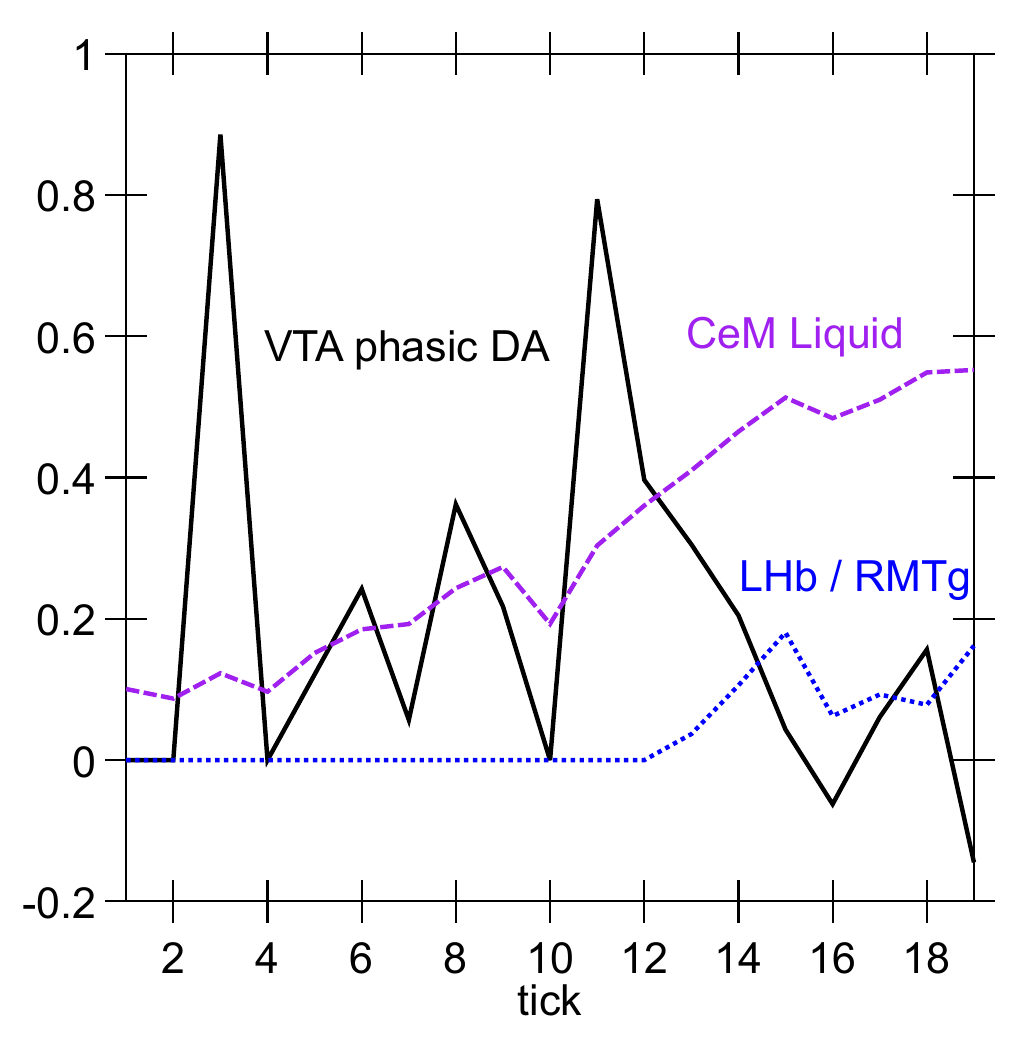}
  \caption{\small Phasic dopamine (in the VTA) driven by progressive approach toward the learned PV location associated with liquid water outcomes, as reflected in increasing CeM activation.  The LHb / RMTg activity reflects learned anticipation of PV outcomes as a result of projections from the ventral striatum (PVi), the latter also serving to directly shunt phasic DA bursting that would otherwise occur at the time of PV outcome receipt.}
  \label{fig.emery_appr_liq_da}
\end{figure}

Thus, a typical run involves emery making a sequence of rotations (scanning behavior) until the target landmark associated with the selected goal comes into view, followed by a sequence of forward approach steps until the PV delivery location is reached, directly in front of the visible landmark.  During this approach process, the increasing proximity to the learned PV outcome location drives learned LV associations in the CeM (and BLA) from both the visual pathway (particularly the intermediate V4 level, which has some degree of feature abstraction, but retains also significant view specificity), and a simple proxy for hippocampal place-cell representations in the HippoRatPos layer.  These increasing CeM LV values then drive phasic dopamine firing in the VTA, in proportion to the local gradient (temporal derivative) (Figure~\ref{fig.emery_appr_liq_da}).  This phasic dopamine signal is then available to train the specific motor actions that lead to increased proximity to the desired goal --- this training ultimately results in the formation of sensory-motor habits that can optimize well-learned goal approach behaviors.  More details on this gdPVLV phasic dopamine firing are provided in a subsequent section.

Once the primary value outcome is achieved, it drives learning in all of the relevant areas, from CeM to BLA to OFC --- these areas all learn to link the sensory cues and other internal representations that are currently active with the corresponding PV values that were experienced (this is why we refer to these representations as LV or learned value representations, even though they also receive a strong primary drive from PV inputs).  This PV receipt also results in the satiation of the associated body state signal(s), and potentially the destabilization of the existing OFC, ACC, and dlPFC goal representations, if the degree of PV experience was sufficiently satisfying.  In the model, we simply assume that each PV outcome experience is sufficiently satisfying, and we start the process over again with a new random set of negative body state representations, which then drive a new round of goal selection, to satisfy those needs.  All the while, the model continues to strengthen its knowledge about expected PV outcomes, and sensory stimuli associated therewith, so that the goal selection process can be more well-informed and accurate.

In summary, this model demonstrates the functioning of the core mechanisms hypothesized in this paper, in the context of an embodied sensory-motor system that can enable many different behavioral scenarios to be examined, in a realistic manner.  Much work remains to be done to explore specific behavioral paradigms and associated neural recording and other relevant neural data that can constrain and inform this model.  Here, we address a few specific examples, but a whole research program is required to more fully test the novel predictions of this overall framework.  We outline some of these predictions in the discussion section as well.

\subsection{Testable Predictions of the Model}

As explicated in the remainder of the paper, our model makes contact with a very broad range of data, at many different levels of analysis, from detailed neuroscience up to salient features of daily human life.  However, there are a few specific predictions that our model makes that are particularly useful for highlighting some of its relatively unique properties, which we discuss here.

{\bf Ventral Striatum (VS) activity should be concentrated at the start and end of tasks, whereas Dorsal Striatum (DS) tracks online action selection during task performance more directly.}  This prediction is based on the fact that VS provides the discrete working memory updating signals for the goal maintenance areas in the OFC and ACC, and this should occur at the transition from goal selection to goal engagement (to lock in the selected goal representations), and then again to deactivate goals upon goal achievement, and enable a new round of goal selection.  In contrast, dorsal areas of the striatum are involved in selecting more specific online action plans as a task unfolds under a given overall goal plan.  Data across two studies using the same T maze paradigm nicely confirms this key prediction.  \incite{AtallahMcCoolHoweEtAl14} recently found that principal neurons in the VS fire phasically at the very start of the maze, and gradually ramp up firing near the end.  In contrast, DS neurons fire at various points during the running of the maze, in particular at the choice point \cite{ThornAtallahHoweEtAl10}.  Note that these striatal data are more diagnostic than direct recordings of OFC and ACC, which are likely to exhibit ongoing and updated firing throughout the task.

{\bf Integrating reward and cost values (e.g., effort, delay, or risk) depends on the ACC and its interconnectivity with the BLA and VS.}  In our model, the ACC provides the nexus for integrating the different value representations encoded elsewhere, to represent an overall utility reflecting rewards and other positive outcomes weighed against costs and other negative outcomes.  This is consistent with several lines of data.  For example, integrity of the ACC appears particularly crucial for choosing effortful options, as ACC lesions compromise the ability to choose rewards with high effort costs \cite{WaltonBannermanAlterescuEtAl03,RudebeckWaltonSmythEtAl06}. 

Dopamine depletion also reduces effortful choices, and rats with lesions of the BLA or NAc also show reduced preferences to choose high reward options paired with effort costs over low reward, less effortful options \cite{SalamoneCousinsSynder97,FlorescoStOngeGhods-SharifiEtAl08}.
Lesioning the connection between BLA and ACC leads to similar difficulties with choosing the high reward arm paired with high effort \cite{FlorescoGhods-Sharifi07,FlorescoStOngeGhods-SharifiEtAl08}.  In the domain of risky decision making, lesioning the top-down connection from ACC to BLA caused increased probability of choosing a large but risky option, and reduced ability to shift choices after omissions, supporting the idea that the ACC is crucial for integrating and updating this risk-based value information \cite{StOngeStopperZahmEtAl12}. Similarly, ACC lesions impaired decision making on a dynamic foraging task that involved flexible integration of risk and reward information \cite{KennerleyWaltonBehrensEtAl06}. In addition, electrophysiological data shows that neurons in the ACC flexibly integrate multiple decision-making attributes, such as reward, risk and effort into a common choice value that is used to guide decisions \cite{KennerleyBehrensWallis11,KennerleyDahmubedLaraEtAl09}.

{\bf OFC-mediated goal progress signals are important for sustaining goal engagement in the absence of other cues, whereas explicit sensory cues can suffice without an OFC.}  We hypothesize that the OFC represents the highest level of the LV goal-progress tracking system, which is uniquely capable of using active maintenance to sustain estimates of progress even in the absence of external sensory inputs.  Consistent with this prediction, OFC lesions lead to increased impulsive choices in some tasks \cite{RudebeckWaltonSmythEtAl06,MobiniBodyHoEtAl02}, but other studies found decreased impulsive choices with OFC lesions \cite{ZeebFlorescoWinstanley10,WinstanleyTheobaldCardinalEtAl04}. Critically, these studies differed in the presence of stimuli that could have acted as cues to ongoing progress toward the goal \cite{FlorescoStOngeGhods-SharifiEtAl08,ZeebFlorescoWinstanley10} --- the OFC was only important when these cues were absent.

{\bf OFC is important for outcome-based action selection, e.g., in outcome devaluation and related paradigms.}  This prediction is well established \cite{BalleineDickinson98,CardinalParkinsonHallEtAl02,YinKnowltonBalleine06,ValentinDickinsonODoherty07,PauliClarkGuentherEtAl12}, and consistent with multiple models, so we only raise it to flag that this literature could provide a rich dataset for more detailed modeling in future work.  The standard finding is that an intact OFC (and associated ventral striatum) is important for an animal to overcome habitual responses that result in delivery of food or other outcomes that are no longer rewarding (e.g., due to an added unpleasant bitter flavor).  In our framework, a hungry animal selects an active goal based on available information, and the constraint satisfaction process in OFC and other areas is critical for choosing a goal that integrates all the current available information.  Thus, in the absence of the OFC, the fact that the food is now aversive cannot be integrated into the goal selection process, and more primitive lower-level systems drive behavior toward the default outcomes given the state (i.e., when hungry, seek food).  Furthermore, even more basic habitual associations between stimuli and responses, encoded in dorsolateral striatum and SMA-level motor control areas, can drive behavior even in the absence of any engaged goal at all.  One interesting question we plan to explore is the extent to which goal-driven behavior can be sustained in the absence of the OFC in mammals, e.g., relying on the BLA and CeM, or if everything truly is purely habitual, as is characterized in the existing literature.

{\bf Value function dissociations between goal selection and goal engaged modes.}  The most important broader predictions from our model, for which we are not aware of direct empirical tests, are about the dissociations in value functions between the goal selection (conservative weighing of costs and benefits) versus goal engaged (dominated by progress toward enaged goal) states.  One possible reflection of this dissociation comes from data on monkeys performing tasks that required 1, 2 or 3 steps of a mildly effortful task prior to obtaining a reward \cite{LiuRichmond00}.  They found that the monkeys starting step 2 of the 3-step task were much more accurate and faster than monkeys starting step 1 of the 2-step task.  This is despite the fact that both of these cases are equally distant from the subsequent reward outcome.  We could interpret the difference as being due to the extra conservative value function at the point of starting the two step task, compared to the goal-engaged {\em momentum} from already being engaged in the three-step task.  Although this data is potentially encouraging, we plan to develop more comprehensive experimental paradigms to more conclusively test this prediction in the future.

In the next section, we explicate in more detail the gdPVLV component of the model, followed by a broad outline of data relevant to the larger-scale map of goal-relevant representations throughout the brain, up through the v/mPFC areas (including OFC and ACC).

\section{Goal-Driven Primary Value, Learned Value (gdPVLV)}

Over the past decade, we have been developing a computational model of the biology of phasic dopamine responses to primary rewards (unconditioned stimuli, USes) and conditioned stimuli (CSes), called PVLV (primary value, learned value) \cite{OReillyFrankHazyEtAl07,HazyFrankOReilly10}.  This model is based on the classic Rescorla-Wagner model (RW; \nopcite{RescorlaWagner72}), extended to address the inability of the RW model to deal with phasic dopamine responses to CSes.  The widely-cited temporal differences (TD) model \cite{HollermanSchultz98,MontagueDayanSejnowski96,SchultzDayanMontague97,Schultz98,SchultzDickinson00,Sutton88,SuttonBarto90,SuttonBarto98,WaeltiDickinsonSchultz01} addressed this problem through a backward chaining process where each step in time attempts to predict the reward prediction at the next step.  However,  this chaining process is inherently unreliable in real world situations \cite{OReillyFrankHazyEtAl07} and does not connect very directly with the known biology.  The PVLV model instead adopts an associative learning mechanism (the LV mechanism) to account for the CS-triggered firing, within an essentially RW paradigm.  This model assigns empirically-supported computational roles for a variety of brain areas, including the medial segment of the central nucleus of the amygdala (CeM), the ventral striatum (VS; principally the Nucleus Accumbens, NAc), the lateral hypothalamic area (LH), and the principle dopamine cell-containing areas the ventral tegmental area (VTA) and the substantia nigra, pars compacta (SNc). 

Here, we outline a reframing of the PVLV algorithm from the goals-first point-of-view, which retains all of the major features and explanatory power of the original model, while providing a broader interpretation of its essential computational features (captured succinctly in Figure~\ref{fig.pvlv_food}).  For clarity, we call this new goal-driven framework {\em gdPVLV}.  Instead of confining the framing to the rather limited stimulus-then-reward Pavlovian paradigm, we now adopt the much broader --- and more ecologically valid --- paradigm of foraging, as illustrated in our model.  Here, landmarks become the ecological versions of conditioned stimuli (CSes), which thus assume the role of driving the LV system in PVLV.  Natural landmarks provide a reliable source of graded perceptual cues for how far the organism is away from its current goal (e.g., finding food if it is hungry).  Going berry picking in a reasonably large and variably berried terrain clearly fits in the category of addictive incremental reward tasks, and could plausibly be the evolutionary prototype for all that follows this template.  The rich literature on foraging strategies in different animals and situations should provide many additional insights into the nature of the goal-driven learning system \cite[e.g.,]{KamilRoitblat85,Pyke84,MenzelGiurfa01}.  Also, this  ``directional'' aspect of dopaminergic function is consistent with the framework of \incite{SalamoneCorrea12}, and the notion that the mesolimbic dopamine system and other limbic areas coordinate approach towards rewarding stimuli \cite{Ikemoto10}.

We start by reformulating the learned value (LV) aspect in PVLV as providing estimates of the probability of specific goal (PV) achievement, through learned connections with perceptual inputs.  Then, we describe the different mechanisms involved in learning from and processing primary values (PVs).  Overall, the core new insight incorporated into the gdPVLV framework is that a goal-driven system needs to perform two distinct types of computations.  The LV computation provides a {\em prospective} or {\em anticipatory} monitoring capacity to estimate how close you are getting to a given goal.  It is strictly an estimate, based on the best available data, but it cannot substitute for actual goal attainment --- it is fallible, but better than nothing, and can help drive action learning and provide feedback about whether it seems reasonable to continue pursuing the goal or not.  In contrast, the PV system represents real concrete outcomes that unequivocally indicate progress toward a goal, including the state of achieving a given goal (which engages goal deactivation mechanisms). The PV system anchors everything that is learned for computing LV values, and there are mechanisms that provide more specific estimates of PV values, so that discrepancies can be detected (including the critical case of detecting when an expected PV fails to materialize at all).


\subsection{Learned Values (LV) for Graded Goal Probability Estimation}

\begin{figure}
  \centering\includegraphics[height=2in]{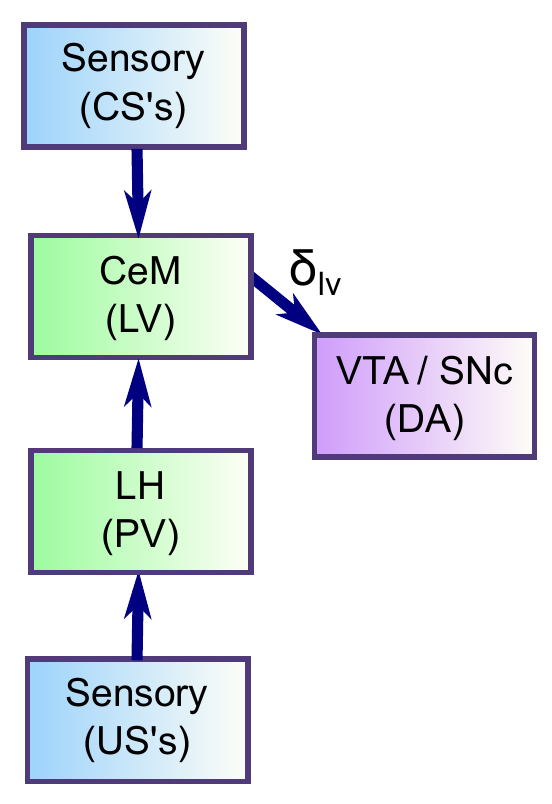}
  \caption{\small LV (learned value) system in the medial central nucleus of the amygdala (CeM) associates stimuli with primary values (PV) driven by learning at the time of PV activation, signaled by bottom-up drive from nuclei such as the lateral hypothalamus (LH) for positive-valenced PVs (e.g., food, water).  The LV system drives phasic bursts or dips in the dopamine nuclei (VTA \& SNc) according to the temporal derivative (difference) in LV activation over time.}
  \label{fig.gdpvlv_lv}
\end{figure}

The core equation for the LV aspect of gdPVLV directly codifies the notion that dopamine bursts and dips (though the latter actually involves additional, separate neural mechanisms as described below) signal changes to an accumulating, perceptually-driven estimate for the probability of goal achievement:
\begin{equation}
  \delta_{lv} = \sum_g g(t) \left( LV_g(t) - LV_g(t-1) \right)
  \label{eq:lv}
\end{equation}
where $LV_g$ can be interpreted as a learned current-best estimate of the probability that specific goal $g$ will occur within the immediate time horizon (and modulated by current goal activation level $g(t)$, which reflects top-down activation from goal maintenance areas modulating responsiveness of the dopamine system).  $LV_g$ is not strictly a properly normalized probability however --- it just needs to provide graded values that have a maximum at time of actual goal satisfaction, and a minimum presumably at a distance infinitely far away from the goal.  Biologically, the phasic dopamine signal has been shown to be normalized and rescaled in order to optimize its dynamic range \cite{ToblerFiorilloSchultz05} --- and this normalization is reflected in the LV-associated dopamine signals as well. The important thing is that $LV_g$ have a reasonable, monotonic relationship to goal proximity. Theoretically, there can be multiple different goals operating at once and so Equation (\ref{eq:lv}) is a summation.  Each separate goal can have its own complicated metric space that calculates separate $LV_g$ values.  Nonetheless, changes to each $LV_g$ ultimately get boiled down to a {\em common currency} at the level of phasic changes in dopamine firing levels, represented by the $\delta_{lv}$ value.

The delta function (a discretized temporal derivative) is evaluated at two different points in time: the current time $t$ relative to a previous time interval (with a behaviorally relevant time scale presumably in the 100's of millisecond range).  Relative to the original PVLV formulation, this $LV_g$ value is equivalent to the $LV_e$ value (e = excitatory) --- in the interest of avoiding subscript overload we drop the e = excitatory / i = inhibitory terminology in this discussion, although the fundamental juxtaposition of competing excitatory and inhibitory drivers for both PV and LV still prevails and is a useful way to think about what is actually going on.

Biologically, at the lowest level of the goal pyramid, we map the appetitive $LV_g$ function to the activation of glutamatergic neurons in the medial segment of the central nucleus of the amygdala (CeM) that respond natively to appetitive primary rewards such as food \cite[e.g., ]{FukudaOno93,LeeGallagherHolland10,KnapskaWalasekNikolaevEtAl06,OnoNishijoUwano95,OnoTamuraNishijoEtAl89}, and through learning come to fire in response to conditioned stimuli (CSs) that come to be associated with those PVs (unconditioned stimuli; USs; \nopcite{OnoNishijoUwano95}).  CeM neurons have been shown to have direct excitatory projections onto the dopamine cells of both VTA and SNc and, even more densely, to the pedunculopontine tegmental (PPTg) nucleus which itself projects strongly to both VTA and SNc \cite{FudgeHaber00,WallaceMagnusonGray92,TakayamaMiura91}.  The PPTg is a complex and heterogeneous structure, still poorly understood, but has been shown to be a potent driver of phasic dopamine cell bursting \cite{FlorescoWestAshEtAl03,GraceBunney84,LodgeGrace06,PanHyland05}.  Available data is consistent with the idea that a highly-tuned dynamical interaction within a subnetwork that includes the CeM, the PPTg, the lateral dorsal tegmental (LDTg) nucleus \cite{LodgeGrace06b} and the dopamine cells themselves is responsible for enabling a pattern of bursting that acts like a positive-rectified temporal derivative of $LV_g$.  There are also direct excitatory projections from lateral hypothalamus (LH) to the CeM \cite[e.g.,]{AmaralVeazeyCowan82,Ottersen80}, and to the VTA/SNc both directly \cite{Phillipson78}, and indirectly via the PPTg  \cite{SembaFibiger92}. Thus, as a source of PV signals occurring at the time of the US, LH afferents provide a local training signal for CeM neurons to pair CSs with the USs they predict \cite{HazyFrankOReilly10,OnoNishijoUwano95}. Figure~\ref{fig.gdpvlv_lv} summarizes the key components of the appetitive LV system as currently proposed.


The overall model for LV (CeM) learning is that it is driven by the conjunction of a newly arriving PV (unconditioned stimulus; US) signal (probably from the LH), along with the phasic dopamine burst triggered by the US.  Critically, in the absence of this specific conjunctive PV learning signal, the LV neurons cannot learn and just continue to report a graded sensory response in proportion to how close the current perceptual inputs are to those stimuli present during prior learning situations when the PV input was active.  
A simple mathematical equation expresses this learning rule:
\begin{equation}
  \Delta w_{ig} =  g \, \epsilon \, \delta_{pv} \left( {PV}_g - {LV}_g \right) x_i \; \; \mbox{iff} \; {PV}_g > 0
  \label{eq:lv_dw}
\end{equation}
where $x_i$ is a sensory input to the LV unit, $g$ is goal activation, $\epsilon$ is a learning rate, $\delta_{pv}$ is the phasic dopamine signal driven at the time of the PV (see below).  $PV_g$ is an actual PV outcome received (or some transform thereof to increase dynamic range), and the {\em iff} (if and only if) statement constrains learning to when the primary value is actually present (all values assumed to be at the present time, so time dependence is assumed but not notated to keep things simpler).  This PV constraint provides grounding for LV firing so that it must be continually ``validated'' by real primary values, and cannot take on a life of its own.

The contribution of phasic dopamine at the time of the PV in this equation ($\delta_{pv}$) is important for two functional reasons, and is consistent with the fact that both the CeM and BLA are richly innervated by dopamine fibers from the VTA \cite{FallonCiofi92,AmaralPricePitkanenEtAl92}.  First, it is necessary to account for the blocking effect \cite{Kamin69,Kamin69a} and conditioned inhibition (\nopcite{RescorlaWagner72}; see \nopcite{WaeltiDickinsonSchultz01} and \nopcite{ToblerDickinsonSchultz03}, respectively  for a full theoretical discussion).  For example, in the blocking effect, an existing learned association predicts the occurrence of the PV, so that the phasic dopamine signal is zero, preventing new learning of a newly introduced stimulus that redundantly predicts the PV.  Second, this term, when negative, enables the weakening of afferent $LV_g$ synaptic weights when they ought to go down as, for example, when an expected PV later comes to reliably fail to materialize due to changed contingencies, generating phasic dopamine dips.

As discussed at length in earlier PVLV papers \cite{HazyFrankOReilly10,OReillyFrankHazyEtAl07}, an important implication of the LV learning equation is that phasic dopamine bursts, once acquired by the LV system (equation~\ref{eq:lv}), must not be allowed to feed back upon themselves and drive further weight strengthening to the LV itself.  If it did, this would give rise to a runaway positive feedback loop not grounded in actual PV values.  Instead, LV-generated dopamine signals can only be allowed to train {\em other} contingencies as, for example, learning to initiate actions in response to CS signals that increase the chances of receiving reward. 

Mathematically, the LV phasic dopamine equation (\ref{eq:lv}) is essentially the simple Y-dot (temporal derivative) formulation explored by Sutton and Barto prior to their developing the full TD algorithm \cite{SuttonBarto81,SuttonBarto90}.  The Y-dot formulation was adopted by us in the second version of our PVLV algorithm \cite{HazyFrankOReilly10} as somewhat of a technical convenience since we had not yet fully worked out a mechanism for phasic dopamine bursting.  Furthermore, we had not yet fully appreciated the powerful significance of this formulation, which the new goal-centric framework and the foraging paradigm serve to highlight. Importantly, unlike Sutton \& Barto's original Y-dot formula as well as their subsequent TD algorithm, both of which propose a single mechanism covering all aspects of reward learning, we emphasize that there are important and distinct computational functions not performed by the LV side of PVLV system, resulting in the need for additional mechanisms on the PV side of the system.

In particular, the LV equation alone cannot handle the problem of knowing when an expected PV (goal achievement) has not occurred.  It is essential that the LV signal be driven by perceptual inputs in advance of any receipt of the PV, but the consequence of this is that the absence of a PV cannot inhibit LV firing.  Instead, a separate system must drive an active signal for the absence of an expected PV --- this is the job of the PV side of the PVLV system, which learns to detect all manner of deviations in the timing, magnitude and probability of PVs relative to expectations --- these deviations from expectation can be important clues about changes in the environment or performance that can affect the goal and action systems.  It is well established that the dopamine response at the time of the US (PV) is indeed sensitive to these kinds of deviations from expectations \cite[e.g.,]{Schultz98,Schultz02,ToblerFiorilloSchultz05}.

\subsection{Detecting Primary Value Discrepancies}

At a purely abstract level, the core PV dopamine driving equation can be thought of as a Rescorla-Wagner comparison between expected and obtained primary values:
\begin{equation}
  \delta_{pv} = \sum_g g \left( PV_g - \PVest_g \right)
  \label{eq:pv}
\end{equation}
where $\PVest_g$ represents an estimate of the $PV$ value expected for every timestep, and the net delta value reflects any discrepancy between the actual and estimated $PVs$ (all values assumed to be at the current time point), driving phasic changes in dopamine firing.

As in the LV, a simple mathematical expression captures the basic learning rule for the weights coming into the $\PVest_g$  representation: 
\begin{equation}
  \Delta w_{ig} = g \epsilon \left( {PV}_g - \PVest_g \right) x_i
  \label{eq:pv_dw}
\end{equation}
which differs from the LV learning equation in lacking the ``iff'' conditional that restricts LV learning to only those times when the PV is present. (It also lacks a dopamine term.)  Thus, $\PVest_g$ representations are constantly learning about the relative presence or absence of PV inputs for each timestep and so enable very precise expectations regarding the timing, probability, and magnitude of individual PVs.  This learning is based on a wide range of sensory inputs that support the ability to generate these expectations (represented by the $x_i$ inputs in the above equation).  Comparing the LV and PV learning equations further, it is clear that there are different demands for each form of learning and that neither can be subsumed by the other.  Relative to previous PVLV versions which used versions of the RW framework for both LV and PV, contrasting the new goal-driven LV equation as a pure Y-dot calculation against the standard RW formulation used in the PV equation makes the distinction particularly clear and compelling.

\begin{figure}
  \centering\includegraphics[width=3in]{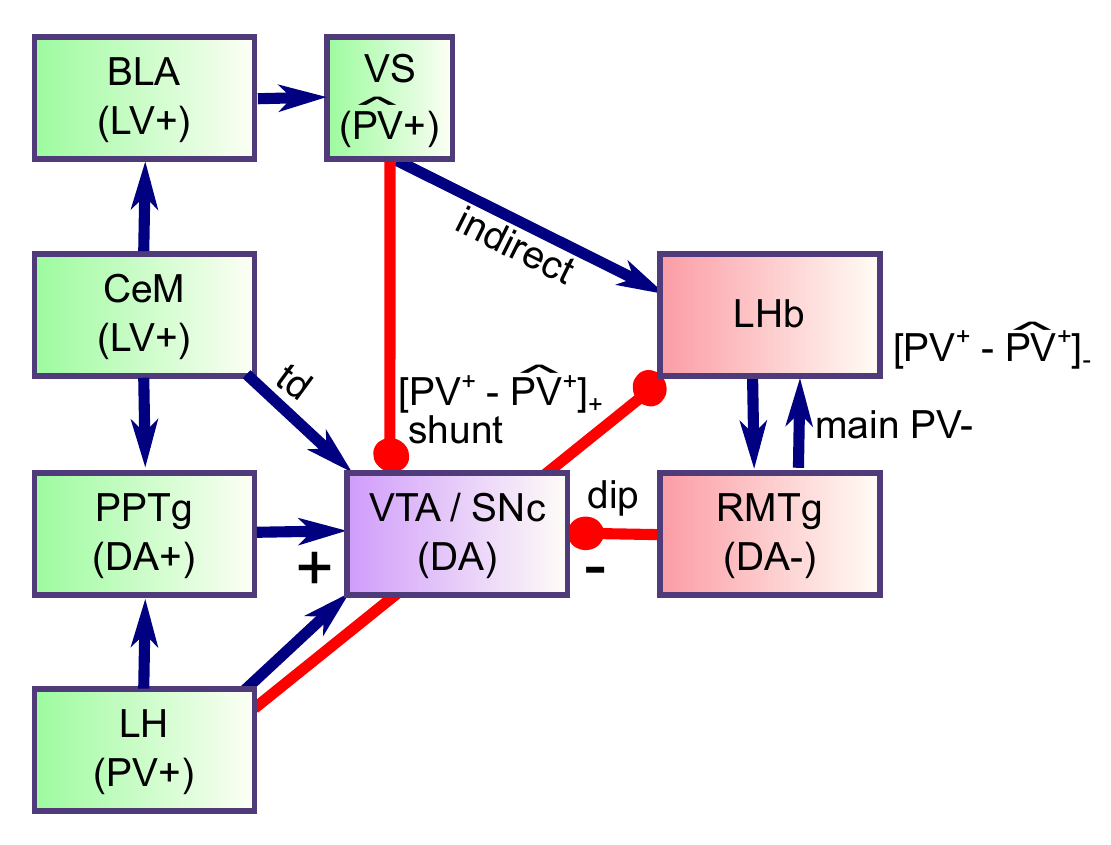}
  \caption{\small Neural pathways involved in the PV system, comparing the actual received PV signal with the estimate $\PVest$.  The ability of the LHb (lateral habenula) to drive dips in dopamine firing can compute $[\pPV - \pPVest]_-$: a positive outcome that doesn't live up to the expectation or is missing altogether.  The direct inhibitory projections from VS to the dopamine areas, compared with direct drive from $\pPV$, can account for $[\pPV - \pPVest]_+$: a dopamine burst to the extent that the expectation is below the actual positive outcome value (and a cancellation when the expectation matches).}
  \label{fig.pv_lhb_posonly}
\end{figure}

Biologically, we map the $\PVest_g$ representations to neurons in the striosomal compartment of the ventral striatum (VS), particularly those of the nucleus accumbens (NAc; Figure~\ref{fig.pv_lhb_posonly}).  Because the ventral striatum preferentially receives afferents from all ``limbic'' areas, importantly including a dense projection from the basolateral amygdala (BLA; \nopcite[e.g.,]{ChararaGrace03,CrittendenGraybiel11,Floresco07,FlorescoBlahaYangEtAl01,HowlandTaepavaraprukPhillips02}), it is clearly well positioned for acquiring $\PVest_g$ representations. The empirical evidence strongly suggests that the brain uses these representations in two distinct ways (involving partially separate pathways) for dealing with discrepancies in the positive vs. negative direction (Figure~\ref{fig.pv_lhb_posonly}).  This is a consequence of the fact that {\em phasic} dopamine cell bursting and constant, low-frequency (~3-5 Hz) {\em tonic} firing are under the control of largely separate afferent systems \cite{GraceBunney84,FlorescoWestAshEtAl03}.  As a result, the prevention of {\em bursts} and the generation of {\em dips} each depend upon largely separate mechanisms as well, as we summarize here:
\begin{itemize}
\item {\bf Shunting Bursts:} Striosomal neurons of the VS make direct, monosynaptic GABA-ergic contacts onto dopamine neurons across all of the VTA and SNc \cite{Gerfen85,GerfenHerkenhamThibault87,JoelWeiner00,SmithBolam90}.  Thus, the (anticipatory) activation of these neurons can directly shunt excitatory inputs produced by PV occurrence, and once $\PVest_g$ representations are fully learned they will prevent any burst-firing that would otherwise occur.  Any PV magnitude greater than that currently coded in a $\PVest_g$ representation can still produce a partial bursting signal that reflects the amount of upside surprise.  However, these shunting GABA-ergic inputs can not produce overt pauses in tonic firing.

\item {\bf Driving Dips:} $\PVest_g$ representations in the VS can also support the detection of expectation violations and enable the triggering of dopamine dips.  The biological machinery for this leverages the neural pathways by which negatively valenced outcomes such as pain can directly produce pauses in dopamine firing.  $\PVest_g$ VS neurons send projections via the pallidum (probably collaterals of the burst-preventing projections just-described, but not necessarily) that drive the lateral habenula (LHb) with expectations about the magnitude and precise timing of PVs \cite{HongHikosaka08,JiShepard07,MatsumotoHikosaka07}.  Via a separate pathway, hypothesized by us to also involve the VS and pallidum \cite{ApicellaLjungbergScarnatiEtAl91,Bromberg-MartinMatsumotoHongEtAl10,HongHikosaka08}, the LHb is also driven by the PV itself.  If the expected PV fails to occur, or is weaker than anticipated, the uncancelled remainder of the $\PVest_g$ signal increases LHb firing, triggering dopamine dips via a projection to the rostromedial tegmental nucleus (RMTg) (also known as the tail of the VTA) \cite{BourdyBarrot12,HongJhouSmithEtAl11,JhouFieldsBaxterEtAl09}.
\end{itemize}

\subsection{Negative Primary Values}

Because the goal-driven framework stipulates that proper goals are fundamentally appetitive in nature, we do not need to be directly concerned here with the complexities of what happens to the dopamine system for negatively-valenced outcomes, such as pain.  As summarized in Figure~\ref{fig.pv_list} and elaborated below, negatively valenced affective states can serve to motivate the selection of appropriate goals to overcome these negative states, but it makes no sense to say that negative-outcome states can represent a desired outcome of any active goal --- indeed, they act rather as ``anti-goals'' in this case.  It is known that negative outcomes (e.g., pain onset, or an aversive air puff to the eye) can reliably drive dopamine dips (at least for most dopamine cells) through direct activation of the RMTg and/or the LHb as described above \cite{JiShepard07,LecourtierDeFrancescoMoghaddam08,MatsumotoHikosaka07,ShepardHolcombGold06,StamatakisStuber12,LammelLimRanEtAl12}.  These dips can contribute to avoidance learning as, for example, to avoid movements or actions that result in pain outcomes.  Furthermore, integrated experiences of pain and other aversive states should logically be accumulated over the course of action and goal engagement, so that those kinds of cost factors can be factored into subsequent goal and action selection in the future. 

It is well documented that negative outcomes drive various activation and arousal mechanisms, including the recent discovery of a small sub-population of previously unrecognized dopamine cells in medial posterior VTA that receive direct glutamatergic input from the LHb and respond to noxious stimuli with brisk and unambiguous bursting
\cite{BrischouxChakrabortyBrierleyEtAl09,LammelIonRoeperEtAl11,LammelLimRanEtAl12}.  In addition, the offset of a sustained painful stimulus can trigger rebound burst firing in the majority-population of dopamine cells that are paused by the pain in the first place \cite{BrischouxChakrabortyBrierleyEtAl09} and there is compelling (if still indirect) evidence that the active avoidance of an expected shock may trigger bursting as well at the time of the expected-but-avoided shock \cite{OlesonGentryChiomaEtAl12}.  Both of these relief-like signals are likely important in providing a {\em positive} reinforcement signal for potential escape/avoidance behaviors \cite{OlesonCheer13}.  Further, the ability to trigger dopamine bursts by this kind of mitigation of aversive primary PVs can sometimes be transferred to a sizeable subset of the LV-type cells that reliably predict the occurrence of the aversive PV in the first place \cite{MatsumotoHikosaka09b}.  This makes sense because the same stimulus that predicts the aversive event logically also will predict its offset and/or can learn to predict its successful prevention in the case of avoidance. These positive reinforcement signals then can play a role in strengthening actions that mitigate the threat.      

Finally, these various activating mechanisms are undoubtedly important for engaging systems critical for threat mitigation, including potentially the proactive strategy of the activating appetitive systems of goal selection so as to find positive ways of escaping or avoiding the negative situation --- e.g., looking for a tree to climb to get away from a tiger.   Once such an appetitive-type goal has become activated, the ``common currency'' for dopamine signaling reflected in the equations and dynamics of gdPVLV as described above can be fully leveraged so as to learn the best actions to take under the circumstances.  We are in the process of incorporating the most important of these additional mechanisms into the gdPVLV framework in order to better accommodate negatively-valenced signals.

\subsection{Relationship between gdPVLV and TD}

Many of the important differences between the PVLV framework and the widely-adopted temporal differences (TD) learning were reviewed in \incite{HazyFrankOReilly10} and remain applicable.  However, the more central role of the Y-dot style temporal derivative in the gdPVLV formulation provided here brings the framework in somewhat closer contact with the TD model, at least at a mathematical level.  In particular, our model of the LV component of gdPVLV will generate dopamine bursts that are essentially equivalent to a TD model with state value estimates $\hat{V(s)}$ that happen to produce a progressively increasing value as one gets closer to the goal.  Furthermore, after the chaining-based TD learning process has converged on a stable set of state-value associations for a stable reward location in a given space, it should in fact produce this gradient of increasing state values as one approaches the goal.  This is indeed the classic result of TD applied to a simple grid-world navigation environment \cite{SuttonBarto90}.

The key difference is that the gdPVLV model does not have to rely on a time-consuming chaining process to learn individual state values --- it instead learns only at the time of PVs, and relies on perceptual generalization functions and other dynamics (e.g., temporal integration in activation-based working memory representations) to produce a graded estimates of distance to the goal.  Because it is biased for this specific form of goal-driven learning, the gdPVLV algorithm is less general than the TD algorithm, but this bias is advantageous for the conditions under which it applies.  This is a standard tradeoff in machine learning algorithms \cite{GemanBienenstockDoursat92}.    Furthermore, as detailed above, the PVLV framework provides specific functional roles for different neural systems that are compatible with the known biology, while the TD framework is more monolithic and does not map obviously onto this menagerie of biological substrates.

\section{A Goal-Processing Pyramid in the Brain}


\begin{figure}
  \centering\includegraphics[width=5in]{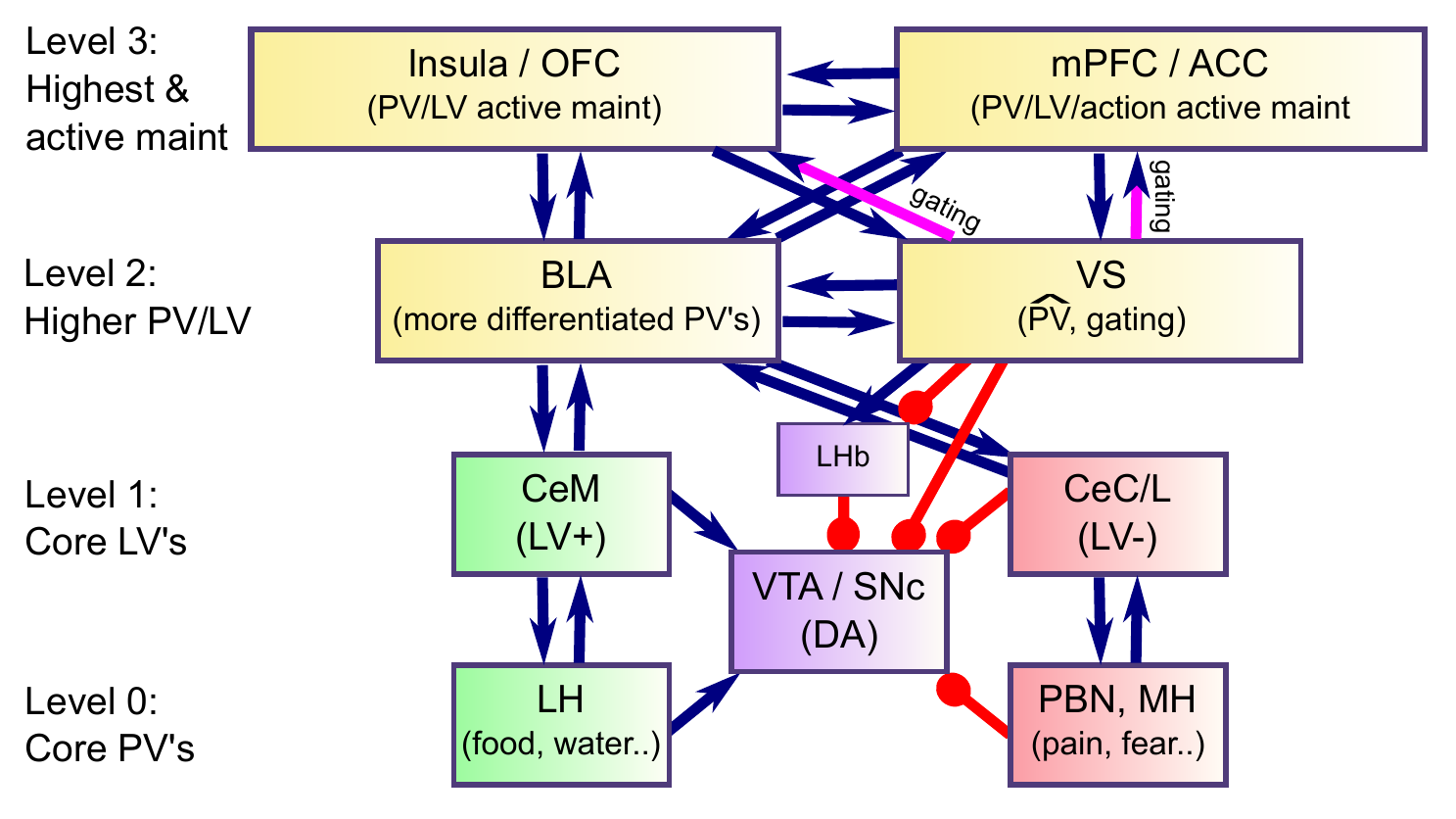}
  \caption{\small The (inverted) pyramid of goals, anchored at the base in the most basic of primary values (PVs), divided into positive and negative cases, handled by different areas e.g., positive = LH (lateral hypothalamus), negative = PBN (parabrachial nucleus) and VMH (ventromedial hypothalamus).  At the next level up are learned value (LV) areas that can associate perceptual stimuli with PVs, to provide a prospective, anticipatory estimate of probability of goal attainment, e.g., CeM (medial central nucleus of the amygdala) and CeC, CeL (capsular and lateral central amygdala).  The next level adds the basolateral amygdala (BLA), which has a much more extensive set of PV representations, and it supports LV sensory learning as well.  BLA feeds into the ventral striatum, where some neurons learn to expect the timing, magnitude, and probability of PVs, contributing to dopamine responses reflecting the discrepancy from these expectations.  At the highest level are the limbic frontal areas, including insula which represents highly differentiated primary values (e.g., different food tastes), OFC (orbital frontal cortex) which represents many different goals, continuing up the medial wall of mPFC (medial PFC) and anterior cingulate cortex (ACC).}
  \label{fig.goal_pyramid}
\end{figure}

\begin{figure}
  \centering\includegraphics[width=4in]{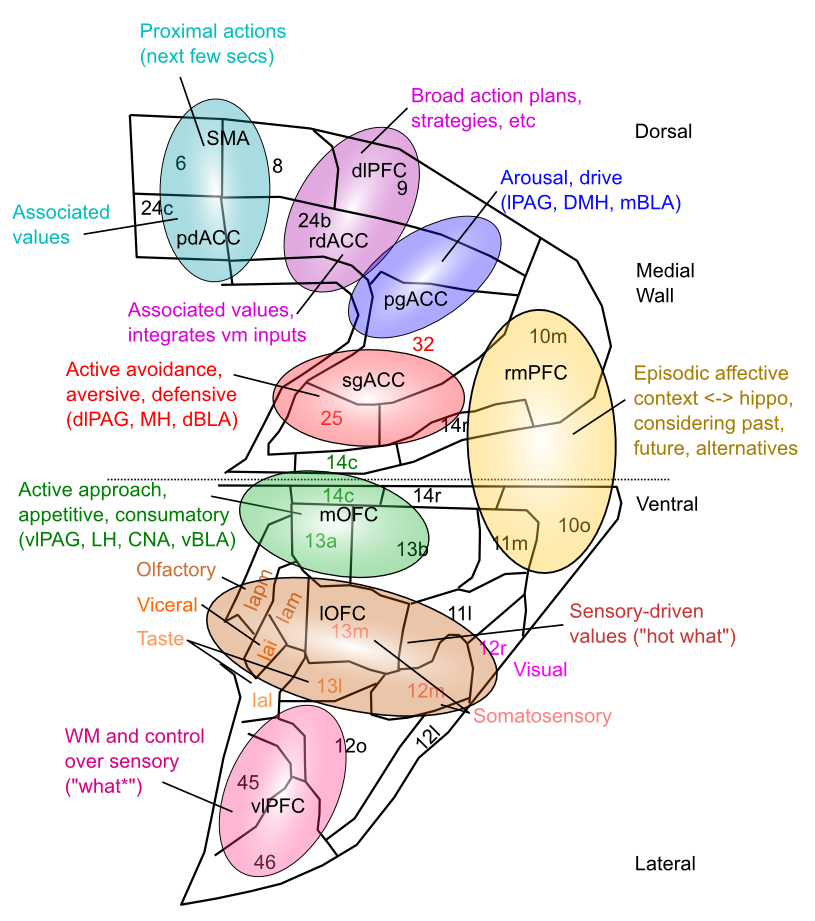}
  \caption{\small Map of goals in v/mPFC, based on the ``follow the money'' strategy of tracing connections down to subcortical brainstem areas by Ongur \& Price.  The lateral OFC (lOFC), which overlaps with the anterior insula, provides a sensory map of values from all the different senses (the insula is important for the olfactory, viceral, and taste signals).  The medial OFC (mOFC) is interconnected with appetitive areas, and thus may represent positive values ($\pPV$, $\pLVd$) (Noonan et al., 2010).  Moving up the medial wall, the subgenual ACC (sgACC) interconnects with aversive areas ($\nPV$, $\nLVd$) (Ongur \& Price, 2000).  Still further up is the pregenual ACC (pgACC), which interconnects with subcortical areas important for arousal, and may represent an intensity dimension (Chou et al., 2003).  The further dorsal and posterior ACC areas may encode more motor-action specific values, including probability of error, effort, uncertainty, conflict, etc (Alexander \& Brown, 2010; Kennerley \& Walton, 2011).  These values can play an important role in shaping the goal selection process.  Finally, the rostral medial PFC (rmPFC) is interconnected with the hippocampus (Carmichael \& Price, 1995), and may be particularly important in marshalling prior episodic memories in the goal selection phase (Cabeza, et al, 1997).  Functional neuroimaging studies (e.g., Roy et al, 2012) also suggest that this area may be important for social goals. } 
  \label{fig.goal_map}
\end{figure}

One of the appealing features of the goals-first perspective is that it provides a consistent explanatory framework for understanding motivated behavior across the full spectrum of mammalian brains and probably at least some other classes as well.  Anatomically, this is manifest as a kind of nested hierarchy of brain areas, which we can refer to as the {\em goal-processing pyramid} (Figure~\ref{fig.goal_pyramid}).  This pyramid is inverted, based at a narrow aspect in the subcortical areas corresponding to the primary value (PV) system described above.  With the LV areas of the CeA and VS added we have the makings of the necessary infrastructure for supporting goal-driven functioning.  When particular motivational states become engaged (e.g., hunger, thirst), subcortical areas specialized for processing relevant PVs will be up-regulated in a primitive form of attention. Similarly, the processing of sensory inputs previously associated with those PVs, such as food locations and similar cues will activate LV representations leading to approach behavior.  This low-level state biased for food-seeking behaviors will  remain engaged so long as you remain hungry, modulated by the phasic dopamine signaling driven by the gdPVLV system.  

At the next level up, the basal lateral amygdala (BLA) is known to function in much the same way as the CeA, in that individual neurons are selective to specific PV events, and learn at the time of a PV to associate active stimulus representations, and subsequently can fire based on these stimulus inputs alone, providing an LV-like signal \cite{BalleineKillcross06,BaxterMurray02,BelovaPatonMorrisonEtAl07,OnoNishijoUwano95,PatonBelovaMorrisonEtAl06,SchoenbaumChibaGallagher99,SchoenbaumSetlowSaddorisEtAl03}.  Critically, however, the BLA is cortex-like histologically and thus produces a much more fine-grained, differentiated map of underlying PV values and their associated LV cues, as compared to the more coarse, basic values encoded in the CeA level.  Further, in addition to its unidirectional driving projection to the CeA, the BLA is bilaterally interconnected with the orbital frontal cortex (OFC) and thus can drive anticipatory PV-like representations there in reaction to LV-like cues in the environment \cite{PauliHazyOReilly12}.  Finally, further distinguishing its role from the CeA, the BLA connects densely and unidirectionally to the VS and thus is positioned to play a critical role in driving gating of the OFC and related v/mPFC areas by the VS. 

At the highest purely-affective level of the goal-processing pyramid are the OFC and anterior cingulate cortex (ACC), which are strongly interconnected with key substrates at lower levels of the pyramid (most notably the BLA and VS) \cite{OngurPrice00,OngurAnPrice98,PriceDrevets10}.  As diagrammed in Figure~\ref{fig.goal_map}, this grounding in lower-level systems contributes to an overall ``map of goals'', but at this cortical level is also much more fine-grained and plastic than the lower levels, supporting ongoing learning of novel and/or changing goal representations.  This is by virtue of it being the main outpost of the goal system within the vast expanse of the neocortex, which is generally thought to be only weakly genetically configured, with considerable developmental learning taking place through activity-dependent plasticity mechanisms.  Also, as regions of frontal cortex, the OFC and ACC can support robust active maintenance of neural firing over time \cite{FrankClaus06,PauliHazyOReilly12}, giving these areas the ability to sustain active goal firing over longer durations.  We discuss these dynamics in more detail below.

Note that overall this functional organization preserves the What vs. How organization for frontal cortex \cite{OReilly10}.  Specifically, the ventral {\em What} pathways throughout the neocortex are thought to encode stimulus identity information, while the dorsal {\em How} pathways transform stimulus information to guide motor control \cite{UngerleiderMishkin82,GoodaleMilner92}.  When mapped onto the v/mPFC areas, the most ventral area, OFC, has much more directly stimulus-driven What-like representations \cite{HollandGallagher04,MurrayODohertySchoenbaum07,NoonanWaltonBehrensEtAl10,WestDesJardinGaleEtAl11}, whereas the most dorsal area, ACC, is much more directly interconnected with the motor control pathways in dlPFC, and is thought to play a more direct role in guiding action selection on the basis of affective values \cite{KouneiherCharronKoechlin09,CroxsonWaltonOReillyEtAl09,AlexanderBrown10,KennerleyWalton11,KollingBehrensMarsEtAl12,ShenhavBotvinickCohen13}.  This specific contrast between OFC and ACC has been emphasized particularly convincingly by Rushworth and colleagues \cite{RushworthBehrensRudebeckEtAl07,RudebeckBehrensKennerleyEtAl08}.  The areas between these two ventral and dorsal extremes may play more integrative roles. For example, the pregenual ACC may represent overall arousal and intensity \cite{ChouScammellGooleyEtAl03,MobbsMarchantHassabisEtAl09,NemethHegedusMolnar88}, while the most anterior rostral medial PFC area may integrate affective information with episodic memories \cite{CarmichaelPrice95,CabezaKapurTulving97,GilbertSpenglerSimonsEtAl06}, and otherwise play an integrative function \cite{RoyShohamyWager12}.  Interestingly, a similar map was developed based on analyzing a range of different human fMRI studies \cite{RoyShohamyWager12}.

Given the nature of the map goals in the v/mPFC as characterized above, we can associate the stimulus-driven OFC areas as being more LV-like overall, in providing stimulus-driven graded representations of progress toward goals.  In contrast, the ACC areas may be more important for representing the affective consequences of different potential courses of action, which then shape the goal selection process.  The ACC and OFC are tightly interconnected, so it is likely that both are involved in both goal selection and ongoing goal engagement and monitoring processes.  It is quite plausible that active maintenance of the goal is distributed across lateral PFC areas as well.  Consistent with many other examples in the brain, there is no single locus for a given goal representation --- it is distributed across multiple different areas which each contribute important content and interconnections to the overall goal state.  We are optimistic that approaching the representations in these areas from a goal-driven perspective may shed new light on their respective overall contributions, and dynamics at different points in the goal selection vs. goal engagement processing.  For example, a network of ACC, OFC, and vlPFC areas have been associated with the degree of motivated reasoning or belief bias exhibited by people \cite{WestenBlagovHarenskiEtAl06,StollstorffVartanianGoel12} --- this is consistent with the idea that these areas reflect an engaged, maintained goal state that is driving subsequent cognition in a goal-directed manner.

\subsection{A Rough Taxonomy of Primary Values}

There are many different taxonomies of emotional and motivational states that could be leveraged to produce a taxonomy of primary values in the human brain \cite{Maslow43,Reiss04}, but a full treatment of this topic is beyond the scope of this paper.  We offer a brief summary here that attempts to put emotional and other motivational states into a goal-centric perspective, elaborating upon the core list provided in Figure~\ref{fig.pv_list}.  As noted above, we distinguish affective {\em states} vs. {\em outcomes}, where affective states can function somewhat like a {\em drive} \cite{Hull43}, and are more diffuse and temporally extended.  

\incite{Reiss04} has developed a list of 16 {\em intrinsic motivations}, which have been established through factor analysis techniques.  They include some listed in Figure~\ref{fig.pv_list}, and are as follows (with some of our own interpretation added, including both Reiss's label for the motive and the intrinsic feeling or affective state): Power/Dominance/Efficacy, Curiosity/Wonder, Independence/Freedom, Status/Self-importance, Social contact/Fun, Vengeance/Vindication, Honor/Loyalty, Idealism/Compassion, Physical exercise/Vitality/Strength, Romance/Lust, Family/Love, Order/Stability, Eating/Satiation (presumably includes thirst), Acceptance/Self-confidence, Tranquility/Safety (lack of fear), Saving/Ownership.  Because these are derived from factor analysis based on human verbal reports, they may miss a few things, and some of the factors could be rotated into different primary dimensions.  For example, the need to share has been argued to be a critical human motivation \cite{Tomasello01}, which can support cultural transmission through direct instruction \cite[e.g.,]{HuangHazyHerdEtAl13} and imitation \cite[e.g.,]{Meltzoff85,Meltzoff88,MeltzoffDecety03}, which are widely regarded as essential for bootstrapping human cognitive abilities.  Hate is missing as an obvious complement to Love, and the goal-related feelings are notably absent as well.  Embarrassment, which has strong physiological correlates, seems likely to be primary, though it could be argued to be the negative side of status or social acceptance.

The following specific affective states merit more detailed discussion, due to their complexity and general issues they illustrate:

{\bf Fear}: Fear is a negative valence state resulting from the perception of a threat to life or other things of value, from predators or other sources of danger.  It can be seen as a negative valence drive-like state akin to hunger, in that it serves to motivate avoidance or aggressive behavior to overcome the fear if possible.  But it is triggered by external stimuli, not internal ones, and is interesting in being largely anticipatory and perceptual in nature, instead of being driven by actual experienced changes in bodily state.  The successful overcoming of fear states is experienced as a positive valenced relief outcome.  Unlike eating, there is not obviously a specific sensory pathway or characteristic phasic stimuli that define the relief outcome, however, so it is not clear if there is a discrete PV state that corresponds to this relief state, or whether it is necessarily a second-order derived outcome reflecting the absence of a formerly negative state.  Also, because it is perceptual and anticipatory, fear is particularly subject to appraisal dynamics, and can often be mitigated by reappraisal and the decision to mentally overcome the state (e.g., finding some courage). The brain areas involved in fear reappraisal are some of the same involved in goal-driven cognition, including the dlPFC which plays a role in fear reappraisal and other types of emotion regulation \cite{HartleyPhelps10}, and the vmPFC, which plays an important role in fear extinction and reversal through connections with the amygdala and striatum \cite{SchillerDelgado10}. Fear can be driven by US-like stimuli (e.g., genetically coded predator detection), other experienced negative outcomes (e.g., pain), and also more LV-like CSs that were associated with prior negative outcomes.  By drawing a clear dividing line between affective states and goal-relevant PV outcomes, we can partition the fear system into this more diffuse state-like side, even though it also has conditioning-like properties as well, which depend on many of the same neural mechanisms captured by the PVLV system.  However, the learning and activation dynamics of the fear system may be quite different than that of the goal-driven PVLV system, particularly with respect to the important connection to the VS gating system described in the next section. 

{\bf Pain}: Pain like fear is a negative valence state, but it is typically thought of as a direct consequence of bodily harm of some sort.  Critically, pain typically emerges as a reaction to insult, and is not necessarily available prospectively to guide goal selection.  In this sense, pain is a negative outcome, not a drive state {\em per se}.  Pain does persist over time, however, and can drive various goals to treat or manage the pain itself, and in this case any relief that is experienced can be viewed as a positive relief PV outcome associated with such goals.  Anticipated pain, if severe enough, is likely to activate the fear system, and there is also evidence that learned pain expectations can mediate activity in pain-sensitive regions through activity in the OFC and striatum \cite{AtlasBolgerLindquistEtAl10}.  Another interesting form of learning in the pain system is the well-known placebo effect, which has been shown to alter activity in several pain-sensitive regions \cite{WatsonEl-DeredyIannettiEtAl09,WagerRillingSmithEtAl04,EippertBingelSchoellEtAl09,PetrovicKalsoPeterssonEtAl02}.

One case of anticipated pain that is very relevant to the goal-driven system is when it functions as a cost-like variable, like effort, where lower levels of pain are regarded as inevitable negative consequences to desired goal achievement (e.g., ``no pain, no gain'').  As such, we would expect that the ACC would represent these lower levels of pain much like effort is represented, whereas higher levels of pain would drive the fear system, creating a complex bifurcation in how pain is represented depending on intensity. Indeed, the sgACC was involved in subjective valuation in a decision-making task involving tradeoffs between different levels of pain and reward, and sgACC-amygdala connectivity was modulated by money only under high pain conditions \cite{ParkKahntRieskamp11}.  As suggested earlier, we may be able to effectively ignore pain signals when in a goal engaged state, even as these costs are being tallied for later goal selection, which is consistent with the strong influence of PFC and striatum on expectations and learning about pain and analgesia \cite{AtlasWhittingtonLindquistEtAl12,WagerAtlasLeottiEtAl11}, and evidence suggesting that the mPFC is dysregulated for chronic pain patients \cite{Apkarian08}.  Finally, pain appears to serve as a concrete anchor to many higher-level forms of negative outcomes that are not viscerally painful, but people describe them as such, and they activate many of the same neural pathways \cite{KrossBermanMischelEtAl11,Eisenberger12}.  Thus, pain may serve as a kind of universal negative PV pathway for many different outcomes, which evolution may have co-opted to take advantage of the low-level neuromodulatory pathways that were already in place.  

{\bf Stress, Uncertainty:} These are another category of negative valence states that need to be addressed --- are these just special cases of the overall fear system, or are there specialized neural systems and intrinsic motivations associated directly with stress and uncertainty?  The evidence strongly favors the latter: the stress system is strongly associated with the HPA (hypothalamic, pituitary, adrenal) axis and the release of corticotrophin-releasing factor (CRF), under top-down regulation by limbic brain areas such as the mPFC, amygdala and NAc \cite{FranklinSaabMansuy12}.  The stress system interacts with the dopamine system, and influences plasticity and interactions with other neuromodulators in several limbic brain regions, the specifics of which are beyond the scope of this article \cite{BelujonGrace11,RussoMurroughHanEtAl12,CabibPuglisi-Allegra12}. Additionally, there may be specialized representations for uncertainty in several areas, including brainstem areas such as the anterodorsal septum \cite{MonosovHikosaka13} and the anterior cingulate cortex \cite{RushworthBehrens08}. Connections between regions of the BNST (bed nucleus of the stria terminalis) and hypothalamic and brainstem regions are also important for the modulation of behavioral effects of anxiety and stress, and this region plays an important role in the integration of stress and reward information through connections with the the VTA. \cite{KimAdhikariLeeEtAl13,JenningsSpartaStamatakisEtAl13,FlavinWinder13}.

{\bf Happiness, Disappointment, Sadness, Frustration, Anger, Boredom, Depression:}  These various goal-associated affective states are particularly interesting in relation to the centrality of these affective states in daily human life.  It is said that maximizing happiness is the broader goal of life, where this form of happiness is clearly distinguished from raw hedonistic pleasure.  In the goal-driven framework, some PV outcomes are experienced as pleasurable (positive primary valence), but the larger context is accomplishing active goals, which leads to happiness as distinct from pleasure. Although dopamine was originally associated directly with hedonic pleasure \cite[e.g.,]{Wise80}, it is now clear that it plays a much more functional role in goal-driven motivated behavior --- a distinction which has been labeled as {\em liking} (hedonic pleasure) vs. {\em wanting} (motivated behavior) \cite{BerridgeRobinson98,BerridgeRobinson03,SalamoneCorrea12}.  The clearest case for dissociating these two states is when accomplishing a goal driven by a strong negative valence situation --- e.g., working hard against many challenges to achieve something difficult and important.  Daily life is arduous and full of negative valence, but the progress and eventual accomplishment result in extreme Happiness and a feeling of life satisfaction.  This is not too far from the so-called American Dream ideal, which may be a cultural construct that closely aligns with the goal-driven structure of our brains.

\begin{figure}
  \centering\includegraphics[width=4in]{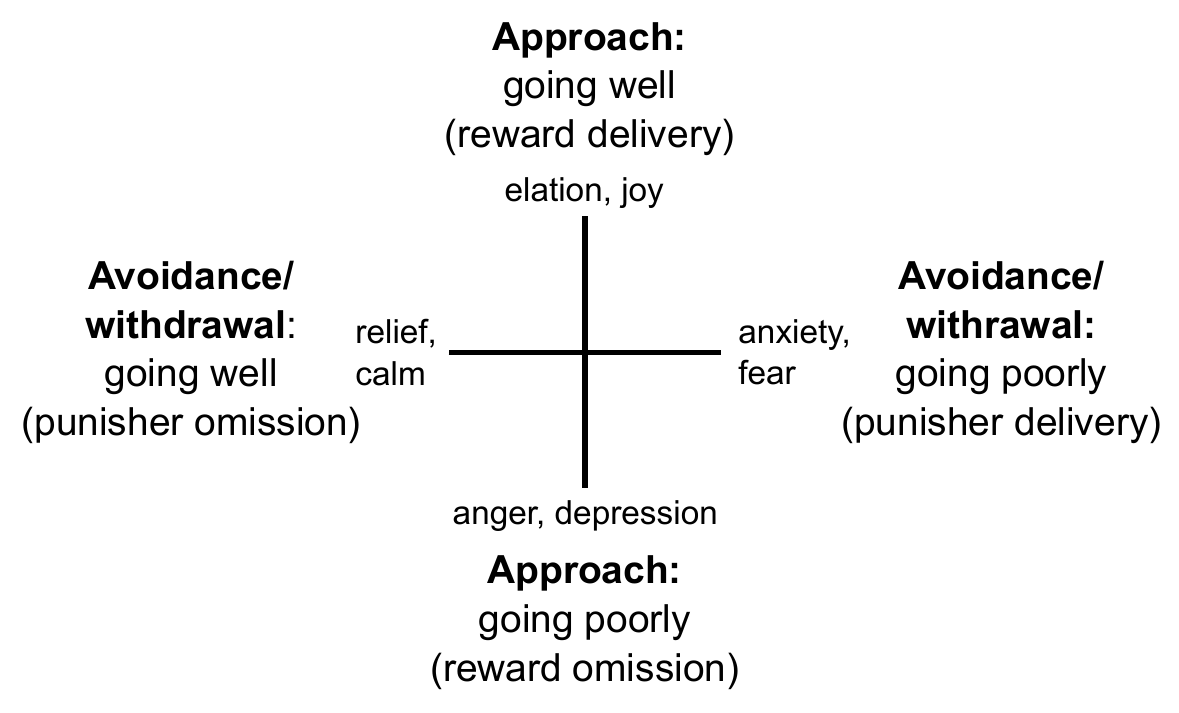}
  \caption{\small Approach vs. Avoidance organization of affective states, according to Carver \& Scheier (1982), merged with the organization proposed by Rolls (2005) based on reward vs. punishment, as proposed by Carver \& Harmon-Jones (2009). } 
  \label{fig.approach_avoid}
\end{figure}

Finally, at the most abstract level, various 2D organizations of some emotional states have been proposed, for example along positive vs. negative affect dimensions \cite{WatsonTellegen85,RussellBarrett99}, or along approach vs. avoidance dimensions (Figure~\ref{fig.approach_avoid}; \nopcite{CarverScheier82}), which can be aligned with reward and punishment axes \cite{Rolls05,CarverHarmonJones09}.  The approach axis aligns well with the goal-driven processing states described above, but we are not sure at this point how to integrate the avoidance axis.  In many cases, avoidance situations can be turned into approach situations, thereby realigning completely with our framework.  It remains unclear how direct punishment avoidance would mobilize cognition and behavior without being mediated proximally by these approach goals --- we return to this issue in the Discussion section.   Interestingly, it does appear that prefrontal cortex has at least some degree of lateralization according to approach vs. avoidance dominance, as contrasted with basic positive vs. negative affect dimensions.  This was established by examining the key affective state of {\em anger}, which is negatively valenced, but drives approach behavior, and can thus dissociate valence from approach vs. avoidance  \cite{CarverHarmonJones09}.  There is also a question as to whether intensity or arousal is coded separately from valence or more specific states --- for example the classic James-Lange theory and modern incarnations thereof argue that we just interpret raw bodily arousal signals or somatic markers \cite[e.g.,]{BecharaDamasioTranelEtAl05,RussellBarrett99}.  The pgACC as mentioned above may provide a locus for such independent coding of arousal, at the same time it is likely that intensity is encoded within each specific affective state.

In summary, it seems clear that there are a multitude of innate motivational or affective states, but the biologically privileged dimensions of this space have yet to be established.  From a goal-driven cognition perspective, the different affective states have important implications for appropriate outcomes and action plans, likely at a very fine-grained level.  Not all negative or positive states are equivalent --- fear, pain, and anger each have very different dynamics and implications, and the brain is likely to take advantage of these differences.  Thus, it is important to avoid the temptation to reduce the dimensionality of this space to the level of e.g., phasic dopamine dynamics --- those are just one element of a much richer overall space.  It is plausible that other neuromodulators may expand the dimensionality of the space \cite[e.g.,]{Doya02,Doya08}, but even there, v/mPFC is a very large space that has plenty of room for very high dimensional representations of affective states and goals.

\section{Discussion}

We have presented a theoretical framework, and initial biologically-based computational model, based around the central idea that goals are primary, and one generally must have an active goal engaged.  The process of selecting a new goal is characterized by a careful weighing of costs and benefits, in part because once a goal is selected and engaged, it comes to dominate the effective value function of the organism, biased toward rewarding progress toward the goal.  This strong dichotomy between the conservative value function at work during goal selection versus the optimistic, goal-dominated value function during goal engaged mode is consistent with a number of anecdotal phenomena, and our broader goal-driven framework and computational model is consistent with a wide range of cognitive neuroscience data.  In the remainder of this discussion, we discuss related computational modeling work, followed by some major unresolved issues, and highlight a few of the broader implications of the framework to understanding clinical disorders and daily life.

\subsection{Other Computational Models of Goal-Driven Cognition}

Computational models that have incorporated a goal-driven component have done so largely within a symbolic processing framework \cite{Sun09,Anderson90,LeakeRam95}, where the goal selection process can be implemented explicitly and without much complication.  For example, in the ACT-R framework \cite{Anderson90,AndersonLebiere98,AndersonBothellByrneEtAl04}, there is an explicit goal buffer, and specific productions load and update this buffer to guide progress toward goal achievement.  And the contents of the goal buffer can drive spreading activation to bias goal-relevant information and processing in other parts of the architecture.  But there is no specific connection between goal states and intrinsic motivations.  In contrast, the CLARION architecture \cite{Sun09} has incorporated many of the intrinsic motivations described by \incite{Reiss04}, to specifically drive an explicit goal selection process, which lies at the center of multiple modules within the architecture.  Once active, the symbolically-represented goals serve to bias the action selection process.  The overall dynamics of the goal selection process seem to be based on a Hull-style drive reduction model \cite{Hull43}, with new goals selected when the drive activation associated with a goal is reduced below a given threshold.  In contrast, our framework is centered around the principle that goals contain a representation of the desired outcome end state, and are thus self-terminating when this end is achieved.  Finally, the goal-driven framework of \incite{LeakeRam95} articulates many of the benefits of a goal-driven approach, within an overall symbolic architecture.

More recently, other modeling approaches are beginning to incorporate aspects of goal-driven behavior.  For example, the internal model hypothesized in the model-based reinforcement learning (RL) approach \cite{DawNivDayan05,Dayan09,RangelHare10} has a lot in common with our construct of a goal-space.  Of particular relevance to the goals-first perspective, \incite{SolwayBotvinick12} have recently incorporated a novel Bayesian approach to essentially run the internal model backwards to find the optimal policy that is most likely to achieve the maximum reward.   In a sense, by going backwards from assumed perfect success they only have to do the math once.  Nonetheless, their approach differs in fundamental ways from the neural network-based, goals-first approach proposed here.  For example, their outcomes are simple scalar quantities of abstract value and thus lack the grounding in real-world representations by which biasing of appropriate subgoals and action plans can actually occur.  Furthermore, the notion of a goal as an actively maintained representation that guides behavior until its target end state is achieved, which is central to our approach, is missing.  In effect, the \incite{SolwayBotvinick12} model only addresses the planning aspect of the process, not the full architecture of goal-driven cognition.

The theoretical framework articulated in \incite{PezzuloCastelfranchi09} covers a lot of territory in common with our model (and includes a comprehensive discussion of other relevant literature), while focusing mainly on the sensory-motor (``cold'') aspects of the space.  One central hypothesis in their framework, shared by a number of other theorists \cite[e.g.,]{Hommel04}, is based on the notion of {\em ideomotor} representations, which are action representations that are activated and controlled by an anticipatory representation of their desired effect \cite{James90}.  The appealing idea is that one can somehow invert the natural learning of action $\rightarrow$ effect sequences, to then drive effect $\rightarrow$ action selection.  However, it is not clear how well that key inversion process actually works in practice for realistically complex motor control \cite{HerbortButz12,JordanRumelhart92}.  The goals-first framework provides a different alternative: actions are driven from the start by engaged goals, which contain at least some encoding of the desired effect.  Learning can tune these action $\rightarrow$ effect representations, but by having the goal representations active from the start, the binding of outcomes to goal representations is more straightforward and effective.  The \incite{PezzuloCastelfranchi09} framework also relies to some extent on the forward projection process to anticipate the outcomes of actions --- this is another process that we think is most often avoided by goal-selection happening first.  We do not believe that non-human animals have much capacity for this kind of look-ahead process (despite the popularity of this idea, and what we consider to be ambiguous evidence; \nopcite{PennyZeidmanBurgess13}).  Furthermore, data suggests that people avoid it to a large extent (even in advanced game playing situations such as chess where this kind of lookahead is strategically quite beneficial; \nopcite{GobetSimon96}).

\subsection{Major Outstanding Issues}

\subsubsection{Avoidance Goals, and Lateralization of the Goal Sytem}

The goals-first framework articulated in this paper is centered around approach goals with positive valence outcomes.  In the case of aversive situations and states, these goals are directed toward overcoming adversity, whether it is hunger, fear or any other aversive state (Figure~\ref{fig.pv_list}).  However, there is a considerable literature on the notion that the frontal cortex is lateralized according to approach (left lateralized) versus avoidance (right lateralized) \cite{CarverWhite94,SuttonDavidson97,Davidson00,CarverHarmonJones09,SpielbergHellerMiller13,KelleyHortensiusHarmon-Jones13}.  The key question is thus: how does avoidance work to drive behavior, if not through positive approach goals?  We can see two distinctive alternatives:
\begin{itemize}
\item Just add a negative sign to everything, so that the PV goal is directly to reduce or eliminate the negative state (as opposed to approaching a positive state), and the LV progress monitoring is proportional to the gradient of reduction of proximity to the negative stimuli (i.e., getting further away from a predator).  This is consistent with various proposals in the literature \cite{Carver06,Elliot06,Higgins97}.  The key difference here is that it is hard to concretely assess the end point associated with this kind of avoidance goal --- you can't get infinitely far away from a predator, and you never know if it might be lurking behind something and just about to pounce.  Similarly, it tends to be more difficult to formulate concrete action plans based on avoidance, without focusing on a specific positive goal --- ``run away'' is much less specific than ``run toward this specific safe location''.  Thus, consistent with subjective experience, avoidance goals, if they exist, are likely to be more free-floating and amorphous, and not as effective in driving cognitive and action systems compared to the more concrete approach goals.
\item Reconceptualize the right hemisphere activation as representing a space of alternative goals, action plans, and associated state information (of both positive and negative valence), that are accumulated and maintained as possible competitors or alternatives to the dominant goals and plans represented in the left hemisphere.  Critically, this right hemisphere information can be actively maintained and updated, without interfering with the current active goals and action plans --- the hemispheric lateralization can play the critical functional role of designating that class of information as ``offline'' for the purposes of behavioral control.  In the case of avoidance goals, the right hemisphere would maintain all the worries and concerns and negative state information, which are essential to keep online and update as things progress, and this information can inform the selection and engagement of new goal states, but these new active goal states are represented always in left hemisphere, driving positive approach goal dynamics as conceptualized in this paper.  This framework is consistent with data showing that the right ventral lateral PFC is monitoring for low-frequency signals that indicate a need to alter the dominant task, e.g., in the stop-signal paradigm \cite{ChathamClausKimEtAl12,MunakataHerdChathamEtAl11}.  Also, an fMRI study localized a secondary task to the right frontal cortex while the dominant task was localized to the left \cite{CharronKoechlin10}.  Furthermore, an electrical stimulation paradigm was able to convert anger (an active approach motivation) to rumination (a passive response to adversity) by increasing right vs.\/ left frontal activation \cite{KelleyHortensiusHarmon-Jones13}.
\end{itemize}

One potential source of data to distinguish among these alternatives is individual differences between anxiety and depression --- higher depression individuals generate fewer approach goals, compared to higher anxiety individuals who generate more avoidance plans \cite{DicksonMacLeod04}.  In either case, it does seem that having an overall avoidance focus compared to an approach focus results in negative overall cognitive and emotional outcomes \cite{Sideridis05,ElliotSheldon97}.  Interestingly, positive outcomes can sometimes be associated with goal abandonment, but this seems to depend on being able to move past the abandoned goal within a larger overall approach goal \cite{WroschScheierMillerEtAl03}.

\subsubsection{Subgoals and Goal Learning}

To this point, we've only considered goals that are grounded directly in primary values (PVs), which we assume to be genetically encoded, and thus a closed-class.  The system would be much more open-ended and powerful, and also much more dangerous, if it could develop novel goal representations.  One important category of such representations are subgoals that reliably lead to achieving a PV, but are not themselves directly associated with a closed-class PV.  There are multiple challenges in getting this to work.  First, the overarching PV goal must somehow drive the learning of any intervening subgoals --- this implies that the overarching PV goal must be engaged first, which is not unreasonable, but nevertheless represents an important constraint.  The incremental dopamine bursts in response to progress on the outer PV goal must then reinforce Go firing for the subgoal.  In effect, the relationship is identical to that of goal and action.  But the PV-based grounding of the VS gating signals does not operate here.

Thus, one overall take is to say that subgoals not directly associated with a PV outcome are indistinguishable from dlPFC action representations, and indeed that is quite possibly where they exclusively reside.  Another alternative is to allow for grounded PVs to support the learning of second-order PV-like neurons in higher areas of the goal pyramid, potentially including the ventral striatum, so that the same PV-based grounding of the VS gating signals can still hold.  The challenge here then becomes delineating the conditions under which this second-order PV learning operates, so as to not become too open-ended and lead to the pursuit of manifestly maladaptive outcomes.  We do know that only neurons in the BLA, but not CeA, are important for second-order conditioning \cite{HatfieldHanConleyEtAl96}, and that learning beyond second-order is extremely limited or nonexistent \cite{DennyRatner70,Dinsmoor01}.  Thus, in contradiction to the TD chaining model, the brain appears to treat higher-order conditioning with extreme caution, and it is unclear at this time whether even the second-order level is sufficiently broadly enabled to play a significant role in subgoal learning.  As we develop our computational models to include a subgoaling ability, we will confront and hopefully at least provisionally resolve these important issues.

\subsubsection{High-level strategic action vs. lower-level perceptual-motor control}

The form of goal-driven processing that is the focus of this paper can be characterized as high-level {\em strategic} processing (deciding among multiple options, e.g., whether to go out to eat or cook dinner), as contrasted with the more continuous {\em tactical} control required for dynamic motor control (e.g., how to actually slice food with a knife while making dinner).  The basal-ganglia / frontal cortical system seems particularly well suited for making the more discrete strategic decisions, while the cerebellum interacting with the parietal cortex seems more well-suited for the dynamic online control aspects.  Indeed, there is evidence for dedicated perceptual-motor goal spaces in various regions in parietal cortex \cite{BuneoAndersen06}, and electrical stimulation in parietal can cause the feeling of having an intention to execute a motor movement \cite{DesmurgetReillyRichardEtAl09}.  We hypothesize that these perceptual-motor goal spaces develop through a combination of appropriate genetically-determined initial connectivity, and scaffolding from affective goal systems.  One suggestive indication that the domains of online perceptual-motor control and higher-level goal and action selection have very different computational demands, is that there is surprisingly little direct communication between the cerebellum and the basal ganglia and their respective paths through the thalamus to the cortex are largely segregated \cite[e.g.,]{PercheronFrancoisTalbiEtAl96}.  Recent work has emphasized some focal areas of interconnectivity however \cite{BostanDumStrick13}, and the nature of these interactions should be instructive for our overall framework.

Another important way in which the basal ganglia and perceptual-motor systems may differ is in the use of positive feedback vs. negative feedback dynamics.  \incite{Powers73} and others \cite[e.g.,]{MillerGalanterPribram60,CarverScheier82} have persuasively argued that a cybernetic-style negative feedback controller model is particularly powerful for online perceptual-motor control.  Here, goals serve as reference values that are compared against current perceptual inputs, and discrepancies or disturbances lead to behavioral actions that minimize this discrepancy.  In contrast, the dopamine-based learning that is central to understanding how the basal ganglia operates (and central to our overall framework) has a strong positive-feedback reinforcement learning dynamic, where, over the course of learning, some representations become stronger and more likely to be activated, while others are attenuated.  More generally, we believe that learning is essential to understanding many aspects of the system, because our initial starting state is so impoverished, and much of our intelligence must be built up slowly over the developmental learning process.  At a very abstract level, having such a big investment in learning suggests that our brains evolved in relatively stable environments where much of what we learn can be applied productively throughout our lives.  In contrast, the hallmark of the negative feedback control systems is their extreme flexibility in the face of all manner of unpredictable disturbances, and some of this flexibility is lost once learning mechanisms are introduced.  These differences are emblematic of the distinctions between strategic and tactical domains, where the former benefits from wisdom and experience, and the latter benefits from flexibility and ability to deal with whatever comes along.  Clearly, both of these dynamics have their tradeoffs, and as in many other cases in the brain, it is likely that different neural systems optimize each of these types of control (i.e., basal ganglia for reinforcement learning, and cerebellum for negative feedback control).  

\subsubsection{Novel problem solving}

Finally, we address perhaps the most challenging problem facing cognitive neuroscience generally, including the goals-first framework: novel, challenging situations that require more advanced problem-solving techniques, which really tap the special cognitive abilities of the human brain.  Fundamentally, the goal-driven framework stipulates that hard problems are solved in goal-space first, and only then are actions undertaken.  Thus, you have to ``see'' the path to the solution first.  This doesn't mean the whole solution needs to be worked out entirely in advance, but at least the shape of the solution and the first step need to be there.  A universal weak solution to hard problems is to break them down into more tractable smaller problems, and again this is something that would be done in goal space first.  All of this requires specific, probably at least partially culturally-transmitted strategies for goal-space processing.   How these strategies are learned and executed constitutes yet another major challenge for this framework.  Finally, even when all else fails and you have to engage in random trial-and-error exploration, the key point is that this exploration occurs in goal space first, and perhaps then also in action space within the context of an engaged goal.  This allows the goal to direct and evaluate the action selection process as progress is either made or not.  Furthermore, if the whole thing turns out to be a success, the resulting learning will make this novel goal-action complex more available as a possible solution to similar predicaments in the future.  A strong bias to generally have a tractable goal guiding behavior would naturally support this goal-first exploration process.

\subsection{Potential Relevance to Clinical Disorders}

Clinically, the goals-first perspective is consistent with the profiles of major clinical disorders including depression (major depressive disorder, MDD), obsessive-compulsive disorder (OCD), and ADHD.  As noted earlier, we can think of depression as characterized an inability to enter the goal-engaged state, resulting in pervasive feelings of fatigue and motivational anhedonia \cite{DemyttenaereDeFruytStahl05,SalamoneCorreaFarrarEtAl07,TreadwayZald11,SalamoneCorrea12}.  This view of MDD (which is only recently taking hold, overcoming the traditional view that it is an inability to experience pleasure derived from impaired dopamine function; \nopcite{WiseRompre89,DisnerBeeversHaighEtAl11}), is consistent with our overall framework at multiple levels.  For example, considerable data implicates the ventral striatum (nucleus accumbens) dopamine system in MDD, along with sub-genual ACC (sgACC), and many other areas of the v/mPFC goal processing system \cite{PriceDrevets10,MaybergBrannanMahurinEtAl97}.  In particular the sgACC, which according to the map of goals presented above is important for encoding negative valence information, appears to be over-activated in MDD.  Our interpretation is that MDD represents a vicious feedback loop between initial negative affective states combined with an inability to activate positive goals to overcome these states, and the resulting negative affect associated with inability to become goal-engaged.  The result is sustained negative affect and motivational anhedonia (i.e., {\em apathy}, which is an important component of MDD and other related disorders), corresponding to excessive activation of the sgACC and the NoGo pathways in the VS/NAcc.  This inability to drive goal activation then also drives an overall deactivation of the dlPFC \cite{PriceDrevets10}, with various sources suggesting a right vs.\/ left lateralization imbalance \cite{RosadiniRossi67,CarverHarmonJones09} consistent with the avoidance goals discussion above.  This model of depression is generally consistent with the learned helplessness model \cite{Maier84,MaierWatkins10,MaierSeligman76}, which was recently shown to involve lateral habenula dopamine modulation \cite{LiPirizMirrioneEtAl11}.  However, for the vicious circle feedback to be sustained enough to produce MDD symptoms, there likely needs to be significant negative valence activation that is strong enough to prevent the natural positive feedback of goal engagement from taking over again relatively soon (and/or a biological disorder with similar effects) --- it is unlikely that a few exposures to uncontrollable shock rise to that level.

OCD is a particularly interesting case for the relationship between aversive states and approach-related goals to overcome them --- the central dynamic in this condition is that ritual behaviors are motivated by aversive intrusive thoughts and fears \cite{AbramowitzTaylorMcKay09}.  As discussed in the context of aversive goals above, it is difficult to know when the aversive condition is over, especially when it involves fears of things like germ contamination or the possibility of losing control and causing harm to others, which are common fears in OCD.  Thus, the aversive states tend to persist, and because the rituals are also not particularly effective in directly overcoming the adversity (even though they are positive concrete actions motivated by them), the whole condition enters a vicious cycle.  This cycle is reinforced through the temporary feelings of relief provided by the ritual behaviors, to the point that it becomes difficult to overcome.  Thus, OCD represents a aberrant attractor state of the goal-driven system, and a variety of sources strongly implicate the OFC and ventral striatum in this disorder \cite{AbramowitzTaylorMcKay09}.  Hopefully, a clearer understanding of these systems can lead to better treatments --- there are various places where this cycle could be broken, and some may be more effective and amenable to treatment than others.

Although ADHD has long been thought to involve deficits in executive function, a more recent perspective emphasizes the prominent role of motivation in this disorder \cite{CastellanosSonuga-BarkeMilhamEtAl06,DovisVanderOordWiersEtAl12,SlusarekVellingBunkEtAl01}.  In this view, people with ADHD have difficulty sustaining activation of a given goal state over time, because they lose motivation and become interested in doing something else.  When brought into the lab and given relatively short tests of executive function under strong external demand characteristics, ADHD patients can often perform quite well. Furthermore, motivational manipulations have relatively strong effects, serving to at least partially ameliorate any observed deficits \cite{DovisVanderOordWiersEtAl12,SlusarekVellingBunkEtAl01}.  The motivational account also helps to make sense of the increase in response time variability observed in these patients \cite{CastellanosSonuga-BarkeMilhamEtAl06,FrankSantamariaOReillyEtAl07} --- sometimes they are engaged in the task, and other times they are not.  One intriguing possibility in this context is that the neuromodulator noradrenaline (NA, also known as norepinephrine or NE), which has been implicated in the exploration vs. exploitation tradeoff in reinforcement learning \cite{Aston-JonesCohen05} and in ADHD \cite{SwansonPerryKoch-KruegerEtAl06,FrankSantamariaOReillyEtAl07}, could reflect, and/or affect, the current level of goal engagement. In the goal-driven framework, exploitation means sticking with the current goal, and exploration means switching to an alternative goal.

In any case, it is likely that when left to their own devices in the real world, with many more alternative goals than in a lab experiment, goal engagement in ADHD patients could be more fleeting, and thus the disorder more severe than the lab experiments would otherwise suggest.  The goal-driven perspective can offer novel avenues for treatment, by focusing on the dynamics of goal engagement and maintenance.  For example, people could be trained to engage in a more deliberate attempt to break longer goals into more tangible, shorter subgoals, that would deliver more continuous graded reward signals, and thus reinforce goal engagement.  More generally, it seems that everyone could benefit from such training, to be more effectively engaged in anything that they want to achieve.  This is supported by work suggesting that implementation intentions specifying when, where and how to achieve a particular goal have a positive effect on goal attainment. \cite{GollwitzerSheeran06}. 

\subsection{Insights into Cognitive Biases and their Mitigation}

The goals-first framework posits that people's evaluations of cost/benefit tradeoffs and other decisional factors may be differentially biased depending on whether they have a goal currently engaged or not.  When a goal is engaged, estimates of costs (and indeed subjective experiences of pain and effort) are likely underestimated, whereas they are likely to be more rationally estimated otherwise.  Furthermore, the dominance of the engaged goal and reluctance to abandon it can give rise to a strong confirmation bias, as mentioned earlier. This confirmation bias is based in part on the active goal activating associated hypotheses and evidence \cite{MorewedgeKahneman10}. Maintained goals can also account for the pervasive phenomena of the sunk cost fallacy, where people irrationally persist in pursuing a given goal despite mounting losses, and motivated reasoning \cite{Kunda90,WestenBlagovHarenskiEtAl06,KahanJenkins-SmithBraman11}, in which people arrive at conclusions that are not logically correct, but satisfy personal goals (such as gaining in-group approval).  Our framework suggests potentially novel angles for addressing these pervasive cognitive biases, by focusing on the source of the problem (the goal engaged state) and attempting to temporarily disengage or otherwise distance the individual from their goals, so as to facilitate a more rational evaluation of the current situation.  Interestingly, available evidence suggests that for disengaged goals, increasing psychological distance results in {\em less} consideration of costs \cite{TropeLiberman10}, which may mean that more proximal choices are more carefully evaluated as concrete candidates for goal selection.  We maintain that the results should go the other way when examining engaged vs.\/ proximal disengaged goals, or as a function of manipulations to distance people from currently engaged goals.  More broadly, the evidence about emotional state and rationality of judgments has been mixed (despite the pervasive belief that depressed people are more rational) \cite{Pham07}, but the bias toward rumination associated with negative mood may result in more carefully considered solutions to challenging problems, and in this way be adaptive \cite{AndrewsThomson09}.

\subsection{Application to Daily Life}

The goal-driven perspective on human cognition can provide numerous practical insights into the nature of human behavior.  For example, one of the hard lessons of life is that it is remarkably difficult to convince others of anything, much less change their behavior in any meaningful way, through the pure application of reason and logic.  Instead, you have to somehow find a way into the deep recesses of the goal hierarchy, and drive the activation of goals that lead to the desired behaviors.  The flip side of this point is captured in the central theme of Buddhist philosophy, which is that the positive feedback dynamics of goal attachments are very strong, and will tend to dominate our lives, unless we consciously work to counteract them.  Thus, we cannot be psychologically free unless we escape the goal engaged state.  And yet, the goal engaged state is fundamental to producing the state of happiness.  Amid these and many other contradictory forces, we must each find our own balance, but perhaps an awareness of the forces at work in our brains will enable people to understand why they feel the way they do, and make better informed decisions about how to proceed.

At a more practical level, the goal-driven framework can help us understand why procrastination is so prevalent.  From this framework, procrastination represents the selection of goals that have a better overall net utility compared to the more onerous thing that you actually need to be doing.  In other words, you select things that are easier and/or result in more immediate, positive outcomes.  When you look at it this way, procrastination is completely rational, and we have emphasized that the goal selection process represents the most rational weighing of options.  However, once you become goal engaged, this careful weighing goes out the window.  Thus, the key step for avoiding procrastination is to somehow trick yourself into getting started on that big task you need to do, so that you can then leverage the goal engaged positive feedback loop.  And the trick to getting started is to make the first step {\em extremely easy and immediate} --- if you're blocked on starting a paper, just set a goal of writing the title.  Or the methods.  Or any little part of it that is the easiest.  Maybe it is making the figures.  In any case, once you get started, then the goal engaged dynamics kick in, and off you go!  Just make sure not to hit any major roadblocks along the way, because those will certainly trigger a new round of goal selection.  Interestingly, recent research shows that the tendency to procrastinate is strongly associated with impulsivity (the two are identical at a latent genetic level) \cite{GustavsonMiyakeHewittEtAl14,Steel07}, which is consistent with the goal selection framework: impulsive decision makers tend to focus on more immediate positive outcomes, and those are exactly the goals that interfere with making progress on a more difficult task, prior to when it is absolutely essential.

One other interesting connection with daily life, and an active area of current research, is in terms of the extensive amount of time that people spend ruminating about possible future goals and plans, and reconsidering the outcomes of prior ones.  Specifically, this is one of the leading hypotheses for what goes on during the so-called {\em default mode} of cognition, which is characterized by a network of brain areas that overlap significantly with those of the goal network (Figure~\ref{fig.goal_map}).  This default mode network appears to be engaged whenever people are not otherwise engaged in an active task, and it could be characterized as a third fundamental state of goal-driven processing: episodically-driven reevaluation and planning of future goal states, so that when the time comes, you will have more thoroughly worked out the possible consequences and outcomes associated with different action plans, and formulated a set of possible goal/plan representations that can be more easily activated in the appropriate circumstances.

\subsection{Conclusions}

Although we have reached the end of the paper, the end of the story (and the moment of goal achievement satisfaction) awaits considerable more work to build on the very preliminary work here in implementing computational models, and determining what works and what doesn't.  Many possible experiments can be conducted to test these ideas, and hopefully the more detailed predictions of our implemented models.

\section{Acknowledgments}

Supported by: MH079485, ONR grant N00014-13-1-0067, ONR D00014-12-C-0638, and by the Intelligence Advanced Research Projects Activity (IARPA) via Department of the Interior (DOI) contract number D10PC20021. The U.S. Government is authorized to reproduce and distribute reprints for Governmental purposes notwithstanding any copyright annotation thereon. The views and conclusions contained hereon are those of the authors and should not be interpreted as necessarily representing the official policies or endorsements, either expressed or implied, of IARPA, DOI, or the U.S. Government.

Thanks to Kai Kruger, Christopher Chatham, Yuko Munakata, Alex Petrov, David Jilk, Peter Pirolli, Brad Minnery, and members of the CCN Lab for their comments.

\newpage

\begin{thebibliography}{}

\bibitem[\protect\citename{Abramowitz}{Taylor}{McKay}{AbramowitzETAL}{2009}{AbramowitzTaylorMcKay09}]{AbramowitzTaylorMcKay09}
Abramowitz, J.~S., Taylor, S., \& McKay, D. (2009).
\newblock Obessive-compulsive disorder.
\newblock {\em The Lancet}, {\em 374\/}, 491--499.

\bibitem[\protect\citename{Ajzen}{}{}{Ajzen}{1991}{Ajzen91}]{Ajzen91}
Ajzen, I. (1991).
\newblock The theory of planned behavior.
\newblock {\em Organizational Behavior and Human Decision Processes}, {\em
  50\/}, 179--211.

\bibitem[\protect\citename{Alexander}{}{Brown}{Alexander-Brown}{2010}{AlexanderBrown10}]{AlexanderBrown10}
Alexander, W.~H., \& Brown, J.~W. (2010).
\newblock Computational models of performance monitoring and cognitive control.
\newblock {\em Topics in Cognitive Science}, {\em 2\/}(4), 658--677.

\bibitem[\protect\citename{Amaral}{Price,
  Pitkanen}{Carmichael}{AmaralETAL}{1992}{AmaralPricePitkanenEtAl92}]{AmaralPricePitkanenEtAl92}
Amaral, D.~G., Price, J.~L., Pitkanen, A., \& Carmichael, S.~T. (1992).
\newblock Anatomical organization of the primate amygdaloid complex.
\newblock In {\em {The Amygdala: Neurobiological Aspects of Emotion, Memory,
  and Mental Dysfunction}} (pp.\ 1--66). New York: Wiley-Liss, 1st edition.

\bibitem[\protect\citename{Amaral}{Veazey}{Cowan}{AmaralETAL}{1982}{AmaralVeazeyCowan82}]{AmaralVeazeyCowan82}
Amaral, D.~G., Veazey, R.~B., \& Cowan, W.~M. (1982).
\newblock Some observations on hypothalamo-amygdaloid connections in the
  monkey.
\newblock {\em Brain Research}, {\em 252\/}(1), 13--27.

\bibitem[\protect\citename{Amitai~Shenhav}{}{}{Amitai~Shenhav}{2013}{ShenhavBotvinickCohen13}]{ShenhavBotvinickCohen13}
Amitai~Shenhav, Matthew M.~Botvinick, J. D.~C. (2013).
\newblock The expected value of control: An integrative theory of anterior
  cingulate cortex function.
\newblock {\em Neuron}, {\em 79\/}, 217--240.

\bibitem[\protect\citename{Amsel}{}{}{Amsel}{1962}{Amsel62}]{Amsel62}
Amsel, A. (1962).
\newblock Frustrative nonreward in partial reinforcement and discrimination
  learning: some recent history and a theoretical extension.
\newblock {\em Psychological review}, {\em 69\/}, 306--328.

\bibitem[\protect\citename{Anderson}{}{}{Anderson}{1990}{Anderson90}]{Anderson90}
Anderson, J.~R. (1990).
\newblock {\em {The Adaptive Character of Thought}}.
\newblock Hillsdale, NJ: Lawrence Erlbaum Associates.

\bibitem[\protect\citename{Anderson}{Bothell, Byrne, Douglass,
  Lebiere}{Qin}{AndersonETAL}{2004}{AndersonBothellByrneEtAl04}]{AndersonBothellByrneEtAl04}
Anderson, J.~R., Bothell, D., Byrne, M.~D., Douglass, S., Lebiere, C., \& Qin,
  Y. (2004).
\newblock An integrated theory of the mind.
\newblock {\em Psychological Review}, {\em 111\/}(4), 1036--1060.

\bibitem[\protect\citename{Anderson}{}{Lebiere}{Anderson-Lebiere}{1998}{AndersonLebiere98}]{AndersonLebiere98}
Anderson, J.~R., \& Lebiere, C. (1998).
\newblock {\em The atomic components of thought.}
\newblock Mahwah, NJ: Erlbaum.

\bibitem[\protect\citename{Andrews}{}{Thomson}{Andrews-Thomson}{2009}{AndrewsThomson09}]{AndrewsThomson09}
Andrews, P.~W., \& Thomson, J.~A. (2009).
\newblock The bright side of being blue: depression as an adaptation for
  analyzing complex problems.
\newblock {\em Psychological review}, {\em 116\/}.

\bibitem[\protect\citename{Apicella}{Ljungberg,
  Scarnati}{Schultz}{ApicellaETAL}{1991}{ApicellaLjungbergScarnatiEtAl91}]{ApicellaLjungbergScarnatiEtAl91}
Apicella, P., Ljungberg, T., Scarnati, E., \& Schultz, W. (1991).
\newblock Responses to reward in monkey dorsal and ventral striatum.
\newblock {\em Experimental Brain Research}, {\em 85\/}(3), 491--500.

\bibitem[\protect\citename{Apkarian}{}{}{Apkarian}{2008}{Apkarian08}]{Apkarian08}
Apkarian, A.~V. (2008).
\newblock Pain perception in relation to emotional learning.
\newblock {\em Current opinion in neurobiology}, {\em 18\/}.

\bibitem[\protect\citename{Aston-Jones}{}{Cohen}{Aston-Jones-Cohen}{2005}{Aston-JonesCohen05}]{Aston-JonesCohen05}
Aston-Jones, G., \& Cohen, J.~D. (2005).
\newblock An integrative theory of locus coeruleus-norepinephrine function:
  adaptive gain and optimal performance.
\newblock {\em Annual review of neuroscience}, {\em 28\/}, 403--450.

\bibitem[\protect\citename{Atallah}{McCool,
  Howe}{Graybiel}{AtallahETAL}{2014}{AtallahMcCoolHoweEtAl14}]{AtallahMcCoolHoweEtAl14}
Atallah, H.~E., McCool, A.~D., Howe, M.~W., \& Graybiel, A.~M. (2014).
\newblock Neurons in the ventral striatum exhibit cell-type specific
  representations of outcome during learning.
\newblock {\em Neuron}.

\bibitem[\protect\citename{Atlas}{Bolger,
  Lindquist}{Wager}{AtlasETAL}{2010}{AtlasBolgerLindquistEtAl10}]{AtlasBolgerLindquistEtAl10}
Atlas, L.~Y., Bolger, N., Lindquist, M.~A., \& Wager, T.~D. (2010).
\newblock Brain mediators of predictive cue effects on perceived pain.
\newblock {\em The Journal of neuroscience}, {\em 30\/}.

\bibitem[\protect\citename{Atlas}{Whittington, Lindquist, Wielgosz,
  Sonty}{Wager}{AtlasETAL}{2012}{AtlasWhittingtonLindquistEtAl12}]{AtlasWhittingtonLindquistEtAl12}
Atlas, L.~Y., Whittington, R.~A., Lindquist, M.~A., Wielgosz, J., Sonty, N., \&
  Wager, T.~D. (2012).
\newblock Dissociable influences of opiates and expectations on pain.
\newblock {\em The Journal of neuroscience}, {\em 32\/}.

\bibitem[\protect\citename{Austin}{}{Vancouver}{Austin-Vancouver}{1996}{AustinVancouver96}]{AustinVancouver96}
Austin, J.~T., \& Vancouver, J.~B. (1996).
\newblock Goal constructs in psychology: Structure, process, and context.
\newblock {\em Psychological Bulletin}, {\em 120\/}, 338--375.

\bibitem[\protect\citename{Balleine}{}{Dickinson}{Balleine-Dickinson}{1998}{BalleineDickinson98}]{BalleineDickinson98}
Balleine, B.~W., \& Dickinson, A. (1998).
\newblock Goal-directed instrumental action: contingency and incentive learning
  and their cortical substrates.
\newblock {\em Neuropharmacology}, {\em 37\/}, 407--19.

\bibitem[\protect\citename{Balleine}{}{Killcross}{Balleine-Killcross}{2006}{BalleineKillcross06}]{BalleineKillcross06}
Balleine, B.~W., \& Killcross, S. (2006).
\newblock Parallel incentive processing: {A}n integrated view of amygdala
  function.
\newblock {\em Trends in Neurosciences}, {\em 29\/}(5), 272--279.

\bibitem[\protect\citename{Bandura}{}{}{Bandura}{1977}{Bandura77}]{Bandura77}
Bandura, A. (1977).
\newblock Self-efficacy: {T}oward a unifying theory of behavioral change.
\newblock {\em Psychological Review}, {\em 84\/}(2), 191--215.

\bibitem[\protect\citename{Bandura}{}{}{Bandura}{2001}{Bandura01}]{Bandura01}
Bandura, A. (2001).
\newblock Social cognitive theory: {A}n agentic perspective.
\newblock {\em Annual Review of Psychology}, {\em 52\/}, 1--26.

\bibitem[\protect\citename{Baxter}{}{Murray}{Baxter-Murray}{2002}{BaxterMurray02}]{BaxterMurray02}
Baxter, M.~G., \& Murray, E.~A. (2002).
\newblock The amygdala and reward.
\newblock {\em Nature Reviews Neuroscience}, {\em 3\/}, 563--572.

\bibitem[\protect\citename{Bechara}{Damasio,
  Tranel}{Damasio}{BecharaETAL}{2005}{BecharaDamasioTranelEtAl05}]{BecharaDamasioTranelEtAl05}
Bechara, A., Damasio, H., Tranel, D., \& Damasio, A.~R. (2005).
\newblock The {Iowa Gambling T}ask and the somatic marker hypothesis: {S}ome
  questions and answers.
\newblock {\em Trends in Cognitive Sciences}, {\em 9\/}(4), 159--162.

\bibitem[\protect\citename{Belova}{Paton,
  Morrison}{Salzman}{BelovaETAL}{2007}{BelovaPatonMorrisonEtAl07}]{BelovaPatonMorrisonEtAl07}
Belova, M.~A., Paton, J.~J., Morrison, S.~E., \& Salzman, C.~D. (2007).
\newblock Expectation modulates neural responses to pleasant and aversive
  stimuli in primate amygdala.
\newblock {\em Neuron}, {\em 55\/}(6), 970--984.

\bibitem[\protect\citename{Belujon}{}{Grace}{Belujon-Grace}{2011}{BelujonGrace11}]{BelujonGrace11}
Belujon, P., \& Grace, A.~A. (2011).
\newblock Hippocampus, amygdala, and stress: interacting systems that affect
  susceptibility to addiction.
\newblock {\em Annals of the New York Academy of Sciences}, {\em 1216\/}.

\bibitem[\protect\citename{Berridge}{}{Robinson}{Berridge-Robinson}{1998}{BerridgeRobinson98}]{BerridgeRobinson98}
Berridge, K.~C., \& Robinson, T.~E. (1998).
\newblock What is the role of dopamine in reward: {H}edonic impact, reward
  learning or incentive salience.
\newblock {\em Brain Research. Brain Research Reviews}, {\em 28\/}(3),
  309--369.

\bibitem[\protect\citename{Berridge}{}{Robinson}{Berridge-Robinson}{2003}{BerridgeRobinson03}]{BerridgeRobinson03}
Berridge, K.~C., \& Robinson, T.~E. (2003).
\newblock Parsing reward.
\newblock {\em Trends in Neurosciences}, {\em 26\/}(9), 507--513.

\bibitem[\protect\citename{Bienenstock}{Cooper}{Munro}{BienenstockETAL}{1982}{BienenstockCooperMunro82}]{BienenstockCooperMunro82}
Bienenstock, E.~L., Cooper, L.~N., \& Munro, P.~W. (1982).
\newblock {Theory for the development of neuron selectivity}: {Orientation
  specificity and binocular interaction in visual cortex.}
\newblock {\em The Journal of Neuroscience}, {\em 2\/}(2), 32--48.

\bibitem[\protect\citename{Bostan}{Dum}{Strick}{BostanETAL}{2013}{BostanDumStrick13}]{BostanDumStrick13}
Bostan, A.~C., Dum, R.~P., \& Strick, P.~L. (2013).
\newblock Cerebellar networks with the cerebral cortex and basal ganglia.
\newblock {\em Trends in Cognitive Sciences}, {\em 17\/}(5), 241--254.

\bibitem[\protect\citename{Bourdy}{}{Barrot}{Bourdy-Barrot}{2012}{BourdyBarrot12}]{BourdyBarrot12}
Bourdy, R., \& Barrot, M. (2012).
\newblock A new control center for dopaminergic systems: pulling the {VTA} by
  the tail.
\newblock {\em Trends in Neurosciences}.

\bibitem[\protect\citename{Bouton}{}{}{Bouton}{2011}{Bouton11}]{Bouton11}
Bouton, M.~E. (2011).
\newblock Learning and the persistence of appetite: {E}xtinction and the
  motivation to eat and overeat.
\newblock {\em Physiology \& Behavior}, {\em 103\/}(1), 51--58.

\bibitem[\protect\citename{Brette}{}{Gerstner}{Brette-Gerstner}{2005}{BretteGerstner05}]{BretteGerstner05}
Brette, R., \& Gerstner, W. (2005).
\newblock Adaptive exponential integrate-and-fire model as an effective
  description of neuronal activity.
\newblock {\em Journal of Neurophysiology}, {\em 94\/}(5), 3637--3642.

\bibitem[\protect\citename{Brischoux}{Chakraborty,
  Brierley}{Ungless}{BrischouxETAL}{2009}{BrischouxChakrabortyBrierleyEtAl09}]{BrischouxChakrabortyBrierleyEtAl09}
Brischoux, F., Chakraborty, S., Brierley, D.~I., \& Ungless, M.~A. (2009).
\newblock Phasic excitation of dopamine neurons in ventral vta by noxious
  stimuli.
\newblock {\em Proceedings of the National Academy of Sciences USA}, {\em
  106\/}(12), 4894--4899.

\bibitem[\protect\citename{Bromberg-Martin}{Matsumoto,
  Hong}{Hikosaka}{Bromberg-MartinETAL}{2010}{Bromberg-MartinMatsumotoHongEtAl10}]{Bromberg-MartinMatsumotoHongEtAl10}
Bromberg-Martin, E.~S., Matsumoto, M., Hong, S., \& Hikosaka, O. (2010).
\newblock A pallidus-habenula-dopamine pathway signals inferred stimulus
  values.
\newblock {\em Journal of {N}europhysiology}, {\em 104\/}(2), 1068--1076.

\bibitem[\protect\citename{Buneo}{}{Andersen}{Buneo-Andersen}{2006}{BuneoAndersen06}]{BuneoAndersen06}
Buneo, C.~A., \& Andersen, R.~A. (2006).
\newblock The posterior parietal cortex: sensorimotor interface for the
  planning and online control of visually guided movements.
\newblock {\em Neuropsychologia}, {\em 44\/}.

\bibitem[\protect\citename{Cabeza}{Kapur}{Tulving}{CabezaETAL}{1997}{CabezaKapurTulving97}]{CabezaKapurTulving97}
Cabeza, R., Kapur, S., \& Tulving, E. (1997).
\newblock Functional neuroanatomy of recall and recognition: a pet study of
  episodic memory.
\newblock {\em Journal of Cognitive Neuroscience}, {\em 9\/}, 254.

\bibitem[\protect\citename{Cabib}{}{Puglisi-Allegra}{Cabib-Puglisi-Allegra}{2012}{CabibPuglisi-Allegra12}]{CabibPuglisi-Allegra12}
Cabib, S., \& Puglisi-Allegra, S. (2012).
\newblock The mesoaccumbens dopamine in coping with stress.
\newblock {\em Neuroscience and biobehavioral reviews}, {\em 36\/}.

\bibitem[\protect\citename{Cardinal}{Parkinson,
  Hall}{Everitt}{CardinalETAL}{2002}{CardinalParkinsonHallEtAl02}]{CardinalParkinsonHallEtAl02}
Cardinal, R.~N., Parkinson, J.~A., Hall, J., \& Everitt, B.~J. (2002).
\newblock Emotion and motivation: the role of the amygdala, ventral striatum,
  and prefrontal cortex.
\newblock {\em Neuroscience and biobehavioral reviews}, {\em 26\/}, 321--52.

\bibitem[\protect\citename{Carmichael}{}{Price}{Carmichael-Price}{1996}{CarmichaelPrice95}]{CarmichaelPrice95}
Carmichael, S.~T., \& Price, J.~L. (1996).
\newblock Limbic connections of the orbital and medial prefrontal cortex in
  macaque monkeys.
\newblock {\em The Journal of comparative neurology}, {\em 363\/}, 615--41.

\bibitem[\protect\citename{Carver}{}{}{Carver}{2006}{Carver06}]{Carver06}
Carver, C. (2006).
\newblock Approach, avoidance, and the self-regulation of affect and action.
\newblock {\em Motivation and Emotion}, 105--110.

\bibitem[\protect\citename{Carver}{}{Harmon-Jones}{Carver-Harmon-Jones}{2009}{CarverHarmonJones09}]{CarverHarmonJones09}
Carver, C.~S., \& Harmon-Jones, E. (2009).
\newblock Anger is an approach-related affect: evidence and implications.
\newblock {\em Psychological bulletin}, {\em 135\/}.

\bibitem[\protect\citename{Carver}{}{Scheier}{Carver-Scheier}{1982}{CarverScheier82}]{CarverScheier82}
Carver, C.~S., \& Scheier, M.~F. (1982).
\newblock Control theory: {A} useful conceptual framework for
  personality-social, clinical, and health psychology.
\newblock {\em Psychological Bulletin}, {\em 92\/}(1), 111--135.

\bibitem[\protect\citename{Carver}{}{Scheier}{Carver-Scheier}{1990}{CarverScheier90}]{CarverScheier90}
Carver, C.~S., \& Scheier, M.~F. (1990).
\newblock Origins and functions of positive and negative affect: A
  control-process view.
\newblock {\em Psychological Review}, {\em 97\/}, 19--35.

\bibitem[\protect\citename{Carver}{}{White}{Carver-White}{1994}{CarverWhite94}]{CarverWhite94}
Carver, C.~S., \& White, T. (1994).
\newblock Behavioral inhibition, behavioral activation, and affective responses
  to impending reward and punishment: The bis/bas scales.
\newblock {\em Journal of Personality and Social Psychology}, {\em 67\/},
  319--333.

\bibitem[\protect\citename{Castellanos}{Sonuga-Barke,
  Milham}{Tannock}{CastellanosETAL}{2006}{CastellanosSonuga-BarkeMilhamEtAl06}]{CastellanosSonuga-BarkeMilhamEtAl06}
Castellanos, F.~X., Sonuga-Barke, E. J.~S., Milham, M.~P., \& Tannock, R.
  (2006).
\newblock Characterizing cognition in adhd: beyond executive dysfunction.
\newblock {\em Trends in cognitive sciences}, {\em 10\/}(3), 117--123.

\bibitem[\protect\citename{Charara}{}{Grace}{Charara-Grace}{2003}{ChararaGrace03}]{ChararaGrace03}
Charara, A., \& Grace, A.~A. (2003).
\newblock Dopamine receptor subtypes selectively modulate excitatory afferents
  from the hippocampus and amygdala to rat nucleus accumbens neurons.
\newblock {\em Neuropsychopharmacology}, {\em 28\/}(8), 1412--1421.

\bibitem[\protect\citename{Charron}{}{Koechlin}{Charron-Koechlin}{2010}{CharronKoechlin10}]{CharronKoechlin10}
Charron, S., \& Koechlin, E. (2010).
\newblock Divided representation of concurrent goals in the human frontal
  lobes.
\newblock {\em Science}, {\em 328\/}, 360--363.

\bibitem[\protect\citename{Chatham}{Claus, Kim, Curran,
  Banich}{Munakata}{ChathamETAL}{2012}{ChathamClausKimEtAl12}]{ChathamClausKimEtAl12}
Chatham, C.~H., Claus, E.~D., Kim, A., Curran, T., Banich, M.~T., \& Munakata,
  Y. (2012).
\newblock Cognitive control reflects context monitoring, not motoric stopping,
  in response inhibition.
\newblock {\em PloS one}, {\em 7\/}.

\bibitem[\protect\citename{Chou}{Scammell, Gooley, Gaus,
  Saper}{Lu}{ChouETAL}{2003}{ChouScammellGooleyEtAl03}]{ChouScammellGooleyEtAl03}
Chou, T.~C., Scammell, T.~E., Gooley, J.~J., Gaus, S.~E., Saper, C.~B., \& Lu,
  J. (2003).
\newblock Critical role of dorsomedial hypothalamic nucleus in a wide range of
  behavioral circadian rhythms.
\newblock {\em The Journal of neuroscience}, {\em 23\/}.

\bibitem[\protect\citename{Cooper}{Intrator,
  Blais}{Shouval}{CooperETAL}{2004}{CooperIntratorBlaisEtAl04}]{CooperIntratorBlaisEtAl04}
Cooper, L.~N., Intrator, N., Blais, B.~S., \& Shouval, H. (2004).
\newblock {\em Theory of cortical plasticity.}
\newblock New Jersey: World Scientific.

\bibitem[\protect\citename{Crittenden}{}{Graybiel}{Crittenden-Graybiel}{2011}{CrittendenGraybiel11}]{CrittendenGraybiel11}
Crittenden, J.~R., \& Graybiel, A.~M. (2011).
\newblock Basal ganglia disorders associated with imbalances in the striatal
  striosome and matrix compartments.
\newblock {\em Frontiers in Neuroanatomy}, {\em 5\/}, Epub.

\bibitem[\protect\citename{Croxson}{Walton, O'Reilly,
  Behrens}{Rushworth}{CroxsonETAL}{2009}{CroxsonWaltonOReillyEtAl09}]{CroxsonWaltonOReillyEtAl09}
Croxson, P.~L., Walton, M.~E., O'Reilly, J.~X., Behrens, T. E.~J., \&
  Rushworth, M. F.~S. (2009).
\newblock Effort-based cost-benefit valuation and the human brain.
\newblock {\em The Journal of {N}euroscience}, {\em 29\/}(14), 4531--4541.

\bibitem[\protect\citename{Daniela}{}{R.}{Daniela-R.}{2010}{SchillerDelgado10}]{SchillerDelgado10}
Daniela, S., \& R., D.~M. (2010).
\newblock Overlapping neural systems mediating extinction reversal and
  regulation of fear.
\newblock {\em Trends in cognitive sciences}, {\em 14\/}(6), 268--276.

\bibitem[\protect\citename{Davidson}{}{}{Davidson}{2000}{Davidson00}]{Davidson00}
Davidson, R.~J. (2000).
\newblock Affective style, psychopathology, and resilience: brain mechanisms
  and plasticity.
\newblock {\em The American psychologist}, {\em 55\/}.

\bibitem[\protect\citename{Daw}{Niv}{Dayan}{DawETAL}{2005}{DawNivDayan05}]{DawNivDayan05}
Daw, N.~D., Niv, Y., \& Dayan, P. (2005).
\newblock Uncertainty-based competition between prefrontal and dorsolateral
  striatal systems for behavioral control.
\newblock {\em Nature Neuroscience}, {\em 8\/}(12), 1704--1711.

\bibitem[\protect\citename{Dayan}{}{}{Dayan}{2009}{Dayan09}]{Dayan09}
Dayan, P. (2009).
\newblock Goal-directed control and its antipodes.
\newblock {\em Neural Networks}, {\em 22\/}(3), 213--219.

\bibitem[\protect\citename{Demyttenaere}{De~Fruyt}{Stahl}{DemyttenaereETAL}{2005}{DemyttenaereDeFruytStahl05}]{DemyttenaereDeFruytStahl05}
Demyttenaere, K., De~Fruyt, J., \& Stahl, S.~M. (2005).
\newblock The many faces of fatigue in major depressive disorder.
\newblock {\em The international journal of neuropsychopharmacology / official
  scientific journal of the Collegium Internationale
  Neuropsychopharmacologicum}, {\em 8\/}.

\bibitem[\protect\citename{Denny}{}{Ratner}{Denny-Ratner}{1970}{DennyRatner70}]{DennyRatner70}
Denny, M.~R., \& Ratner, S.~C. (1970).
\newblock {\em Comparative psychology: Research in animal behavior, revised
  edition.}
\newblock Homewood, Illinois: The Dorsey Press.

\bibitem[\protect\citename{Desmurget}{Reilly, Richard, Szathmari,
  Mottolese}{Sirigu}{DesmurgetETAL}{2009}{DesmurgetReillyRichardEtAl09}]{DesmurgetReillyRichardEtAl09}
Desmurget, M., Reilly, K.~T., Richard, N., Szathmari, A., Mottolese, C., \&
  Sirigu, A. (2009).
\newblock Movement intention after parietal cortex stimulation in humans.
\newblock {\em Science}, {\em 324\/}, 811--813.

\bibitem[\protect\citename{Dickson}{}{MacLeod}{Dickson-MacLeod}{2004}{DicksonMacLeod04}]{DicksonMacLeod04}
Dickson, J.~M., \& MacLeod, A.~K. (2004).
\newblock Approach and avoidance goals and plans: {T}heir relationship to
  anxiety and depression.
\newblock {\em Cognitive Therapy and Research}, {\em 28\/}(3), 415--432.

\bibitem[\protect\citename{Dinsmoor}{}{}{Dinsmoor}{2001}{Dinsmoor01}]{Dinsmoor01}
Dinsmoor, J.~A. (2001).
\newblock Stimuli inevitably generated by behavior that avoids electric shock
  are inherently reinforcing.
\newblock {\em Journal of the Experimental Analysis of Behavior}, {\em 75\/},
  311--333.

\bibitem[\protect\citename{Disner}{Beevers,
  Haigh}{Beck}{DisnerETAL}{2011}{DisnerBeeversHaighEtAl11}]{DisnerBeeversHaighEtAl11}
Disner, S.~G., Beevers, C.~G., Haigh, E. A.~P., \& Beck, A.~T. (2011).
\newblock Neural mechanisms of the cognitive model of depression.
\newblock {\em Nature reviews}, {\em 12\/}.

\bibitem[\protect\citename{Dovis}{Van~der Oord,
  Wiers}{Prins}{DovisETAL}{2012}{DovisVanderOordWiersEtAl12}]{DovisVanderOordWiersEtAl12}
Dovis, S., Van~der Oord, S., Wiers, R.~W., \& Prins, P. J.~M. (2012).
\newblock Can motivation normalize working memory and task persistence in
  children with attention-deficit/hyperactivity disorder? the effects of money
  and computer-gaming.
\newblock {\em Journal of abnormal child psychology}, {\em 40\/}.

\bibitem[\protect\citename{Doya}{}{}{Doya}{2002}{Doya02}]{Doya02}
Doya, K. (2002).
\newblock Metalearning and neuromodulation.
\newblock {\em Neural Networks}, {\em 15\/}, 495--506.

\bibitem[\protect\citename{Doya}{}{}{Doya}{2008}{Doya08}]{Doya08}
Doya, K. (2008).
\newblock Modulators of decision making.
\newblock {\em Nature neuroscience}, {\em 11\/}.

\bibitem[\protect\citename{Eippert}{Bingel, Schoell, Yacubian, Klinger,
  Lorenz}{Buchel}{EippertETAL}{2009}{EippertBingelSchoellEtAl09}]{EippertBingelSchoellEtAl09}
Eippert, F., Bingel, U., Schoell, E.~D., Yacubian, J., Klinger, R., Lorenz, J.,
  \& Buchel, C. (2009).
\newblock Activation of the opioidergic descending pain control system
  underlies placebo analgesia.
\newblock {\em Neuron}, {\em 63\/}.

\bibitem[\protect\citename{Eisenberger}{}{}{Eisenberger}{2012}{Eisenberger12}]{Eisenberger12}
Eisenberger, N.~I. (2012).
\newblock The pain of social disconnection: examining the shared neural
  underpinnings of physical and social pain.
\newblock {\em Nature reviews}, {\em 13\/}.

\bibitem[\protect\citename{Elliot}{}{}{Elliot}{2006}{Elliot06}]{Elliot06}
Elliot, A.~J. (2006).
\newblock The hierarchical model of approach-avoidance motivation.
\newblock {\em Motivation and Emotion}, {\em 30\/}(2), 111--116.

\bibitem[\protect\citename{Elliot}{}{Sheldon}{Elliot-Sheldon}{1997}{ElliotSheldon97}]{ElliotSheldon97}
Elliot, A.~J., \& Sheldon, K.~M. (1997).
\newblock Avoidance achievement motivation: a personal goals analysis.
\newblock {\em Journal of personality and social psychology}, {\em 73\/}.

\bibitem[\protect\citename{Fallon}{}{Ciofi}{Fallon-Ciofi}{1992}{FallonCiofi92}]{FallonCiofi92}
Fallon, J.~H., \& Ciofi, P. (1992).
\newblock Distribution of monoamines within the amygdala.
\newblock In J.~P. Aggleton (Ed.), {\em {The Amygdala: Neurobiological Aspects
  of Emotion, Memory, and Mental Dysfunction}} (pp.\ 97--114). New York:
  Wiley-Liss, 1st edition.

\bibitem[\protect\citename{Flavin}{}{Winder}{Flavin-Winder}{2013}{FlavinWinder13}]{FlavinWinder13}
Flavin, S.~A., \& Winder, D.~G. (2013).
\newblock Noradrenergic control of the bed nucleus of the stria terminalis in
  stress and reward.
\newblock {\em Neuropharmacology}, {\em 70\/}.

\bibitem[\protect\citename{Floresco}{}{}{Floresco}{2007}{Floresco07}]{Floresco07}
Floresco, S.~B. (2007).
\newblock Dopaminergic regulation of limbic-striatal interplay.
\newblock {\em Journal of Psychiatry \& Neuroscience}, {\em 32\/}(6), 400--411.

\bibitem[\protect\citename{Floresco}{Blaha,
  Yang}{Phillips}{FlorescoETAL}{2001}{FlorescoBlahaYangEtAl01}]{FlorescoBlahaYangEtAl01}
Floresco, S.~B., Blaha, C.~D., Yang, C.~R., \& Phillips, A.~G. (2001).
\newblock Dopamine {D1 and NMDA} receptors mediate potentiation of basolateral
  amygdala-evoked firing of nucleus accumbens neurons.
\newblock {\em The Journal of Neuroscience}, {\em 21\/}(16), 6370--6376.

\bibitem[\protect\citename{Floresco}{}{Ghods-Sharifi}{Floresco-Ghods-Sharifi}{2007}{FlorescoGhods-Sharifi07}]{FlorescoGhods-Sharifi07}
Floresco, S.~B., \& Ghods-Sharifi, S. (2007).
\newblock Amygdala-prefrontal cortical circuitry regulates effort-based
  decision making.
\newblock {\em Cerebral cortex}, {\em 17\/}.

\bibitem[\protect\citename{Floresco}{St~Onge,
  Ghods-Sharifi}{Winstanley}{FlorescoETAL}{2008}{FlorescoStOngeGhods-SharifiEtAl08}]{FlorescoStOngeGhods-SharifiEtAl08}
Floresco, S.~B., St~Onge, J.~R., Ghods-Sharifi, S., \& Winstanley, C.~A.
  (2008).
\newblock Cortico-limbic-striatal circuits subserving different forms of
  cost-benefit decision making.
\newblock {\em Cognitive, Affective \& Behavioral Neuroscience}, {\em 8\/}.

\bibitem[\protect\citename{Floresco}{West, Ash,
  Moore}{Grace}{FlorescoETAL}{2003}{FlorescoWestAshEtAl03}]{FlorescoWestAshEtAl03}
Floresco, S.~B., West, A.~R., Ash, B., Moore, H., \& Grace, A.~A. (2003).
\newblock Afferent modulation of dopamine neuron firing differentially
  regulates tonic and phasic dopamine transmission.
\newblock {\em Nature Neuroscience}, {\em 6\/}, 968--973.

\bibitem[\protect\citename{Frank}{}{}{Frank}{2005}{Frank05}]{Frank05}
Frank, M.~J. (2005).
\newblock Dynamic dopamine modulation in the basal ganglia: A
  neurocomputational account of cognitive deficits in medicated and
  non-medicated {P}arkinsonism.
\newblock {\em Journal of Cognitive Neuroscience}, {\em 17\/}, 51--72.

\bibitem[\protect\citename{Frank}{}{Claus}{Frank-Claus}{2006}{FrankClaus06}]{FrankClaus06}
Frank, M.~J., \& Claus, E.~D. (2006).
\newblock Anatomy of a decision: Striato-orbitofrontal interactions in
  reinforcement learning, decision making, and reversal.
\newblock {\em Psychological Review}, {\em 113\/}(2), 300--326.

\bibitem[\protect\citename{Frank}{Santamaria,
  O'Reilly}{Willcutt}{FrankETAL}{2007}{FrankSantamariaOReillyEtAl07}]{FrankSantamariaOReillyEtAl07}
Frank, M.~J., Santamaria, A., O'Reilly, R.~C., \& Willcutt, E. (2007).
\newblock Testing computational models of dopamine and noradrenaline
  dysfunction in attention deficit/hyperactivity disorder.
\newblock {\em Neuropsychopharmacology : official publication of the American
  College of Neuropsychopharmacology}, {\em 32\/}, 1583--1599.

\bibitem[\protect\citename{Franklin}{Saab}{Mansuy}{FranklinETAL}{2012}{FranklinSaabMansuy12}]{FranklinSaabMansuy12}
Franklin, T.~B., Saab, B.~J., \& Mansuy, I.~M. (2012).
\newblock Neural mechanisms of stress resilience and vulnerability.
\newblock {\em Neuron}, {\em 75\/}.

\bibitem[\protect\citename{Fudge}{}{Haber}{Fudge-Haber}{2000}{FudgeHaber00}]{FudgeHaber00}
Fudge, J.~L., \& Haber, S.~N. (2000).
\newblock The central nucleus of the amygdala projection to dopamine
  subpopulations in primates.
\newblock {\em Neuroscience}, {\em 97\/}, 479--494.

\bibitem[\protect\citename{Fukuda}{}{Ono}{Fukuda-Ono}{1993}{FukudaOno93}]{FukudaOno93}
Fukuda, M., \& Ono, T. (1993).
\newblock Amygdala-hypothalamic control of feeding behavior in monkey: Single
  cell responses before and after reversible blockade of temporal cortex or
  amygdala projections.
\newblock {\em Behav Brain Research}, {\em 55\/}(2), 233--141.

\bibitem[\protect\citename{Fuster}{}{Alexander}{Fuster-Alexander}{1971}{FusterAlexander71}]{FusterAlexander71}
Fuster, J.~M., \& Alexander, G.~E. (1971).
\newblock Neuron activity related to short-term memory.
\newblock {\em Science}, {\em 173\/}, 652--654.

\bibitem[\protect\citename{Geman}{Bienenstock}{Doursat}{GemanETAL}{1992}{GemanBienenstockDoursat92}]{GemanBienenstockDoursat92}
Geman, S., Bienenstock, E.~L., \& Doursat, R. (1992).
\newblock Neural networks and the bias/variance dilemma.
\newblock {\em Neural Computation}, {\em 4\/}, 1--58.

\bibitem[\protect\citename{Gerfen}{}{}{Gerfen}{1985}{Gerfen85}]{Gerfen85}
Gerfen, C.~R. (1985).
\newblock The neostriatal mosaic: I. compartmental organization of projections
  of the striatonigral system in the rat.
\newblock {\em Journal of Comparative Neurology}, {\em 236\/}, 454--476.

\bibitem[\protect\citename{Gerfen}{Herkenham}{Thibault}{GerfenETAL}{1987}{GerfenHerkenhamThibault87}]{GerfenHerkenhamThibault87}
Gerfen, C.~R., Herkenham, M., \& Thibault, J. (1987).
\newblock The neostriatal mosaic: Ii. patch- and matrix-directed mesostriatal
  dopaminergic and non-dopaminergic systems.
\newblock {\em Journal of Neuroscience}, {\em 7\/}(12), 3915--3934.

\bibitem[\protect\citename{Gilbert}{Spengler, Simons, Steele, Lawrie,
  Frith}{Burgess}{GilbertETAL}{2006}{GilbertSpenglerSimonsEtAl06}]{GilbertSpenglerSimonsEtAl06}
Gilbert, S.~J., Spengler, S., Simons, J.~S., Steele, J.~D., Lawrie, S.~M.,
  Frith, C.~D., \& Burgess, P.~W. (2006).
\newblock Functional specialization within rostral prefrontal cortex (area 10):
  a meta-analysis.
\newblock {\em Journal of cognitive neuroscience}, {\em 18\/}(6), 932--948.

\bibitem[\protect\citename{Gobet}{}{Simon}{Gobet-Simon}{1996}{GobetSimon96}]{GobetSimon96}
Gobet, F., \& Simon, H.~A. (1996).
\newblock the roles of recognition processes and look-ahead search in
  time-constrained expert problems solving: Evidence from grand-master-level
  chess.
\newblock {\em Psychological Science}, {\em 7\/}, 52.

\bibitem[\protect\citename{Goldman-Rakic}{}{}{Goldman-Rakic}{1987}{GoldmanRakic87}]{GoldmanRakic87}
Goldman-Rakic, P.~S. (1987).
\newblock Circuitry of primate prefrontal cortex and regulation of behavior by
  representational memory.
\newblock {\em Handbook of Physiology --- The Nervous System}, {\em 5\/},
  373--417.

\bibitem[\protect\citename{Gollwitzer}{}{}{Gollwitzer}{1993}{Gollwitzer93}]{Gollwitzer93}
Gollwitzer, P.~M. (1993).
\newblock Goal achievement: The role of intentions.
\newblock {\em European Review of Social Psychology}, {\em 4\/}, 141--185.

\bibitem[\protect\citename{Gollwitzer}{}{Sheeran}{Gollwitzer-Sheeran}{2006}{GollwitzerSheeran06}]{GollwitzerSheeran06}
Gollwitzer, P.~M., \& Sheeran, P. (2006).
\newblock Implementation intentions and goal achievement: A meta-analysis of
  effects and processes.
\newblock {\em Advances in experimental social psychology}, {\em 38\/},
  69--119.

\bibitem[\protect\citename{Goodale}{}{Milner}{Goodale-Milner}{1992}{GoodaleMilner92}]{GoodaleMilner92}
Goodale, M.~A., \& Milner, A.~D. (1992).
\newblock Separate visual pathways for perception and action.
\newblock {\em Trends in Neurosciences}, {\em 15\/}(1), 20--25.

\bibitem[\protect\citename{Grace}{}{Bunney}{Grace-Bunney}{1984}{GraceBunney84}]{GraceBunney84}
Grace, A.~A., \& Bunney, B.~S. (1984).
\newblock The control of firing pattern in nigral dopamine neurons: burst
  firing.
\newblock {\em The Journal of Neuroscience}, {\em 4\/}(11), 2877--2890.

\bibitem[\protect\citename{Gustavson}{Miyake,
  Hewitt}{Friedman}{GustavsonETAL}{2014}{GustavsonMiyakeHewittEtAl14}]{GustavsonMiyakeHewittEtAl14}
Gustavson, D.~E., Miyake, A., Hewitt, J.~K., \& Friedman, N.~P. (2014).
\newblock Genetic influences on procrastination overlap completely with those
  on impulsivity: Implications for the evolutionary origin of procrastination.
\newblock {\em Psychological Science}.

\bibitem[\protect\citename{Hartley}{}{Phelps}{Hartley-Phelps}{2010}{HartleyPhelps10}]{HartleyPhelps10}
Hartley, C.~A., \& Phelps, E.~A. (2010).
\newblock Changing fear: the neurocircuitry of emotion regulation.
\newblock {\em Neuropsychopharmacology}, {\em 35\/}.

\bibitem[\protect\citename{Hatfield}{Han,
  Conley}{Holland}{HatfieldETAL}{1996}{HatfieldHanConleyEtAl96}]{HatfieldHanConleyEtAl96}
Hatfield, T., Han, J.~S., Conley, M., \& Holland, P. (1996).
\newblock Neurotoxic lesions of basolateral, but not central, amygdala
  interfere with {P}avlovian second-order conditioning and reinforcer
  devaluation effects.
\newblock {\em The Journal of Neuroscience}, {\em 16\/}, 5256--5265.

\bibitem[\protect\citename{Hazy}{Frank}{O'Reilly}{HazyETAL}{2006}{HazyFrankOReilly06}]{HazyFrankOReilly06}
Hazy, T.~E., Frank, M.~J., \& O'Reilly, R.~C. (2006).
\newblock Banishing the homunculus: Making working memory work.
\newblock {\em Neuroscience}, {\em 139\/}, 105--118.

\bibitem[\protect\citename{Hazy}{Frank}{O'Reilly}{HazyETAL}{2007}{HazyFrankOReilly07}]{HazyFrankOReilly07}
Hazy, T.~E., Frank, M.~J., \& O'Reilly, R.~C. (2007).
\newblock Towards an executive without a homunculus: Computational models of
  the prefrontal cortex/basal ganglia system.
\newblock {\em Philosophical Transactions of the Royal Society of London.
  Series B, Biological Sciences}, {\em 362\/}(1), 105--118.

\bibitem[\protect\citename{Hazy}{Frank}{O'Reilly}{HazyETAL}{2010}{HazyFrankOReilly10}]{HazyFrankOReilly10}
Hazy, T.~E., Frank, M.~J., \& O'Reilly, R.~C. (2010).
\newblock Neural mechanisms of acquired phasic dopamine responses in learning.
\newblock {\em Neuroscience and Biobehavioral Reviews}, {\em 34\/}(5),
  701--720.

\bibitem[\protect\citename{Herbort}{}{Butz}{Herbort-Butz}{2012}{HerbortButz12}]{HerbortButz12}
Herbort, O., \& Butz, M.~V. (2012).
\newblock Too good to be true? ideomotor theory from a computational
  perspective.
\newblock {\em Frontiers in psychology}, {\em 3\/}.

\bibitem[\protect\citename{Higgins}{}{}{Higgins}{1997}{Higgins97}]{Higgins97}
Higgins, E.~T. (1997).
\newblock Beyond pleasure and pain.
\newblock {\em The American psychologist}, {\em 52\/}.

\bibitem[\protect\citename{Higgins}{}{}{Higgins}{2006}{Higgins06}]{Higgins06}
Higgins, E.~T. (2006).
\newblock Value from hedonic experience and engagement.
\newblock {\em Psychological review}, {\em 113\/}.

\bibitem[\protect\citename{Holland}{}{Gallagher}{Holland-Gallagher}{2004}{HollandGallagher04}]{HollandGallagher04}
Holland, P.~C., \& Gallagher, M. (2004).
\newblock Amygdala-frontal interactions and reward expectancy.
\newblock {\em Current opinion in neurobiology}, {\em 14\/}(2), 148--155.

\bibitem[\protect\citename{Hollerman}{}{Schultz}{Hollerman-Schultz}{1998}{HollermanSchultz98}]{HollermanSchultz98}
Hollerman, J.~R., \& Schultz, W. (1998).
\newblock Dopamine neurons report an error in the temporal prediction of reward
  during learning.
\newblock {\em Nature Neuroscience}, {\em 1\/}(4), 304--309.

\bibitem[\protect\citename{Hommel}{}{}{Hommel}{2004}{Hommel04}]{Hommel04}
Hommel, B. (2004).
\newblock Event files: feature binding in and across perception and action.
\newblock {\em Trends in cognitive sciences}, {\em 8\/}(11), 494--500.

\bibitem[\protect\citename{Hong}{}{Hikosaka}{Hong-Hikosaka}{2008}{HongHikosaka08}]{HongHikosaka08}
Hong, S., \& Hikosaka, O. (2008).
\newblock The globus pallidus sends reward-related signals to the lateral
  habenula.
\newblock {\em Neuron}, {\em 60\/}(4), 720--729.

\bibitem[\protect\citename{Hong}{Jhou, Smith,
  Saleem}{Hikosaka}{HongETAL}{2011}{HongJhouSmithEtAl11}]{HongJhouSmithEtAl11}
Hong, S., Jhou, T.~C., Smith, M., Saleem, K.~S., \& Hikosaka, O. (2011).
\newblock Negative reward signals from the lateral habenula to dopamine neurons
  are mediated by rostromedial tegmental nucleus in primates.
\newblock {\em The Journal of neuroscience}, {\em 31\/}.

\bibitem[\protect\citename{Howland}{Taepavarapruk}{Phillips}{HowlandETAL}{2002}{HowlandTaepavaraprukPhillips02}]{HowlandTaepavaraprukPhillips02}
Howland, J.~G., Taepavarapruk, P., \& Phillips, A.~G. (2002).
\newblock Glutamate receptor-dependent modulation of dopamine efflux in the
  nucleus accumbens by basolateral, but not central, nucleus of the amygdala in
  rats.
\newblock {\em The Journal of Neuroscience}, {\em 22\/}(3), 1137--1145.

\bibitem[\protect\citename{Huang}{Hazy,
  Herd}{O'Reilly}{HuangETAL}{2013}{HuangHazyHerdEtAl13}]{HuangHazyHerdEtAl13}
Huang, T.-R., Hazy, T.~E., Herd, S.~A., \& O'Reilly, R.~C. (2013).
\newblock Assembling old tricks for new tasks: {A} neural model of
  instructional learning and control.
\newblock {\em Journal of Cognitive Neuroscience}, {\em 25\/}(6), 843--851.

\bibitem[\protect\citename{Hull}{}{}{Hull}{1943}{Hull43}]{Hull43}
Hull, C.~L. (1943).
\newblock {\em Principles of behavior}.
\newblock Appleton.

\bibitem[\protect\citename{Ikemoto}{}{}{Ikemoto}{2010}{Ikemoto10}]{Ikemoto10}
Ikemoto, S. (2010).
\newblock Brain reward circuitry beyond the mesolimbic dopamine system: a
  neurobiological theory.
\newblock {\em Neuroscience and Biobehavioral Reviews}, {\em 35\/}(2),
  129--150.

\bibitem[\protect\citename{James}{}{}{James}{1890}{James90}]{James90}
James, W. (1890).
\newblock {\em The principles of psychology}.
\newblock New York: Henry Holt.

\bibitem[\protect\citename{Jennings}{Sparta, Stamatakis, Ung, Pleil,
  Kash}{Stuber}{JenningsETAL}{2013}{JenningsSpartaStamatakisEtAl13}]{JenningsSpartaStamatakisEtAl13}
Jennings, J.~H., Sparta, D.~R., Stamatakis, A.~M., Ung, R.~L., Pleil, K.~E.,
  Kash, T.~L., \& Stuber, G.~D. (2013).
\newblock Distinct extended amygdala circuits for divergent motivational
  states.
\newblock {\em Nature}, {\em 496\/}.

\bibitem[\protect\citename{Jhou}{Fields, Baxter,
  Saper}{Holland}{JhouETAL}{2009}{JhouFieldsBaxterEtAl09}]{JhouFieldsBaxterEtAl09}
Jhou, T.~C., Fields, H.~L., Baxter, M.~G., Saper, C.~B., \& Holland, P.~C.
  (2009).
\newblock The rostromedial tegmental nucleus (rmtg), a gabaergic afferent to
  midbrain dopamine neurons, encodes aversive stimuli and inhibits motor
  responses.
\newblock {\em Neuron}, {\em 61\/}, 786--800.

\bibitem[\protect\citename{Ji}{}{Shepard}{Ji-Shepard}{2007}{JiShepard07}]{JiShepard07}
Ji, H., \& Shepard, P.~D. (2007).
\newblock Lateral habenula stimulation inhibits rat midbrain dopamine neurons
  through a gaba-a receptor-mediated mechanism.
\newblock {\em Journal of Neuroscience}, {\em 27\/}(26), 6923--6930.

\bibitem[\protect\citename{Joel}{}{Weiner}{Joel-Weiner}{2000}{JoelWeiner00}]{JoelWeiner00}
Joel, D., \& Weiner, I. (2000).
\newblock The connections of the dopaminergic system with the striatum in rats
  and primates: an analysis with respect to the functional and compartmental
  organization of the striatum.
\newblock {\em Neuroscience}, {\em 96\/}, 451--474.

\bibitem[\protect\citename{Kahan}{Jenkins-Smith}{Braman}{KahanETAL}{2011}{KahanJenkins-SmithBraman11}]{KahanJenkins-SmithBraman11}
Kahan, D., Jenkins-Smith, H., \& Braman, D. (2011).
\newblock Cultural cognition of scientific consensus.
\newblock {\em Journal of Risk Research}, {\em 14\/}, 147--74.

\bibitem[\protect\citename{Kamil}{}{Roitblat}{Kamil-Roitblat}{1985}{KamilRoitblat85}]{KamilRoitblat85}
Kamil, A.~C., \& Roitblat, H.~L. (1985).
\newblock The ecology of foraging behavior: implications for animal learning
  and memory.
\newblock {\em Annual review of psychology}, {\em 36\/}, 141--169.

\bibitem[\protect\citename{Kamin}{}{}{Kamin}{1969a}{Kamin69}]{Kamin69}
Kamin, L.~J. (1969a).
\newblock Predictability, surprise, attention and conditioning.
\newblock In B.~A. Campbell, \& R.~M. Church (Eds.), {\em {Punishment and
  Aversive Behavior}}. New York: Appleton-Century-Crofts.

\bibitem[\protect\citename{Kamin}{}{}{Kamin}{1969b}{Kamin69a}]{Kamin69a}
Kamin, L.~J. (1969b).
\newblock Selective association and conditioning.
\newblock In N.~J. Mackintosh, \& W.~K. Honig (Eds.), {\em {Fundamental Issues
  in Instrumental Learning}} (pp.\ 42--64). Dalhousie: Dalhousie Univ. Press.

\bibitem[\protect\citename{Kelley}{Hortensius}{Harmon-Jones}{KelleyETAL}{2013}{KelleyHortensiusHarmon-Jones13}]{KelleyHortensiusHarmon-Jones13}
Kelley, N.~J., Hortensius, R., \& Harmon-Jones, E. (2013).
\newblock When anger leads to rumination: induction of relative right frontal
  cortical activity with transcranial direct current stimulation increases
  anger-related rumination.
\newblock {\em Psychological science}, {\em 24\/}, 475--481.

\bibitem[\protect\citename{Kennerley}{Dahmubed,
  Lara}{Wallis}{KennerleyETAL}{2009}{KennerleyDahmubedLaraEtAl09}]{KennerleyDahmubedLaraEtAl09}
Kennerley, S.~W., Dahmubed, A.~F., Lara, A.~H., \& Wallis, J.~D. (2009).
\newblock Neurons in the frontal lobe encode the value of multiple decision
  variables.
\newblock {\em Journal of cognitive neuroscience}, {\em 21\/}.

\bibitem[\protect\citename{Kennerley}{}{Walton}{Kennerley-Walton}{2011}{KennerleyWalton11}]{KennerleyWalton11}
Kennerley, S.~W., \& Walton, M.~E. (2011).
\newblock Decision making and reward in frontal cortex: complementary evidence
  from neurophysiological and neuropsychological studies.
\newblock {\em Behavioral Neuroscience}, {\em 125\/}(3), 297--317.

\bibitem[\protect\citename{Kennerley}{Walton, Behrens,
  Buckley}{Rushworth}{KennerleyETAL}{2006}{KennerleyWaltonBehrensEtAl06}]{KennerleyWaltonBehrensEtAl06}
Kennerley, S.~W., Walton, M.~E., Behrens, T. E.~J., Buckley, M.~J., \&
  Rushworth, M. F.~S. (2006).
\newblock Optimal decision making and the anterior cingulate cortex.
\newblock {\em Nature neuroscience}, {\em 9\/}(7), 940--947.

\bibitem[\protect\citename{Kim}{Adhikari, Lee, Marshel, Kim, Mallory, Lo, Pak,
  Mattis, Lim, Malenka, Warden, Neve,
  Tye}{Deisseroth}{KimETAL}{2013}{KimAdhikariLeeEtAl13}]{KimAdhikariLeeEtAl13}
Kim, S.-Y., Adhikari, A., Lee, S.~Y., Marshel, J.~H., Kim, C.~K., Mallory,
  C.~S., Lo, M., Pak, S., Mattis, J., Lim, B.~K., Malenka, R.~C., Warden,
  M.~R., Neve, R., Tye, K.~M., \& Deisseroth, K. (2013).
\newblock Diverging neural pathways assemble a behavioural state from separable
  features in anxiety.
\newblock {\em Nature}, {\em 496\/}.

\bibitem[\protect\citename{Klinger}{}{}{Klinger}{1975}{Klinger75}]{Klinger75}
Klinger, E. (1975).
\newblock Consequences of commitment to and disengagement from incentives.
\newblock {\em Psychological Review}, {\em 82\/}, 1--25.

\bibitem[\protect\citename{Knapska}{Walasek, Nikolaev, Neuhusser-Wespy, Lipp,
  Kaczmarek}{Werka}{KnapskaETAL}{2006}{KnapskaWalasekNikolaevEtAl06}]{KnapskaWalasekNikolaevEtAl06}
Knapska, E., Walasek, G., Nikolaev, E., Neuhusser-Wespy, F., Lipp, H.-P.,
  Kaczmarek, L., \& Werka, T. (2006).
\newblock Differential involvement of the central amygdala in appetitive versus
  aversive learning.
\newblock {\em Learning \& Memory}, {\em 13\/}(2), 192--200.

\bibitem[\protect\citename{Kolling}{Behrens,
  Mars}{Rushworth}{KollingETAL}{2012}{KollingBehrensMarsEtAl12}]{KollingBehrensMarsEtAl12}
Kolling, N., Behrens, T., Mars, R., \& Rushworth, M. (2012).
\newblock Neural mechanisms of foraging.
\newblock {\em Science}, {\em 336\/}(6077), 95--98.

\bibitem[\protect\citename{Kouneiher}{Charron}{Koechlin}{KouneiherETAL}{2009}{KouneiherCharronKoechlin09}]{KouneiherCharronKoechlin09}
Kouneiher, F., Charron, S., \& Koechlin, E. (2009).
\newblock Motivation and cognitive control in the human prefrontal cortex.
\newblock {\em Nature neuroscience}, {\em 12\/}(9), 659--669.

\bibitem[\protect\citename{Kriete}{Noelle,
  Cohen}{O'Reilly}{KrieteETAL}{2013}{KrieteNoelleCohenEtAl13}]{KrieteNoelleCohenEtAl13}
Kriete, T., Noelle, D.~C., Cohen, J.~D., \& O'Reilly, R.~C. (2013).
\newblock Indirection and symbol-like processing in the prefrontal cortex and
  basal ganglia.
\newblock {\em Proceedings of the National Academy of Sciences of the United
  States of America}, {\em 110\/}(41).

\bibitem[\protect\citename{Kross}{Berman, Mischel,
  Smith}{Wager}{KrossETAL}{2011}{KrossBermanMischelEtAl11}]{KrossBermanMischelEtAl11}
Kross, E., Berman, M.~G., Mischel, W., Smith, E.~E., \& Wager, T.~D. (2011).
\newblock Social rejection shares somatosensory representations with physical
  pain.
\newblock {\em Proceedings of the National Academy of Sciences of the United
  States of America}, {\em 108\/}.

\bibitem[\protect\citename{Kunda}{}{}{Kunda}{1990}{Kunda90}]{Kunda90}
Kunda, Z. (1990).
\newblock The case for motivated reasoning.
\newblock {\em Psychological Bulletin}, {\em 108\/}, 480--498.

\bibitem[\protect\citename{Lammel}{Ion,
  Roeper}{Malenka}{LammelETAL}{2011}{LammelIonRoeperEtAl11}]{LammelIonRoeperEtAl11}
Lammel, S., Ion, D.~I., Roeper, J., \& Malenka, R.~C. (2011).
\newblock Projection-specific modulation of dopamine neuron synapses by
  aversive and rewarding stimuli.
\newblock {\em Neuron}, {\em 70\/}, 855--862.

\bibitem[\protect\citename{Lammel}{Lim, Ran, Huang, Betley, Tye,
  Deisseroth}{Malenka}{LammelETAL}{2012}{LammelLimRanEtAl12}]{LammelLimRanEtAl12}
Lammel, S., Lim, B.~K., Ran, C., Huang, K.~W., Betley, M.~J., Tye, K.~M.,
  Deisseroth, K., \& Malenka, R.~C. (2012).
\newblock Input-specific control of reward and aversion in the ventral
  tegmental area.
\newblock {\em Nature}, Epub ahead of print.

\bibitem[\protect\citename{Leake}{}{Ram}{Leake-Ram}{1995}{LeakeRam95}]{LeakeRam95}
Leake, D.~B., \& Ram, A. (1995).
\newblock Learning, goals, and learning goals: A perspective on goal-driven
  learning.
\newblock {\em Artificial Intelligence Review}, {\em 9\/}, 387--422.

\bibitem[\protect\citename{Lecourtier}{DeFrancesco}{Moghaddam}{LecourtierETAL}{2008}{LecourtierDeFrancescoMoghaddam08}]{LecourtierDeFrancescoMoghaddam08}
Lecourtier, L., DeFrancesco, A., \& Moghaddam, B. (2008).
\newblock Differential tonic influence of lateral habenula on prefrontal cortex
  and nucleus accumbens dopamine release.
\newblock {\em European Journal of Neuroscience}, {\em 27\/}, 1755--1762.

\bibitem[\protect\citename{Lee}{Gallagher}{Holland}{LeeETAL}{2010}{LeeGallagherHolland10}]{LeeGallagherHolland10}
Lee, H.~J., Gallagher, M., \& Holland, P.~C. (2010).
\newblock The central amygdala projection to the substantia nigra reflects
  prediction error information in appetitive conditioning.
\newblock {\em Learning \& Memory}, {\em 17\/}(10), 531--538.

\bibitem[\protect\citename{Lewin}{}{}{Lewin}{1928}{Lewin28}]{Lewin28}
Lewin, K. (1928).
\newblock Wille, vorsatz und bedurfnis.
\newblock {\em Psychologische Forschung}, {\em 7\/}, 330--385.

\bibitem[\protect\citename{Li}{Piriz, Mirrione, Chung, Proulx, Schulz,
  Henn}{Malinow}{LiETAL}{2011}{LiPirizMirrioneEtAl11}]{LiPirizMirrioneEtAl11}
Li, B., Piriz, J., Mirrione, M., Chung, C., Proulx, C.~D., Schulz, D., Henn,
  F., \& Malinow, R. (2011).
\newblock Synaptic potentiation onto habenula neurons in the learned
  helplessness model of depression.
\newblock {\em Nature}, {\em 470\/}(7335), 535--539.

\bibitem[\protect\citename{Liu}{}{Richmond}{Liu-Richmond}{2000}{LiuRichmond00}]{LiuRichmond00}
Liu, Z., \& Richmond, B.~J. (2000).
\newblock Response differences in monkey te and perirhinal cortex: stimulus
  association related to reward schedules.
\newblock {\em Journal of neurophysiology}, {\em 83\/}, 1677.

\bibitem[\protect\citename{Locke}{}{}{Locke}{1968}{Locke68}]{Locke68}
Locke, E.~A. (1968).
\newblock Toward a theory of task motivation and incentives.
\newblock {\em Organizational Behavior and Human Decision Processes}, {\em
  3\/}, 157--189.

\bibitem[\protect\citename{Locke}{}{Latham}{Locke-Latham}{2002}{LockeLatham02}]{LockeLatham02}
Locke, E.~A., \& Latham, G.~P. (2002).
\newblock Building a practically useful theory of goal setting and task
  motivation: A 35-year odyssey.
\newblock {\em American Psychologis}, {\em 57\/}(9), 705--717.

\bibitem[\protect\citename{Lodge}{}{Grace}{Lodge-Grace}{2006a}{LodgeGrace06}]{LodgeGrace06}
Lodge, D.~J., \& Grace, A.~A. (2006a).
\newblock The hippocampus modulates dopamine neuron responsivity by regulating
  the intensity of phasic neuron activation.
\newblock {\em Neuropsychopharmacology}, {\em 31\/}(7), 1356--1361.

\bibitem[\protect\citename{Lodge}{}{Grace}{Lodge-Grace}{2006b}{LodgeGrace06b}]{LodgeGrace06b}
Lodge, D.~J., \& Grace, A.~A. (2006b).
\newblock The laterodorsal tegmentum is essential for burst firing of ventral
  tegmental area dopamine neurons.
\newblock {\em Proceedings of the National Academy of Sciences U.S.A.}, {\em
  103\/}(13), 5167--5172.

\bibitem[\protect\citename{Maier}{}{}{Maier}{1984}{Maier84}]{Maier84}
Maier, S.~F. (1984).
\newblock Learned helplessness and animal models of depression.
\newblock {\em Progress in neuro-psychopharmacology \& biological psychiatry},
  {\em 8\/}.

\bibitem[\protect\citename{Maier}{}{Seligman}{Maier-Seligman}{1976}{MaierSeligman76}]{MaierSeligman76}
Maier, S.~F., \& Seligman, M. E.~P. (1976).
\newblock Learned helplesness: Theory and evidence.
\newblock {\em Journal of Experimental Psychology: General}, {\em 105\/},
  3--46.

\bibitem[\protect\citename{Maier}{}{Watkins}{Maier-Watkins}{2010}{MaierWatkins10}]{MaierWatkins10}
Maier, S.~F., \& Watkins, L.~R. (2010).
\newblock Role of the medial prefrontal cortex in coping and resilience.
\newblock {\em Brain research}, {\em 1355\/}, 52--60.

\bibitem[\protect\citename{Masicampo}{}{Baumeister}{Masicampo-Baumeister}{2011}{MasicampoBaumeister11}]{MasicampoBaumeister11}
Masicampo, E.~J., \& Baumeister, R.~F. (2011).
\newblock Unfulfilled goals interfere with tasks that require executive
  functions.
\newblock {\em Journal of Experimental Social Psychology}, {\em 47\/},
  300--311.

\bibitem[\protect\citename{Maslow}{}{}{Maslow}{1943}{Maslow43}]{Maslow43}
Maslow, A.~H. (1943).
\newblock A theory of human motivation.
\newblock {\em Psychological Review}, {\em 50\/}, 370--396.

\bibitem[\protect\citename{Matsumoto}{}{Hikosaka}{Matsumoto-Hikosaka}{2007}{MatsumotoHikosaka07}]{MatsumotoHikosaka07}
Matsumoto, M., \& Hikosaka, O. (2007).
\newblock Lateral habenula as a source of negative reward signals in dopamine
  neurons.
\newblock {\em Nature}, {\em 447\/}, 1111--1115.

\bibitem[\protect\citename{Matsumoto}{}{Hikosaka}{Matsumoto-Hikosaka}{2009}{MatsumotoHikosaka09b}]{MatsumotoHikosaka09b}
Matsumoto, M., \& Hikosaka, O. (2009).
\newblock Two types of dopamine neuron distinctly convey positive and negative
  motivational signals.
\newblock {\em Nature}, {\em 459\/}, 837--842.

\bibitem[\protect\citename{Mayberg}{Brannan, Mahurin, Jerabek, Brickman,
  Tekell, Silva, McGinnis, Glass,
  Martin}{Fox}{MaybergETAL}{1997}{MaybergBrannanMahurinEtAl97}]{MaybergBrannanMahurinEtAl97}
Mayberg, H.~S., Brannan, S.~K., Mahurin, R.~K., Jerabek, P.~A., Brickman,
  J.~S., Tekell, J.~L., Silva, J.~A., McGinnis, S., Glass, T.~G., Martin,
  C.~C., \& Fox, P.~T. (1997).
\newblock Cingulate function in depression: a potential predictor of treatment
  response.
\newblock {\em Neuroreport}, {\em 8\/}(4), 1057--1061.

\bibitem[\protect\citename{Meltzoff}{}{}{Meltzoff}{1985}{Meltzoff85}]{Meltzoff85}
Meltzoff, A.~N. (1985).
\newblock Immediate and deferred imitation in 14- and 24-month-old infants.
\newblock {\em Child Development}, {\em 56\/}, 62--72.

\bibitem[\protect\citename{Meltzoff}{}{}{Meltzoff}{1988}{Meltzoff88}]{Meltzoff88}
Meltzoff, A.~N. (1988).
\newblock Infant imitation and memory: nine-month-olds in immediate and
  deferred tests.
\newblock {\em Child development}, {\em 59\/}, 217--225.

\bibitem[\protect\citename{Meltzoff}{}{Decety}{Meltzoff-Decety}{2003}{MeltzoffDecety03}]{MeltzoffDecety03}
Meltzoff, A.~N., \& Decety, J. (2003).
\newblock What imitation tells us about social cognition: a rapprochement
  between developmental psychology and cognitive neuroscience.
\newblock {\em Philosophical transactions of the Royal Society of London.
  Series B, Biological sciences}, {\em 358\/}, 491--500.

\bibitem[\protect\citename{Menzel}{}{Giurfa}{Menzel-Giurfa}{2001}{MenzelGiurfa01}]{MenzelGiurfa01}
Menzel, R., \& Giurfa, M. (2001).
\newblock Cognitive architecture of a mini-brain: the honeybee.
\newblock {\em Trends in cognitive sciences}, {\em 5\/}.

\bibitem[\protect\citename{M.I.}{}{D.E.}{M.I.-D.E.}{1992}{JordanRumelhart92}]{JordanRumelhart92}
M.I., J., \& D.E., R. (1992).
\newblock Forward models: Supervised learning with a distal teacher.
\newblock {\em Cognitive science}, {\em 16\/}(3), 307--354.

\bibitem[\protect\citename{Miller}{}{Cohen}{Miller-Cohen}{2001}{MillerCohen01}]{MillerCohen01}
Miller, E.~K., \& Cohen, J.~D. (2001).
\newblock An integrative theory of prefrontal cortex function.
\newblock {\em Annual Review of Neuroscience}, {\em 24\/}, 167--202.

\bibitem[\protect\citename{Miller}{Galanter}{Pribram}{MillerETAL}{1960}{MillerGalanterPribram60}]{MillerGalanterPribram60}
Miller, G.~A., Galanter, E., \& Pribram, K.~H. (1960).
\newblock {\em Plans and the structure of behavior}.
\newblock New York: Holt.

\bibitem[\protect\citename{Mobbs}{Marchant, Hassabis, Seymour, Tan, Gray,
  Petrovic,
  Dolan}{Frith}{MobbsETAL}{2009}{MobbsMarchantHassabisEtAl09}]{MobbsMarchantHassabisEtAl09}
Mobbs, D., Marchant, J.~L., Hassabis, D., Seymour, B., Tan, G., Gray, M.,
  Petrovic, P., Dolan, R.~J., \& Frith, C.~D. (2009).
\newblock From threat to fear: the neural organization of defensive fear
  systems in humans.
\newblock {\em The Journal of neuroscience}, {\em 29\/}.

\bibitem[\protect\citename{Mobini}{Body, Ho, Bradshaw, Szabadi,
  Deakin}{Anderson}{MobiniETAL}{2002}{MobiniBodyHoEtAl02}]{MobiniBodyHoEtAl02}
Mobini, S., Body, S., Ho, M.-Y., Bradshaw, C.~M., Szabadi, E., Deakin, J.
  F.~W., \& Anderson, I.~M. (2002).
\newblock Effects of lesions of the orbitofrontal cortex on sensitivity to
  delayed and probabilistic reinforcement.
\newblock {\em Psychopharmacology}, {\em 160\/}, 290--8.

\bibitem[\protect\citename{Monosov}{}{Hikosaka}{Monosov-Hikosaka}{2013}{MonosovHikosaka13}]{MonosovHikosaka13}
Monosov, I.~E., \& Hikosaka, O. (2013).
\newblock Selective and graded coding of reward uncertainty by neurons in the
  primate anterodorsal septal region.
\newblock {\em Nature neuroscience}.

\bibitem[\protect\citename{Montague}{Dayan}{Sejnowski}{MontagueETAL}{1996}{MontagueDayanSejnowski96}]{MontagueDayanSejnowski96}
Montague, P.~R., Dayan, P., \& Sejnowski, T.~J. (1996).
\newblock A framework for mesencephalic dopamine systems based on predictive
  {H}ebbian learning.
\newblock {\em The Journal of Neuroscience}, {\em 16\/}(5), 1936--1947.

\bibitem[\protect\citename{Morewedge}{}{Kahneman}{Morewedge-Kahneman}{2010}{MorewedgeKahneman10}]{MorewedgeKahneman10}
Morewedge, C.~K., \& Kahneman, D. (2010).
\newblock Associative processes in intuitive judgment.
\newblock {\em Trends in cognitive sciences}, {\em 14\/}.

\bibitem[\protect\citename{Munakata}{Herd, Chatham, Depue,
  Banich}{O'Reilly}{MunakataETAL}{2011}{MunakataHerdChathamEtAl11}]{MunakataHerdChathamEtAl11}
Munakata, Y., Herd, S.~A., Chatham, C.~H., Depue, B.~E., Banich, M.~T., \&
  O'Reilly, R.~C. (2011).
\newblock A unified framework for inhibitory control.
\newblock {\em Trends in Cognitive Sciences}, {\em 15\/}(10), 453--459.

\bibitem[\protect\citename{Murray}{O'Doherty}{Schoenbaum}{MurrayETAL}{2007}{MurrayODohertySchoenbaum07}]{MurrayODohertySchoenbaum07}
Murray, E.~A., O'Doherty, J.~P., \& Schoenbaum, G. (2007).
\newblock What we know and do not know about the functions of the orbitofrontal
  cortex after 20 years of cross-species studies.
\newblock {\em The Journal of neuroscience}, {\em 27\/}, 8166--8169.

\bibitem[\protect\citename{Narayanan}{Horst}{Laubach}{NarayananETAL}{2006}{NarayananHorstLaubach06}]{NarayananHorstLaubach06}
Narayanan, N.~S., Horst, N.~K., \& Laubach, M. (2006).
\newblock Reversible inactivations of rat medial prefrontal cortex impair the
  ability to wait for a stimulus.
\newblock {\em Neuroscience}, {\em 139\/}(3), 865--876.

\bibitem[\protect\citename{Newell}{}{Simon}{Newell-Simon}{1976}{NewellSimon76}]{NewellSimon76}
Newell, A., \& Simon, H.~A. (1976).
\newblock Computer science as empirical inquiry: {Symbols} and search.
\newblock {\em Communications of the ACM}, {\em 19\/}, 113--126.

\bibitem[\protect\citename{Noonan}{Walton, Behrens, Sallet,
  Buckley}{Rushworth}{NoonanETAL}{2010}{NoonanWaltonBehrensEtAl10}]{NoonanWaltonBehrensEtAl10}
Noonan, M.~P., Walton, M.~E., Behrens, T. E.~J., Sallet, J., Buckley, M.~J., \&
  Rushworth, M. F.~S. (2010).
\newblock Separate value comparison and learning mechanisms in macaque medial
  and lateral orbitofrontal cortex.
\newblock {\em Proceedings of the National Academy of Sciences of the United
  States of America}, {\em 107\/}.

\bibitem[\protect\citename{Nemeth}{Hegedus}{Molnar}{NemethETAL}{1988}{NemethHegedusMolnar88}]{NemethHegedusMolnar88}
Nemeth, G., Hegedus, K., \& Molnar, L. (1988).
\newblock Akinetic mutism associated with bicingular lesions:
  clinicopathological and functional anatomical correlates.
\newblock {\em European archives of psychiatry and neurological sciences}, {\em
  237\/}.

\bibitem[\protect\citename{Oleson}{}{Cheer}{Oleson-Cheer}{2013}{OlesonCheer13}]{OlesonCheer13}
Oleson, E.~B., \& Cheer, J.~F. (2013).
\newblock On the role of subsecond dopamine release in conditioned avoidance.
\newblock {\em Frontiers in Neuroscience}, {\em 7\/}, 1--9 (online--only).

\bibitem[\protect\citename{Oleson}{Gentry,
  Chioma}{Cheer}{OlesonETAL}{2012}{OlesonGentryChiomaEtAl12}]{OlesonGentryChiomaEtAl12}
Oleson, E.~B., Gentry, R.~N., Chioma, V.~C., \& Cheer, J.~F. (2012).
\newblock Subsecond dopamine release in the nucleus accumbens predicts
  conditioned punishment and its successful avoidance.
\newblock {\em The Journal of Neuroscience}, {\em 32\/}(42), 14804--14808.

\bibitem[\protect\citename{Ongur}{An}{Price}{OngurETAL}{1998}{OngurAnPrice98}]{OngurAnPrice98}
Ongur, D., An, X., \& Price, J.~L. (1998).
\newblock Prefrontal cortical projections to the hypothalamus in macaque
  monkeys.
\newblock {\em The Journal of Comparative Neurology}, {\em 401\/}(4), 480--505.

\bibitem[\protect\citename{Ong{\"u}r}{}{Price}{Ong{\"u}r-Price}{2000}{OngurPrice00}]{OngurPrice00}
Ong{\"u}r, D., \& Price, J.~L. (2000).
\newblock The organization of networks within the orbital and medial prefrontal
  cortex of rats, monkeys and humans.
\newblock {\em Cerebral Cortex}, {\em 10\/}(3), 206--219.

\bibitem[\protect\citename{Ono}{Nishijo}{Uwano}{OnoETAL}{1995}{OnoNishijoUwano95}]{OnoNishijoUwano95}
Ono, T., Nishijo, H., \& Uwano, T. (1995).
\newblock Amygdala role in conditioned associative learning.
\newblock {\em Progress in Neurobiology}, {\em 46\/}, 401--422.

\bibitem[\protect\citename{Ono}{Tamura, Nishijo,
  Nakamura}{Tabuchi}{OnoETAL}{1989}{OnoTamuraNishijoEtAl89}]{OnoTamuraNishijoEtAl89}
Ono, T., Tamura, R., Nishijo, H., Nakamura, K., \& Tabuchi, E. (1989).
\newblock Contribution of amygdalar and lateral hypothalamic neurons to visual
  information processing of food and nonfood in monkey.
\newblock {\em Physiology \& Behavior}, {\em 45\/}(2), 411--421.

\bibitem[\protect\citename{O'Reilly}{}{}{O'Reilly}{2006}{OReilly06}]{OReilly06}
O'Reilly, R. (2006).
\newblock Biologically based computational models of high-level cognition.
\newblock {\em Science}, {\em 314\/}(5796), 91--94.

\bibitem[\protect\citename{O'Reilly}{}{}{O'Reilly}{1996}{OReilly96}]{OReilly96}
O'Reilly, R.~C. (1996).
\newblock {Biologically Plausible Error-driven Learning using Local Activation
  Differences}: {The Generalized Recirculation Algorithm}.
\newblock {\em Neural Computation}, {\em 8\/}(5), 895--938.

\bibitem[\protect\citename{O'Reilly}{}{}{O'Reilly}{2001}{OReilly01}]{OReilly01}
O'Reilly, R.~C. (2001).
\newblock Generalization in interactive networks: {T}he benefits of inhibitory
  competition and {H}ebbian learning.
\newblock {\em Neural Computation}, {\em 13\/}(6), 1199--1242.

\bibitem[\protect\citename{O'Reilly}{}{}{O'Reilly}{2010}{OReilly10}]{OReilly10}
O'Reilly, R.~C. (2010).
\newblock The {\em what} and {\em how} of prefrontal cortical organization.
\newblock {\em Trends in Neurosciences}, {\em 33\/}(8), 355--361.

\bibitem[\protect\citename{O'Reilly}{Braver}{Cohen}{O'ReillyETAL}{1999}{OReillyBraverCohen99}]{OReillyBraverCohen99}
O'Reilly, R.~C., Braver, T.~S., \& Cohen, J.~D. (1999).
\newblock A biologically based computational model of working memory.
\newblock In A. Miyake, \& P. Shah (Eds.), {\em {Models of Working Memory:
  Mechanisms of Active Maintenance and Executive Control.}} (pp.\ 375--411).
  New York: Cambridge University Press.

\bibitem[\protect\citename{O'Reilly}{}{Frank}{O'Reilly-Frank}{2006}{OReillyFrank06}]{OReillyFrank06}
O'Reilly, R.~C., \& Frank, M.~J. (2006).
\newblock Making working memory work: A computational model of learning in the
  prefrontal cortex and basal ganglia.
\newblock {\em Neural Computation}, {\em 18\/}(2), 283--328.

\bibitem[\protect\citename{O'Reilly}{Frank,
  Hazy}{Watz}{O'ReillyETAL}{2007}{OReillyFrankHazyEtAl07}]{OReillyFrankHazyEtAl07}
O'Reilly, R.~C., Frank, M.~J., Hazy, T.~E., \& Watz, B. (2007).
\newblock {PVLV}: The primary value and learned value {P}avlovian learning
  algorithm.
\newblock {\em Behavioral Neuroscience}, {\em 121\/}, 31--49.

\bibitem[\protect\citename{O'Reilly}{}{Munakata}{O'Reilly-Munakata}{2000}{OReillyMunakata00}]{OReillyMunakata00}
O'Reilly, R.~C., \& Munakata, Y. (2000).
\newblock {\em {Computational Explorations in Cognitive Neuroscience}:
  {Understanding the Mind by Simulating the Brain}}.
\newblock Cambridge, MA: The MIT Press.

\bibitem[\protect\citename{O'Reilly}{Munakata, Frank,
  Hazy}{Contributors}{O'ReillyETAL}{2012}{OReillyMunakataFrankEtAl12}]{OReillyMunakataFrankEtAl12}
O'Reilly, R.~C., Munakata, Y., Frank, M.~J., Hazy, T.~E., \& Contributors
  (2012).
\newblock {\em {Computational Cognitive Neuroscience}}.
\newblock Wiki Book, 1st Edition, URL: \url{http://ccnbook.colorado.edu}.

\bibitem[\protect\citename{O'Reilly}{Wyatte, Herd,
  Mingus}{Jilk}{O'ReillyETAL}{2013}{OReillyWyatteHerdEtAl13}]{OReillyWyatteHerdEtAl13}
O'Reilly, R.~C., Wyatte, D., Herd, S., Mingus, B., \& Jilk, D.~J. (2013).
\newblock Recurrent processing during object recognition.
\newblock {\em Frontiers in Psychology}, {\em 4\/}, 124.

\bibitem[\protect\citename{Ottersen}{}{}{Ottersen}{1980}{Ottersen80}]{Ottersen80}
Ottersen, O.~P. (1980).
\newblock Afferent connections to the amygdaloid complex of the rat and cat:
  {II. A}fferents from the hypothalamus and the basal telencephalon.
\newblock {\em The Journal of Comparative Neurology}, {\em 194\/}(1), 267--289.

\bibitem[\protect\citename{Pan}{}{Hyland}{Pan-Hyland}{2005}{PanHyland05}]{PanHyland05}
Pan, W.-X., \& Hyland, B.~I. (2005).
\newblock Pedunculopontine tegmental nucleus controls conditioned responses of
  midbrain dopamine neurons in behaving rats.
\newblock {\em J. Neurosci.}, {\em 25\/}(19), 4725--4732.

\bibitem[\protect\citename{Park}{Kahnt,
  Rieskamp}{Heekeren}{ParkETAL}{2011}{ParkKahntRieskamp11}]{ParkKahntRieskamp11}
Park, S.~Q., Kahnt, T., Rieskamp, J., \& Heekeren, H.~R. (2011).
\newblock Neurobiology of value integration: when value impacts valuation.
\newblock {\em The Journal of neuroscience}, {\em 31\/}(25).

\bibitem[\protect\citename{Paton}{Belova,
  Morrison}{Salzman}{PatonETAL}{2006}{PatonBelovaMorrisonEtAl06}]{PatonBelovaMorrisonEtAl06}
Paton, J.~J., Belova, M.~A., Morrison, S.~E., \& Salzman, C.~D. (2006).
\newblock The primate amygdala represents the positive and negative value of
  visual stimuli during learning.
\newblock {\em Nature}, {\em 439\/}(7078), 865--870.

\bibitem[\protect\citename{Pauli}{Clark, Guenther,
  O'Reilly}{Rudy}{PauliETAL}{2012a}{PauliClarkGuentherEtAl12}]{PauliClarkGuentherEtAl12}
Pauli, W.~M., Clark, A.~D., Guenther, H., O'Reilly, R.~C., \& Rudy, J.~W.
  (2012a).
\newblock Inhibiting pkm$\zeta$ reveals dorsal lateral and dorsal medial
  striatum store the different memories needed to support adaptive behavior.
\newblock {\em Learning and Memory}, {\em 19\/}, 307--14.

\bibitem[\protect\citename{Pauli}{Hazy}{O'Reilly}{PauliETAL}{2012b}{PauliHazyOReilly12}]{PauliHazyOReilly12}
Pauli, W.~M., Hazy, T.~E., \& O'Reilly, R.~C. (2012b).
\newblock Expectancy, ambiguity, and behavioral flexibility: separable and
  complementary roles of the orbital frontal cortex and amygdala in processing
  reward expectancies.
\newblock {\em Journal of Cognitive Neuroscience}, {\em 24\/}(2), 351--366.

\bibitem[\protect\citename{Penny}{Zeidman}{Burgess}{PennyETAL}{2013}{PennyZeidmanBurgess13}]{PennyZeidmanBurgess13}
Penny, W.~D., Zeidman, P., \& Burgess, N. (2013).
\newblock Forward and backward inference in spatial cognition.
\newblock {\em PLoS computational biology}, {\em 9\/}.

\bibitem[\protect\citename{Percheron}{Francois, Talbi,
  Yelnik}{Fenelon}{PercheronETAL}{1996}{PercheronFrancoisTalbiEtAl96}]{PercheronFrancoisTalbiEtAl96}
Percheron, G., Francois, C., Talbi, B., Yelnik, J., \& Fenelon, G. (1996).
\newblock The primate motor thalamus.
\newblock {\em Brain Research}, {\em 22\/}(2), 93--181.

\bibitem[\protect\citename{Petrovic}{Kalso,
  Petersson}{Ingvar}{PetrovicETAL}{2002}{PetrovicKalsoPeterssonEtAl02}]{PetrovicKalsoPeterssonEtAl02}
Petrovic, P., Kalso, E., Petersson, K.~M., \& Ingvar, M. (2002).
\newblock Placebo and opioid analgesia-- imaging a shared neuronal network.
\newblock {\em Science}, {\em 295\/}.

\bibitem[\protect\citename{Pezzulo}{}{Castelfranchi}{Pezzulo-Castelfranchi}{2009}{PezzuloCastelfranchi09}]{PezzuloCastelfranchi09}
Pezzulo, G., \& Castelfranchi, C. (2009).
\newblock Thinking as the control of imagination: a conceptual framework for
  goal-directed systems.
\newblock {\em Psychological research}, {\em 73\/}.

\bibitem[\protect\citename{Pham}{}{}{Pham}{2007}{Pham07}]{Pham07}
Pham, M.~T. (2007).
\newblock Emotion and rationality: A critical review and interpretation of
  empirical evidence.
\newblock {\em Review of General Psychology}, {\em 11\/}, 155--178.

\bibitem[\protect\citename{Phillipson}{}{}{Phillipson}{1978}{Phillipson78}]{Phillipson78}
Phillipson, O.~T. (1978).
\newblock Afferent projections to {A10} dopaminergic neurones in the rat as
  shown by the retrograde transport of horseradish peroxidase.
\newblock {\em Neuroscience Letters}, {\em 9\/}(4), 353--359.

\bibitem[\protect\citename{Powers}{}{}{Powers}{1973}{Powers73}]{Powers73}
Powers, W.~T. (1973).
\newblock {\em Behavior: The control of perception}.
\newblock Hawthorne.

\bibitem[\protect\citename{Price}{}{Drevets}{Price-Drevets}{2010}{PriceDrevets10}]{PriceDrevets10}
Price, J.~L., \& Drevets, W.~C. (2010).
\newblock Neurocircuitry of mood disorders.
\newblock {\em Neuropsychopharmacology}, {\em 35\/}.

\bibitem[\protect\citename{Pyke}{}{}{Pyke}{1984}{Pyke84}]{Pyke84}
Pyke, G.~H. (1984).
\newblock Optimal foraging theory: A critical review.
\newblock {\em Annual review of ecology and systematics}, {\em 15\/}, 523--575.

\bibitem[\protect\citename{Pylyshyn}{}{}{Pylyshyn}{2000}{Pylyshyn00}]{Pylyshyn00}
Pylyshyn, Z.~W. (2000).
\newblock Situating vision in the world.
\newblock {\em Trends in cognitive sciences}, {\em 4\/}, 197--207.

\bibitem[\protect\citename{Rangel}{}{Hare}{Rangel-Hare}{2010}{RangelHare10}]{RangelHare10}
Rangel, A., \& Hare, T. (2010).
\newblock Neural computations associated with goal-directed choice.
\newblock {\em Current Opinion in Neurobiology}, {\em 20\/}(2), 262--279.

\bibitem[\protect\citename{Reiss}{}{}{Reiss}{2004}{Reiss04}]{Reiss04}
Reiss, S. (2004).
\newblock Multifaceted nature of intrinsic motivation: The theory of 16 basic
  desires.
\newblock {\em Review of General Psychology}, {\em 8\/}, 179--193.

\bibitem[\protect\citename{Rescorla}{}{Wagner}{Rescorla-Wagner}{1972}{RescorlaWagner72}]{RescorlaWagner72}
Rescorla, R.~A., \& Wagner, A.~R. (1972).
\newblock A theory of {P}avlovian conditioning: {V}ariation in the
  effectiveness of reinforcement and non-reinforcement.
\newblock In A.~H. Black, \& W.~F. Prokasy (Eds.), {\em {Classical Conditioning
  II: Theory and Research}} (pp.\ 64--99). New York: Appleton-Century-Crofts.

\bibitem[\protect\citename{Rizzolati}{Camarda, Fogassi, Gentilucci,
  Luppino}{Matelli}{RizzolatiETAL}{1988}{RizzolatiCamardaFogassiEtAl88}]{RizzolatiCamardaFogassiEtAl88}
Rizzolati, G., Camarda, R., Fogassi, L., Gentilucci, M., Luppino, G., \&
  Matelli, M. (1988).
\newblock Functional organization of inferior area 6 in the macaque monkey. ii.
  area f5 and the control of distal movements.
\newblock {\em Experimental Brain Research}, {\em 71\/}, 491--507.

\bibitem[\protect\citename{Rogers}{}{Monsell}{Rogers-Monsell}{1995}{RogersMonsell95}]{RogersMonsell95}
Rogers, R.~D., \& Monsell, S. (1995).
\newblock Costs of a predictable switch between simple cognitive tasks.
\newblock {\em Journal of Experimental Psychology: General}, {\em 124\/}(2),
  207--231.

\bibitem[\protect\citename{Rolls}{}{}{Rolls}{2005}{Rolls05}]{Rolls05}
Rolls, E.~T. (2005).
\newblock {\em Emotion explained}.
\newblock Oxford University Press.

\bibitem[\protect\citename{Rosadini}{}{Rossi}{Rosadini-Rossi}{1967}{RosadiniRossi67}]{RosadiniRossi67}
Rosadini, G., \& Rossi, G.~F. (1967).
\newblock On the suggested cerebral dominance for consciousness.
\newblock {\em Brain}, {\em 90\/}, 101--112.

\bibitem[\protect\citename{Rougier}{Noelle, Braver,
  Cohen}{O'Reilly}{RougierETAL}{2005}{RougierNoelleBraverEtAl05}]{RougierNoelleBraverEtAl05}
Rougier, N.~P., Noelle, D., Braver, T.~S., Cohen, J.~D., \& O'Reilly, R.~C.
  (2005).
\newblock Prefrontal cortex and the flexibility of cognitive control: {R}ules
  without symbols.
\newblock {\em Proceedings of the National Academy of Sciences}, {\em
  102\/}(20), 7338--7343.

\bibitem[\protect\citename{Roy}{Shohamy}{Wager}{RoyETAL}{2012}{RoyShohamyWager12}]{RoyShohamyWager12}
Roy, M., Shohamy, D., \& Wager, T.~D. (2012).
\newblock Ventromedial prefrontal-subcortical systems and the generation of
  affective meaning.
\newblock {\em Trends in Cognitive Sciences}, {\em 16\/}(3), 147--156.

\bibitem[\protect\citename{Rudebeck}{Behrens, Kennerley, Baxter, Buckley,
  Walton}{Rushworth}{RudebeckETAL}{2008}{RudebeckBehrensKennerleyEtAl08}]{RudebeckBehrensKennerleyEtAl08}
Rudebeck, P.~H., Behrens, T.~E., Kennerley, S.~W., Baxter, M.~G., Buckley,
  M.~J., Walton, M.~E., \& Rushworth, M. F.~S. (2008).
\newblock Frontal cortex subregions play distinct roles in choices between
  actions and stimuli.
\newblock {\em The Journal of neuroscience}, {\em 28\/}, 13775--13785.

\bibitem[\protect\citename{Rudebeck}{Walton, Smyth,
  Bannerman}{Rushworth}{RudebeckETAL}{2006}{RudebeckWaltonSmythEtAl06}]{RudebeckWaltonSmythEtAl06}
Rudebeck, P.~H., Walton, M.~E., Smyth, A.~N., Bannerman, D.~M., \& Rushworth,
  M. F.~S. (2006).
\newblock Separate neural pathways process different decision costs.
\newblock {\em Nature {N}euroscience}, {\em 9\/}(9), 1161--1168.

\bibitem[\protect\citename{Rushworth}{}{Behrens}{Rushworth-Behrens}{2008}{RushworthBehrens08}]{RushworthBehrens08}
Rushworth, M. F.~S., \& Behrens, T. E.~J. (2008).
\newblock Choice, uncertainty and value in prefrontal and cingulate cortex.
\newblock {\em Nature neuroscience}, {\em 11\/}.

\bibitem[\protect\citename{Rushworth}{Behrens,
  Rudebeck}{Walton}{RushworthETAL}{2007}{RushworthBehrensRudebeckEtAl07}]{RushworthBehrensRudebeckEtAl07}
Rushworth, M. F.~S., Behrens, T. E.~J., Rudebeck, P.~H., \& Walton, M.~E.
  (2007).
\newblock Contrasting roles for cingulate and orbitofrontal cortex in decisions
  and social behaviour.
\newblock {\em Trends in Cognitive Sciences}, {\em 11\/}(4), 168--176.

\bibitem[\protect\citename{Russell}{}{Barrett}{Russell-Barrett}{1999}{RussellBarrett99}]{RussellBarrett99}
Russell, J.~A., \& Barrett, L.~F. (1999).
\newblock Core affect, prototypical emotional episodes, and other things called
  emotion: dissecting the elephant.
\newblock {\em Journal of personality and social psychology}, {\em 76\/},
  805--819.

\bibitem[\protect\citename{Russo}{Murrough, Han,
  Charney}{Nestler}{RussoETAL}{2012}{RussoMurroughHanEtAl12}]{RussoMurroughHanEtAl12}
Russo, S.~J., Murrough, J.~W., Han, M.-H., Charney, D.~S., \& Nestler, E.~J.
  (2012).
\newblock Neurobiology of resilience.
\newblock {\em Nature neuroscience}, {\em 15\/}(11).

\bibitem[\protect\citename{Saddoris}{Gallagher}{Schoenbaum}{SaddorisETAL}{2005}{SaddorisGallagherSchoenbaum05}]{SaddorisGallagherSchoenbaum05}
Saddoris, M.~P., Gallagher, M., \& Schoenbaum, G. (2005).
\newblock Rapid associative encoding in basolateral amygdala depends on
  connections with orbitofrontal cortex.
\newblock {\em Neuron}, {\em 46\/}(2), 321--331.

\bibitem[\protect\citename{Salamone}{}{Correa}{Salamone-Correa}{2012}{SalamoneCorrea12}]{SalamoneCorrea12}
Salamone, J.~D., \& Correa, M. (2012).
\newblock The mysterious motivational functions of mesolimbic dopamine.
\newblock {\em Neuron}, {\em 76\/}(3), 470--485.

\bibitem[\protect\citename{Salamone}{Correa,
  Farrar}{Mingote}{SalamoneETAL}{2007}{SalamoneCorreaFarrarEtAl07}]{SalamoneCorreaFarrarEtAl07}
Salamone, J.~D., Correa, M., Farrar, A., \& Mingote, S.~M. (2007).
\newblock Effort-related functions of nucleus accumbens dopamine and associated
  forebrain circuits.
\newblock {\em Psychopharmacology}, {\em 191\/}, 461--482.

\bibitem[\protect\citename{Salamone}{Cousins}{Snyder}{SalamoneETAL}{1997}{SalamoneCousinsSynder97}]{SalamoneCousinsSynder97}
Salamone, J.~D., Cousins, M.~S., \& Snyder, B.~J. (1997).
\newblock Behavioral functions of nucleus accumbens dopamine: empirical and
  conceptual problems with the anhedonia hypothesis.
\newblock {\em Neuroscience and biobehavioral reviews}, {\em 21\/}(3),
  341--359.

\bibitem[\protect\citename{Schoenbaum}{}{}{Schoenbaum}{2004}{Schoenbaum04}]{Schoenbaum04}
Schoenbaum, G. (2004).
\newblock Affect, action, and ambiguity and the amygdala-orbitofrontal circuit.
  {F}ocus on ``{C}ombined unilateral lesions of the amygdala and orbital
  prefrontal cortex impair affective processing in rhesus monkey''.
\newblock {\em Journal of Neurophysiology}, {\em 91\/}(5), 1938--1939.

\bibitem[\protect\citename{Schoenbaum}{Chiba}{Gallagher}{SchoenbaumETAL}{1999}{SchoenbaumChibaGallagher99}]{SchoenbaumChibaGallagher99}
Schoenbaum, G., Chiba, A.~A., \& Gallagher, M. (1999).
\newblock Neural encoding in orbitofrontal cortex and basolateral amygdala
  during olfactory discrimination learning.
\newblock {\em Journal of Neuroscience}, {\em 19\/}, 1876--84.

\bibitem[\protect\citename{Schoenbaum}{Chiba}{Gallagher}{SchoenbaumETAL}{2000}{SchoenbaumChibaGallagher00}]{SchoenbaumChibaGallagher00}
Schoenbaum, G., Chiba, A.~A., \& Gallagher, M. (2000).
\newblock Changes in functional connectivity in orbitofrontal cortex and
  basolateral amygdala during learning and reversal training.
\newblock {\em The Journal of Neuroscience}, {\em 20\/}(13), 5179--5189.

\bibitem[\protect\citename{Schoenbaum}{Roesch,
  Stalnaker}{Takahashi}{SchoenbaumETAL}{2009}{SchoenbaumRoeschStalnakerEtAl09}]{SchoenbaumRoeschStalnakerEtAl09}
Schoenbaum, G., Roesch, M.~R., Stalnaker, T.~A., \& Takahashi, Y.~K. (2009).
\newblock A new perspective on the role of the orbitofrontal cortex in adaptive
  behaviour.
\newblock {\em Nature Reviews Neuroscience}, {\em 10\/}(12), 885--892.

\bibitem[\protect\citename{Schoenbaum}{}{Setlow}{Schoenbaum-Setlow}{2001}{SchoenbaumSetlow01}]{SchoenbaumSetlow01}
Schoenbaum, G., \& Setlow, B. (2001).
\newblock Integrating orbitofrontal cortex into prefrontal theory: {C}ommon
  processing themes across species and subdivisions.
\newblock {\em Learning \& Memory}, {\em 8\/}, 134--147.

\bibitem[\protect\citename{Schoenbaum}{Setlow,
  Saddoris}{Gallagher}{SchoenbaumETAL}{2003}{SchoenbaumSetlowSaddorisEtAl03}]{SchoenbaumSetlowSaddorisEtAl03}
Schoenbaum, G., Setlow, B., Saddoris, M.~P., \& Gallagher, M. (2003).
\newblock Encoding predicted outcome and acquired value in orbitofrontal cortex
  during cue sampling depends upon input from basolateral amygdala.
\newblock {\em Neuron}, {\em 39\/}, 855--867.

\bibitem[\protect\citename{Schultz}{}{}{Schultz}{1998}{Schultz98}]{Schultz98}
Schultz, W. (1998).
\newblock Predictive reward signal of dopamine neurons.
\newblock {\em Journal of Neurophysiology}, {\em 80\/}(1), 1--27.

\bibitem[\protect\citename{Schultz}{}{}{Schultz}{2002}{Schultz02}]{Schultz02}
Schultz, W. (2002).
\newblock Getting formal with dopamine and reward.
\newblock {\em Neuron}, {\em 36\/}, 241--263.

\bibitem[\protect\citename{Schultz}{Dayan}{Montague}{SchultzETAL}{1997}{SchultzDayanMontague97}]{SchultzDayanMontague97}
Schultz, W., Dayan, P., \& Montague, P.~R. (1997).
\newblock A neural substrate of prediction and reward.
\newblock {\em Science}, {\em 275\/}(5306), 1593--1599.

\bibitem[\protect\citename{Schultz}{}{Dickinson}{Schultz-Dickinson}{2000}{SchultzDickinson00}]{SchultzDickinson00}
Schultz, W., \& Dickinson, A. (2000).
\newblock Neuronal coding of prediction errors.
\newblock {\em Annual Review of Neuroscience}, {\em 23\/}, 473--500.

\bibitem[\protect\citename{Semba}{}{Fibiger}{Semba-Fibiger}{1992}{SembaFibiger92}]{SembaFibiger92}
Semba, K., \& Fibiger, H.~C. (1992).
\newblock Afferent connections of the laterodorsal and the pedunculopontine
  tegmental nuclei in the rat: a retro- and antero-grade transport and
  immunohistochemical study.
\newblock {\em The Journal of Comparative Neurology}, {\em 323\/}(3), 387--410.

\bibitem[\protect\citename{Shepard}{Holcomb}{Gold}{ShepardETAL}{2006}{ShepardHolcombGold06}]{ShepardHolcombGold06}
Shepard, P.~D., Holcomb, H.~H., \& Gold, J.~M. (2006).
\newblock The presence of absence: Habenular regulation of dopamine neurons and
  the encoding of negative outcomes.
\newblock {\em Schizophrenia Bulletin}, {\em 32\/}(3), 417--421.

\bibitem[\protect\citename{Sideridis}{}{}{Sideridis}{2005}{Sideridis05}]{Sideridis05}
Sideridis, G.~D. (2005).
\newblock Goal orientation, academic achievement, and depression: Evidence in
  favor of a revised goal theory framework.
\newblock {\em Journal of Educational Psychology}, {\em 97\/}(3), 366.

\bibitem[\protect\citename{Simonson}{}{Tversky}{Simonson-Tversky}{1992}{SimonsonTversky92}]{SimonsonTversky92}
Simonson, I., \& Tversky, A. (1992).
\newblock Choice in context: Tradeoff contrast and extremeness aversion.
\newblock {\em Journal of Marketing Research}, {\em 29\/}(3).

\bibitem[\protect\citename{Slusarek}{Velling,
  Bunk}{Eggers}{SlusarekETAL}{2001}{SlusarekVellingBunkEtAl01}]{SlusarekVellingBunkEtAl01}
Slusarek, M., Velling, S., Bunk, D., \& Eggers, C. (2001).
\newblock Motivational effects on inhibitory control in children with adhd.
\newblock {\em Journal of the American Academy of Child and Adolescent
  Psychiatry}, {\em 40\/}.

\bibitem[\protect\citename{Smith}{}{Bolam}{Smith-Bolam}{1990}{SmithBolam90}]{SmithBolam90}
Smith, A.~D., \& Bolam, J.~P. (1990).
\newblock The neural network of the basal ganglia as revealed by the study of
  synaptic connections of identified neurones.
\newblock {\em Trends in Neuroscience}, {\em 13\/}, 259--265.

\bibitem[\protect\citename{Snyder}{Banich}{Munakata}{SnyderETAL}{2011}{SnyderBanichMunakata11}]{SnyderBanichMunakata11}
Snyder, H.~R., Banich, M.~T., \& Munakata, Y. (2011).
\newblock Choosing our words: retrieval and selection processes recruit shared
  neural substrates in left ventrolateral prefrontal cortex.
\newblock {\em Journal of cognitive neuroscience}, {\em 23\/}.

\bibitem[\protect\citename{Snyder}{Hutchison, Nyhus, Curran,
  Banich}{Munakata}{SnyderETAL}{2010}{SnyderHutchisonNyhusEtAl10}]{SnyderHutchisonNyhusEtAl10}
Snyder, H.~R., Hutchison, N., Nyhus, E., Curran, T., Banich, M.~T., \&
  Munakata, Y. (2010).
\newblock Neural inhibition enables selection during language processing.
\newblock {\em Proceedings of the National Academy of Sciences}, {\em 107\/},
  16483--16488.

\bibitem[\protect\citename{Solway}{}{Botvinick}{Solway-Botvinick}{2012}{SolwayBotvinick12}]{SolwayBotvinick12}
Solway, A., \& Botvinick, M.~M. (2012).
\newblock Goal-directed decision making as probabilistic inference: {A}
  computational framework and potential neural correlates.
\newblock {\em Psychological Review}, {\em 119\/}(1), 120--154.

\bibitem[\protect\citename{Spielberg}{Heller}{Miller}{SpielbergETAL}{2013}{SpielbergHellerMiller13}]{SpielbergHellerMiller13}
Spielberg, J.~M., Heller, W., \& Miller, G.~A. (2013).
\newblock Hierarchical brain networks active in approach and avoidance goal
  pursuit.
\newblock {\em Frontiers in human neuroscience}, {\em 7\/}.

\bibitem[\protect\citename{St~Onge}{Stopper,
  Zahm}{Floresco}{St~OngeETAL}{2012}{StOngeStopperZahmEtAl12}]{StOngeStopperZahmEtAl12}
St~Onge, J.~R., Stopper, C.~M., Zahm, D.~S., \& Floresco, S.~B. (2012).
\newblock Separate prefrontal-subcortical circuits mediate different components
  of risk-based decision making.
\newblock {\em The Journal of neuroscience}, {\em 32\/}.

\bibitem[\protect\citename{Stamatakis}{}{Stuber}{Stamatakis-Stuber}{2012}{StamatakisStuber12}]{StamatakisStuber12}
Stamatakis, A.~M., \& Stuber, G.~D. (2012).
\newblock Activation of lateral habenula inputs to the ventral midbrain
  promotes behavioral avoidance.
\newblock {\em Nature Neuroscience}, {\em 15\/}(8), 1105--1107.

\bibitem[\protect\citename{Steel}{}{}{Steel}{2007}{Steel07}]{Steel07}
Steel, P. (2007).
\newblock The nature of procrastination: a meta-analytic and theoretical review
  of quintessential self-regulatory failure.
\newblock {\em Psychological bulletin}, {\em 133\/}.

\bibitem[\protect\citename{Steven}{Timothy}{Jonathan}{StevenETAL}{2011}{KennerleyBehrensWallis11}]{KennerleyBehrensWallis11}
Steven, K., Timothy, B., \& Jonathan, W. (2011).
\newblock Double dissociation of value computations in orbitofrontal and
  anterior cingulate neurons.
\newblock {\em Nat Neurosci}, {\em 14\/}(12), 1581--1589.

\bibitem[\protect\citename{Stollstorff}{Vartanian}{Goel}{StollstorffETAL}{2012}{StollstorffVartanianGoel12}]{StollstorffVartanianGoel12}
Stollstorff, M., Vartanian, O., \& Goel, V. (2012).
\newblock Levels of conflict in reasoning modulate right lateral prefrontal
  cortex.
\newblock {\em Brain research}, {\em 1428\/}.

\bibitem[\protect\citename{Sun}{}{}{Sun}{2009}{Sun09}]{Sun09}
Sun, R. (2009).
\newblock Motivational representations within a computational cognitive
  architecture.
\newblock {\em Cognitive Computation}, {\em 1\/}, 91--103.

\bibitem[\protect\citename{Sutton}{}{}{Sutton}{1988}{Sutton88}]{Sutton88}
Sutton, R.~S. (1988).
\newblock Learning to predict by the method of temporal differences.
\newblock {\em Machine Learning}, {\em 3\/}, 9--44.

\bibitem[\protect\citename{Sutton}{}{Barto}{Sutton-Barto}{1981}{SuttonBarto81}]{SuttonBarto81}
Sutton, R.~S., \& Barto, A. (1981).
\newblock Toward a modern theory of adaptive networks: {E}xpectation and
  prediction.
\newblock {\em Psychological Review}, {\em 88\/}(2), 135--170.

\bibitem[\protect\citename{Sutton}{}{Barto}{Sutton-Barto}{1990}{SuttonBarto90}]{SuttonBarto90}
Sutton, R.~S., \& Barto, A.~G. (1990).
\newblock Time-derivative models of {P}avlovian reinforcement.
\newblock In J.~W. Moore, \& M. Gabriel (Eds.), {\em {Learning and
  Computational Neuroscience}} (pp.\ 497--537). Cambridge, MA: MIT Press.

\bibitem[\protect\citename{Sutton}{}{Barto}{Sutton-Barto}{1998}{SuttonBarto98}]{SuttonBarto98}
Sutton, R.~S., \& Barto, A.~G. (1998).
\newblock {\em {Reinforcement Learning: An Introduction.}}
\newblock Cambridge, MA: MIT Press.

\bibitem[\protect\citename{Sutton}{}{Davidson}{Sutton-Davidson}{1997}{SuttonDavidson97}]{SuttonDavidson97}
Sutton, S.~K., \& Davidson, R.~J. (1997).
\newblock Prefrontal brain asymmetry: A biological substrate of the behavioral
  approach and inhibition systems.
\newblock {\em Psychological Science}, {\em 8\/}, 204--210.

\bibitem[\protect\citename{Svebak}{}{Murgatroyd}{Svebak-Murgatroyd}{1985}{SvebakMurgatroyd85}]{SvebakMurgatroyd85}
Svebak, S., \& Murgatroyd, S. (1985).
\newblock Metamotivational dominance: A multimethod validation of reversal
  theory constructs.
\newblock {\em Journal of Personality and Social Psychology}, {\em 48\/},
  107--116.

\bibitem[\protect\citename{Swanson}{Perry, Koch-Krueger, Katner,
  Svensson}{Bymaster}{SwansonETAL}{2006}{SwansonPerryKoch-KruegerEtAl06}]{SwansonPerryKoch-KruegerEtAl06}
Swanson, C.~J., Perry, K.~W., Koch-Krueger, S., Katner, J., Svensson, K.~A., \&
  Bymaster, F.~P. (2006).
\newblock Effect of the attention deficit/hyperactivity disorder drug
  atomoxetine on extracellular concentrations of norepinephrine and dopamine in
  several brain regions of the rat.
\newblock {\em Neuropharmacology}, {\em 50\/}, 755--760.

\bibitem[\protect\citename{Takayama}{}{Miura}{Takayama-Miura}{1991}{TakayamaMiura91}]{TakayamaMiura91}
Takayama, K., \& Miura, M. (1991).
\newblock Glutamate-immunoreactive neurons of the central amygdaloid nucleus
  projecting to the subretrofacial nucleus of {SHR} and {WKY} rats: {A}
  double-labeling study.
\newblock {\em Neuroscience Letters}, {\em 134\/}(1), 62--66.

\bibitem[\protect\citename{Thorn}{Atallah,
  Howe}{Graybiel}{ThornETAL}{2010}{ThornAtallahHoweEtAl10}]{ThornAtallahHoweEtAl10}
Thorn, C.~A., Atallah, H., Howe, M., \& Graybiel, A.~M. (2010).
\newblock Differential dynamics of activity changes in dorsolateral and
  dorsomedial striatal loops during learning.
\newblock {\em Neuron}, {\em 66\/}(5), 781--795.

\bibitem[\protect\citename{Thorndike}{}{}{Thorndike}{1911}{Thorndike11}]{Thorndike11}
Thorndike, E.~L. (1911).
\newblock {\em Animal intelligence: Experimental studies}.
\newblock MacMillan Press.

\bibitem[\protect\citename{Tobler}{Dickinson}{Schultz}{ToblerETAL}{2003}{ToblerDickinsonSchultz03}]{ToblerDickinsonSchultz03}
Tobler, P.~N., Dickinson, A., \& Schultz, W. (2003).
\newblock Coding of predicted reward omission by dopamine neurons in a
  conditioned inhibition paradigm.
\newblock {\em Journal of Neuroscience}, {\em 23\/}, 10402--10.

\bibitem[\protect\citename{Tobler}{Fiorillo}{Schultz}{ToblerETAL}{2005}{ToblerFiorilloSchultz05}]{ToblerFiorilloSchultz05}
Tobler, P.~N., Fiorillo, C.~D., \& Schultz, W. (2005).
\newblock Adaptive coding of reward value by dopamine neurons.
\newblock {\em Science}, {\em 307\/}(5715), 1642--1645.

\bibitem[\protect\citename{Tolman}{}{}{Tolman}{1948}{Tolman48}]{Tolman48}
Tolman, E. (1948).
\newblock Cognitive maps in rats and men.
\newblock {\em Psychological Review}, {\em 55\/}(4), 189--208.

\bibitem[\protect\citename{Tolman}{}{}{Tolman}{1932}{Tolman32}]{Tolman32}
Tolman, E.~C. (1932).
\newblock {\em {Purposive Behavior in Animals and Men}}.
\newblock Century.

\bibitem[\protect\citename{Tomasello}{}{}{Tomasello}{2001}{Tomasello01}]{Tomasello01}
Tomasello, M. (2001).
\newblock {\em {The Cultural Origins of Human Cognition}}.
\newblock Cambridge, MA: Harvard University Press.

\bibitem[\protect\citename{Treadway}{}{Zald}{Treadway-Zald}{2011}{TreadwayZald11}]{TreadwayZald11}
Treadway, M.~T., \& Zald, D.~H. (2011).
\newblock Reconsidering anhedonia in depression: {L}essons from translational
  neuroscience.
\newblock {\em Neuroscience and Biobehavioral Reviews}, {\em 35\/}.

\bibitem[\protect\citename{Trope}{}{Liberman}{Trope-Liberman}{2010}{TropeLiberman10}]{TropeLiberman10}
Trope, Y., \& Liberman, N. (2010).
\newblock Construal-level theory of psychological distance.
\newblock {\em Psychological review}, {\em 117\/}.

\bibitem[\protect\citename{Ungerleider}{}{Mishkin}{Ungerleider-Mishkin}{1982}{UngerleiderMishkin82}]{UngerleiderMishkin82}
Ungerleider, L.~G., \& Mishkin, M. (1982).
\newblock Two cortical visual systems.
\newblock In D.~J. Ingle, M.~A. Goodale, \& R.~J.~W. Mansfield (Eds.), {\em
  {The Analysis of Visual Behavior}} (pp.\ 549--586). Cambridge, MA: MIT Press.

\bibitem[\protect\citename{Urakubo}{Honda,
  Froemke}{Kuroda}{UrakuboETAL}{2008}{UrakuboHondaFroemkeEtAl08}]{UrakuboHondaFroemkeEtAl08}
Urakubo, H., Honda, M., Froemke, R.~C., \& Kuroda, S. (2008).
\newblock Requirement of an allosteric kinetics of {NMDA} receptors for spike
  timing-dependent plasticity.
\newblock {\em The Journal of Neuroscience}, {\em 28\/}(13), 3310--3323.

\bibitem[\protect\citename{Valentin}{Dickinson}{O'Doherty}{ValentinETAL}{2007}{ValentinDickinsonODoherty07}]{ValentinDickinsonODoherty07}
Valentin, V.~V., Dickinson, A., \& O'Doherty, J.~P. (2007).
\newblock Determining the neural substrates of goal-directed learning in the
  human brain.
\newblock {\em The Journal of neuroscience}, {\em 27\/}, 4019--4026.

\bibitem[\protect\citename{Vygotsky}{}{}{Vygotsky}{1978}{Vygotsky78}]{Vygotsky78}
Vygotsky, L.~S. (1978).
\newblock Interaction between learning and development (m. lopez- morillas,
  trans.).
\newblock In M. Cole, V. John-Steiner, S. Scribner, \& E. Souberman (Eds.),
  {\em Mind in society: The development of higher psychological processes}
  (pp.\ 79--91). Cambridge, MA: Harvard University Press.

\bibitem[\protect\citename{Waelti}{Dickinson}{Schultz}{WaeltiETAL}{2001}{WaeltiDickinsonSchultz01}]{WaeltiDickinsonSchultz01}
Waelti, P., Dickinson, A., \& Schultz, W. (2001).
\newblock Dopamine responses comply with basic assumptions of formal learning
  theory.
\newblock {\em Nature}, {\em 412\/}, 43--48.

\bibitem[\protect\citename{Wager}{Atlas,
  Leotti}{Rilling}{WagerETAL}{2011}{WagerAtlasLeottiEtAl11}]{WagerAtlasLeottiEtAl11}
Wager, T.~D., Atlas, L.~Y., Leotti, L.~A., \& Rilling, J.~K. (2011).
\newblock Predicting individual differences in placebo analgesia: contributions
  of brain activity during anticipation and pain experience.
\newblock {\em The Journal of neuroscience}, {\em 31\/}.

\bibitem[\protect\citename{Wager}{Rilling, Smith, Sokolik, Casey, Davidson,
  Kosslyn,
  Rose}{Cohen}{WagerETAL}{2004}{WagerRillingSmithEtAl04}]{WagerRillingSmithEtAl04}
Wager, T.~D., Rilling, J.~K., Smith, E.~E., Sokolik, A., Casey, K.~L.,
  Davidson, R.~J., Kosslyn, S.~M., Rose, R.~M., \& Cohen, J.~D. (2004).
\newblock Placebo-induced changes in fmri in the anticipation and experience of
  pain.
\newblock {\em Science}, {\em 303\/}.

\bibitem[\protect\citename{Wallace}{Magnuson}{Gray}{WallaceETAL}{1992}{WallaceMagnusonGray92}]{WallaceMagnusonGray92}
Wallace, D.~M., Magnuson, D.~J., \& Gray, T.~S. (1992).
\newblock Organization of amygdaloid projections to brainstem dopaminergic,
  noradrenergic, and adrenergic cell groups in the rat.
\newblock {\em Brain Research Bulletin}, {\em 28\/}, 447--454.

\bibitem[\protect\citename{Walton}{Bannerman,
  Alterescu}{Rushworth}{WaltonETAL}{2003}{WaltonBannermanAlterescuEtAl03}]{WaltonBannermanAlterescuEtAl03}
Walton, M.~E., Bannerman, D.~M., Alterescu, K., \& Rushworth, M. F.~S. (2003).
\newblock Functional specialization within medial frontal cortex of the
  anterior cingulate for evaluating effort-related decisions.
\newblock {\em The Journal of neuroscience : the official journal of the
  Society for Neuroscience}, {\em 23\/}, 6475.

\bibitem[\protect\citename{Watson}{El-Deredy, Iannetti, Lloyd, Tracey, Vogt,
  Nadeau}{Jones}{WatsonETAL}{2009}{WatsonEl-DeredyIannettiEtAl09}]{WatsonEl-DeredyIannettiEtAl09}
Watson, A., El-Deredy, W., Iannetti, G.~D., Lloyd, D., Tracey, I., Vogt, B.~A.,
  Nadeau, V., \& Jones, A. K.~P. (2009).
\newblock Placebo conditioning and placebo analgesia modulate a common brain
  network during pain anticipation and perception.
\newblock {\em Pain}, {\em 145\/}.

\bibitem[\protect\citename{Watson}{}{Tellegen}{Watson-Tellegen}{1985}{WatsonTellegen85}]{WatsonTellegen85}
Watson, D., \& Tellegen, A. (1985).
\newblock Toward a consensual structure of mood.
\newblock {\em Psychological bulletin}, {\em 98\/}, 219--235.

\bibitem[\protect\citename{West}{DesJardin,
  Gale}{Malkova}{WestETAL}{2011}{WestDesJardinGaleEtAl11}]{WestDesJardinGaleEtAl11}
West, E.~A., DesJardin, J.~T., Gale, K., \& Malkova, L. (2011).
\newblock Transient inactivation of orbitofrontal cortex blocks reinforcer
  devaluation in macaques.
\newblock {\em The Journal of neuroscience}, {\em 31\/}.

\bibitem[\protect\citename{Westen}{Blagov, Harenski,
  Kilts}{Hamann}{WestenETAL}{2006}{WestenBlagovHarenskiEtAl06}]{WestenBlagovHarenskiEtAl06}
Westen, D., Blagov, P.~S., Harenski, K., Kilts, C., \& Hamann, S. (2006).
\newblock Neural bases of motivated reasoning: an fmri study of emotional
  constraints on partisan political judgment in the 2004 u.s. presidential
  election.
\newblock {\em Journal of cognitive neuroscience}, {\em 18\/}, 1947--1958.

\bibitem[\protect\citename{Winstanley}{Theobald,
  Cardinal}{Robbins}{WinstanleyETAL}{2004}{WinstanleyTheobaldCardinalEtAl04}]{WinstanleyTheobaldCardinalEtAl04}
Winstanley, C.~A., Theobald, D. E.~H., Cardinal, R.~N., \& Robbins, T.~W.
  (2004).
\newblock Contrasting roles of basolateral amygdala and orbitofrontal cortex in
  impulsive choice.
\newblock {\em The Journal of neuroscience : the official journal of the
  Society for Neuroscience}, {\em 24\/}(20), 4718--4722.

\bibitem[\protect\citename{Wise}{}{}{Wise}{1980}{Wise80}]{Wise80}
Wise, R.~A. (1980).
\newblock Action of drugs of abuse on brain reward systems.
\newblock {\em Pharmacology, biochemistry, and behavior}, {\em 13 Suppl 1\/},
  213--223.

\bibitem[\protect\citename{Wise}{}{Rompre}{Wise-Rompre}{1989}{WiseRompre89}]{WiseRompre89}
Wise, R.~A., \& Rompre, P.-P. (1989).
\newblock Brain dopamine and reward.
\newblock {\em Annual Review of Psychology}, {\em 40\/}, 191--225.

\bibitem[\protect\citename{Wolpert}{}{Kawato}{Wolpert-Kawato}{1998}{WolpertKawato98}]{WolpertKawato98}
Wolpert, D., \& Kawato, M. (1998).
\newblock Multiple paired forward and inverse models for motor control.
\newblock {\em Neural networks : the official journal of the International
  Neural Network Society}, {\em 11\/}(7-8), 1317--1329.

\bibitem[\protect\citename{Wrosch}{Scheier, Miller,
  Schulz}{Carver}{WroschETAL}{2003}{WroschScheierMillerEtAl03}]{WroschScheierMillerEtAl03}
Wrosch, C., Scheier, M.~F., Miller, G.~E., Schulz, R., \& Carver, C.~S. (2003).
\newblock Adaptive self-regulation of unattainable goals: goal disengagement,
  goal reengagement, and subjective well-being.
\newblock {\em Personality \& social psychology bulletin}, {\em 29\/}.

\bibitem[\protect\citename{Wulfram}{}{Richard}{Wulfram-Richard}{2009}{GerstnerNaud09}]{GerstnerNaud09}
Wulfram, G., \& Richard, N. (2009).
\newblock How good are neuron models?
\newblock {\em Science}, {\em 326\/}(5951), 379--380.

\bibitem[\protect\citename{Yin}{Knowlton}{Balleine}{YinETAL}{2006}{YinKnowltonBalleine06}]{YinKnowltonBalleine06}
Yin, H.~H., Knowlton, B.~J., \& Balleine, B.~W. (2006).
\newblock Inactivation of dorsolateral striatum enhances sensitivity to changes
  in the action-outcome contingency in instrumental conditioning.
\newblock {\em Behavioural brain research}, {\em 166\/}(2), 189--196.

\bibitem[\protect\citename{Zeeb}{Floresco}{Winstanley}{ZeebETAL}{2010}{ZeebFlorescoWinstanley10}]{ZeebFlorescoWinstanley10}
Zeeb, F.~D., Floresco, S.~B., \& Winstanley, C.~A. (2010).
\newblock Contributions of the orbitofrontal cortex to impulsive choice:
  interactions with basal levels of impulsivity, dopamine signalling, and
  reward-related cues.
\newblock {\em Psychopharmacology}, {\em 211\/}.

\bibitem[\protect\citename{Zeigarnik}{}{}{Zeigarnik}{1927}{Zeigarnik27}]{Zeigarnik27}
Zeigarnik, B. (1927).
\newblock Über das behalten von erledigten und unerledigten handlungen.
\newblock {\em Psychologische Forschung}, {\em 9\/}, 1--85.

\end{thebibliography}

\newpage
\section{Appendix: Computational Model Details}

This appendix provides more information about the emery foraging model.  The purpose of this information is to give more detailed insight into the model's function beyond the level provided in the main text, but with a model of this complexity, the only way to really understand it is to explore the model itself.  It is available for download at \verb\http://grey.colorado.edu/CompCogNeuro/index.php/CCN_Repository\.  And the best way to understand this model is to understand the framework in which it is implemented, which is explained in great detail, with many running simulations explaining specific elements of functionality, at \verb\http://ccnbook.colorado.edu\.

\subsection{Model Training}

The model was trained in a sequence of stages, designed to capture in the end some rough approximation of the relevant functionality present in a rat performing in a simple plus maze environment.  Each individual training stage is not intended to be a completely realistic model of how this functionality actually develops in a rat, however --- that is a larger and important research agenda that we are hoping to make progress on.  Each stage of training typically turns on an increasing number of layers in the network, as needed to accomplish that stage.  The tags for each stage are the configuration id's of the relevant stage as actually implemented in the model.

\begin{itemize}
\item {\bf vision\_train}: This drives random motor exploration in the environment to train up the visual system (V1, V4, Object) so that the output layer of the visual system reports what is being seen (these are the only layers turned on for this stage).  The target outputs are generated by considering the orientation and position of the rat, and thus inferring what it should be looking at.  This produces highly accurate Object representations that can recognize each of the landmarks based on both shape and color information.  See \incite{OReillyWyatteHerdEtAl13} and the objrec simulation in \verb\http://ccnbook.colorado.edu\ for more information about how the visual system learns.  In brief, it learns increasingly spatially invariant and featurally complex representations as it proceeds up the pathway.  Given the simplicity of these stimuli, we did not need an IT layer.

\item {\bf match\_train:} This trains the Match layer to compute whether the Object and Target layers have a matching representation, which is then a useful abstraction for the motor program to operate from.  This randomly presents inputs to the Object and Target layers, and trains the appropriate Match response.  We have found that the motor system does learn without the benefit of this Match layer, but it is much faster and more reliable with it.

\item {\bf motor\_train:} This trains the dlPFC\_Hid layer to generate proper SMA\_out motor actions (left = turn left, middle = go forward, right = turn right) based on visual input relative to target, such that rat will rotate until a match occurs, and then it moves forward to approach.  We assume that this kind of approach-a-target motor program is something that rats learn relatively quickly in life (if they are not born with it in some way) --- we are currently developing models that explore how such motor programs could be learned in a more ecologically-valid fashion.

\item {\bf explore\_env:} This turns on all of the gdPVLV layers, and the OFC and ACC layers, and does initial exploration of the environment for the affective system.  It trains the OFC, ACC, BLA, and CeM layers to know which rewards occur where, and what visual inputs they're associated with. A key outcome of this is that the PosLV\_OFC units learn to activate the proper Target layer for a given target PV outcome.

\item {\bf goal\_learn:}  This is the final stage which was described in detail in the main text.  It alternates between goal selection and goal enaged modes, and demonstrates basic full loop operation of deciding what it wants, and going to get it, based on all the above prior learning.
\end{itemize}

\subsection{Model Algorithms}

The model was implemented using the Leabra framework, which is described in detail in \incite{OReillyMunakataFrankEtAl12}, \incite{OReillyMunakata00}, \incite{OReilly01}, and summarized here.  See Table~\ref{tab.sim_params} for a listing of parameter values, nearly all of which are at their default settings.  These same parameters and equations have been used to simulate over 40 different models in \incite{OReillyMunakataFrankEtAl12} and \incite{OReillyMunakata00}, and a number of other research models.  Thus, the model can be viewed as an instantiation of a systematic modeling framework using standardized mechanisms, instead of constructing new mechanisms for each model.  

This version of Leabra contains three primary differences from the original \cite{OReillyMunakata00}: the activation function is slightly different, in a way that allows units to more accurately reflect their graded excitatory input drive, the inhibition function is much simpler and more biologically realistic, and the learning rule takes a more continuous form involving contrasts between values integrated over different time frames (i.e., with different time constants), which also produces a combination of error-driven and self-organizing learning within the same simple mathematical framework.  These modifications are described in detail in an updated version of the \incite{OReillyMunakata00} textbook, in \incite{OReillyMunakataFrankEtAl12}.  This new learning algorithm goes by the acronym of XCAL (temporally eXtended Contrastive Attractor Learning), and it replaces the combination of Contrastive Hebbian Learning (CHL) and standard Hebbian learning used in the original Leabra framework.

\subsubsection{Pseudocode}

The pseudocode for Leabra is given here, showing exactly how the pieces of the algorithm described in more detail in the subsequent sections fit together. The individual steps are repeated for each event (trial), which can be broken down into a {\em minus} and {\em plus} phase, followed by a synaptic weight updating function. Generally speaking, the minus phase represents the system's expectation for a given input and the plus phase represents the observation of the outcome. The difference between these two phases is then used to compute the updating function that drives learning. Furthermore, each phase contains a period of {\em settling} (measured in {\em cycles}) during which the activation values of each unit are updated taking into account the previous state of the network. Some units are {\em clamped}, or have fixed activation values and are not subject to this updating rule (e.g., V1 input in the minus phase, V1 input and Output in the plus phase).

Outer loop: For each event (trial) in an epoch:
\begin{enumerate}
\item Iterate over minus and plus phases of settling for each event.
 \begin{enumerate}
 \item At start of settling, for all units:
  \begin{enumerate}
  \item Initialize all state variables (activation, $V_m$, etc).
  \item Clamp external patterns (V1 input in minus phase, V1 input \& Output in plus phase).
  \end{enumerate}
 \item During each cycle of settling, for all non-clamped units:
  \begin{enumerate}
  \item Compute excitatory netinput ($g_e(t)$ or $\eta_j$,
   eq~\ref{eq.net_in_avg}).
  \item Compute FFFB inhibition for each layer, based on average net input and activation levels within the layer (eq~\ref{eq.fffb})
  \item Compute point-neuron activation combining excitatory input and inhibition (eq~\ref{eq.vm}).
  \item Update time-averaged activation values (short, medium, long) for use in learning.
  \end{enumerate}
 \end{enumerate}
 \item After both phases update the weights, for all connections:
 \begin{enumerate}
 \item Compute XCAL learning as function of short, medium, and long time averages.
 \item Increment the weights according to net weight change.
 \end{enumerate}
\end{enumerate}

\subsubsection{Point Neuron Activation Function} 

\begin{table}
 \centering
 \begin{tabular}{ll|ll} \hline
Parameter & Value & Parameter & Value \\ \hline
$E_l$ & 0.30 & $\overline{g_l}$ & 0.10 \\
$E_i$ & 0.25 & $\overline{g_i}$ & 1.00 \\
$E_e$ & 1.00 & $\overline{g_e}$ & 1.00 \\
$V_{rest}$ & 0.30 & $\Theta$  & 0.50 \\
$\tau$ & .3 & $\gamma$ & 80 \\ \hline
 \end{tabular}
 \caption{\small Parameters for the simulation (see equations in text
  for explanations of parameters). All are standard default parameter values.}
 \label{tab.sim_params}
\end{table}

Leabra uses a {\em point neuron} activation function that models the electrophysiological properties of real neurons, while simplifying their geometry to a single point. This function is nearly as simple computationally as the standard sigmoidal activation function, but the more biologically-based implementation makes it considerably easier to model inhibitory competition, as described below. Further, using this function enables cognitive models to be more easily related to more physiologically detailed simulations, thereby facilitating bridge-building between biology and cognition. We use normalized units where the unit of time is 1 msec, the unit of electrical potential is 0.1 V (with an offset of -0.1 for membrane potentials and related terms, such that their normal range stays within the $[0, 1]$ normalized bounds), and the unit of current is $1.0x10^{-8}$.

The membrane potential $V_m$ is updated as a function of ionic conductances $g$ with reversal (driving) potentials $E$ as follows:
\begin{equation}
 \Delta V_m(t) = \tau \sum_c g_c(t) \overline{g_c} (E_c - V_m(t))
 \label{eq.vm}
\end{equation}
with 3 channels ($c$) corresponding to: $e$ excitatory input; $l$ leak current; and $i$ inhibitory input. Following electrophysiological convention, the overall conductance is decomposed into a time-varying component $g_c(t)$ computed as a function of the dynamic state of the network, and a constant $\overline{g_c}$ that controls the relative influence of the different conductances. The equilibrium potential can be written in a simplified form by setting the excitatory driving potential ($E_e$) to 1 and the leak and inhibitory driving potentials ($E_l$ and $E_i$) of 0:
\begin{equation}
 V_m^\infty = \frac{g_e \overline{g_e}} {g_e
  \overline{g_e} + g_l \overline{g_l} + g_i \overline{g_i}} 
\end{equation}
which shows that the neuron is computing a balance between excitation and the opposing forces of leak and inhibition. This equilibrium form of the equation can be understood in terms of a Bayesian decision making framework \cite{OReillyMunakata00}.

The excitatory net input/conductance $g_e(t)$ or $\eta_j$ is computed as the proportion of open excitatory channels as a function of sending activations times the weight values:
\begin{equation}
 \eta_j = g_e(t) = \langle x_i \wij \rangle = \oneo{n} \sum_i x_i \wij
 \label{eq.net_in_avg}
\end{equation}
The inhibitory conductance is computed via the kWTA function described in the next section, and leak is a constant.

In its discrete spiking mode, Leabra implements exactly the AdEx (adaptive exponential) model \cite{BretteGerstner05}, which has been found through various competitions to provide an excellent fit to the actual firing properties of cortical pyramidal neurons \cite{GerstnerNaud09}, while remaining simple and efficient to implement. However, we typically use a rate-code approximation to discrete firing, which produces smoother more deterministic activation dynamics, while capturing the overall firing rate behavior of the discrete spiking model.

We recently discovered that our previous strategy of computing a rate-code graded activation value directly from the membrane potential is problematic, because the mapping between $V_m$ and mean firing rate is not a one-to-one function in the AdEx model. Instead, we have found that a very accurate approximation to the discrete spiking rate can be obtained by comparing the excitatory net input directly with the effective computed amount of net input required to get the neuron firing over threshold ($g_e^{\Theta}$), where the threshold is indicated by $\Theta$:
\begin{equation}
g_e^{\Theta} = \frac{g_i \overline{g}_i (E_i - V_m^{\Theta}) +
 \overline{g}_l(E_l - V_m^{\Theta})} {\overline{g}_e (V_m^{\Theta} - E_e)}
\end{equation}
\begin{equation}
 y_j(t) \propto g_e(t) - g_e^{\Theta}
\end{equation}
where $y_j(t)$ is the firing rate output of the unit.

We continue to use the Noisy X-over-X-plus-1 (NXX1) function, which starts out with a nearly linear function, followed by a saturating nonlinearity:
\begin{equation}
 y_j(t) = \oneo{\left(1 + \oneo{\gamma [g_e(t) - g_e^{\Theta}]_+} \right)}
\end{equation}
where $\gamma$ is a gain parameter, and $[x]_+$ is a threshold function that returns 0 if $x<0$ and $x$ if $x>0$. Note that if it returns 0, we assume $y_j(t) = 0$, to avoid dividing by 0. As it is, this function has a very sharp threshold, which interferes with graded learning learning mechanisms (e.g., gradient descent). To produce a less discontinuous deterministic function with a softer threshold, the function is convolved with a Gaussian noise kernel ($\mu=0$, $\sigma=.005$), which reflects the intrinsic processing noise of biological neurons:
\begin{equation}
 y^*_j(x) = \int_{-\infty}^{\infty} \oneo{\sqrt{2 \pi} \sigma}
 e^{-z^2/(2 \sigma^2)} y_j(z-x) dz
 \label{eq.conv}
\end{equation}
where $x$ represents the $[g_e(t) - g_e^{\Theta}]_+$ value, and $y^*_j(x)$ is the noise-convolved activation for that value. In the simulation, this function is implemented using a numerical lookup table.

There is just one last problem with the equations as written above: They don't evolve over time in a graded fashion.  In contrast, the Vm value does evolve in a graded fashion by virtue of being iteratively computed, where it incrementally approaches the equilibrium value over a number of time steps of updating.  Instead the activation produced by the above equations goes directly to its equilibrium value very quickly, because it is calculated based on excitatory conductance and does not take into account the sluggishness with which changes in conductance lead to changes in membrane potentials (due to capacitance).

To introduce graded iterative dynamics into the activation function, we just use the activation value ($y^*(x)$) from the above equation as a ''driving force'' to an iterative temporally-extended update equation:
\begin{equation}
  y_j(t) = y_j(t-1) + dt_{vm} \left(y_j^*(t) - y_j(t-1) \right)
 \label{eq.y_iter}
\end{equation}
This causes the actual final rate code activation output at the current time $t$, $y(t)$ to iteratively approach the driving value given by $y^*(x)$, with the same time constant $dt_{vm}$ that is used in updating the membrane potential.  In practice this works extremely well, better than any prior activation function used with Leabra.

\subsubsection{FFFB Inhibition}

Leabra computes a layer-level inhibition conductance value based on a combination of feed-forward (FF) and feed-back (FB) dynamics.  This is an advance over the more explicit kWTA (k-Winners-Take-All) function that was used previously, though it achieves roughly the same overall kWTA behavior, with a much simpler, more efficient, and biologically plausible formulation.  The FF component is based directly on the average excitatory net input coming into the layer ($<\eta>$), and the FB component is based on the average activation of units within the layer ($<act>$).  Remarkably, fixed gain factors on each of these terms, together with simple time integration of the FB term to prevent oscillations, produces results that are overall comparable to the kWTA dynamics, except that the activations of units in the layer retain more of a proportional response to their overall level of excitatory drive, which is desirable in many cases.

FFFB is conceptually just the sum of the FF and FB components, each with their own ff and fb gain factors, with an overall gain factor (gi) applied to both:
\begin{equation}
  g_i = \mbox{gi} \left( \mbox{ff} [<\eta> - \mbox{ff0}]_+ + \mbox{fb} <act> \right)
  \label{eq.fffb}
\end{equation}
where $[ x ]_+$ indicates the positive part of whatever it contains -- anything negative truncates to zero.  It is important to have a small offset on the FF component, parameterized by ff0 which is typically .1 --- this delays the onset of inhibition and allows the neurons to get a little bit active first.  To minimize oscillations, the feedback component needs to be time integrated, with a fast time constant of .7 -- just a simple exponential approach to the driving fb inhibition value was used:
\begin{equation}
  fb_i(t) = fb_i(t-1) + dt \left(\ \mbox{fb} <act> - fb_i(t-1) \right)
  \label{eq.fbi}
\end{equation}
Typically ff is set to 1.0, fb is 0.5, and the overall gain (gi) is manipulated to achieve desired activity levels -- typically it is around 2.2 or so.

\subsubsection{XCAL Learning} 

The full treatment of the new XCAL version of learning in Leabra is presented in \incite{OReillyMunakataFrankEtAl12}, but the basic equations and a brief motivation for them are presented here.

In the original Leabra algorithm, learning was the sum of two terms: an error-driven component and a Hebbian self-organizing component. In the new XCAL formulation, the error-driven and self-organizing factors emerge out of a single learning rule, which was derived from a biologically detailed model of synaptic plasticity by Urakubo et al.~\cite{UrakuboHondaFroemkeEtAl08}, and is closely related to the Bienenstock, Cooper \& Munro (BCM) algorithm \cite{BienenstockCooperMunro82}. In BCM, a Hebbian-like sender-receiver activation product term is modulated by the extent to which the receiving unit is above or below a long-term running average activation value:
\begin{equation}
 \Delta_{bcm} \wij = xy (y - \langle y^2 \rangle)
 \label{eq:bcm}
\end{equation}
($x$ = sender activation, $y$ = receiver activation, and $\langle y^2 \rangle$ = long-term average of squared receiver activation). The long-term average value acts like a dynamic plasticity threshold, and causes less-active units to increase their weights, while more-active units tend to decrease theirs (i.e., a classic homeostatic function). This form of learning resembles Hebbian learning in several respects, but can learn higher-order statistics, whereas Hebbian learning is more constrained to extract low-order correlational statistics. Furthermore, the BCM model may provide a better account of various experimental data, such as monocular deprivation experiments \cite{CooperIntratorBlaisEtAl04}.

\begin{figure}[ht!]
 \centering\includegraphics[height=2in]{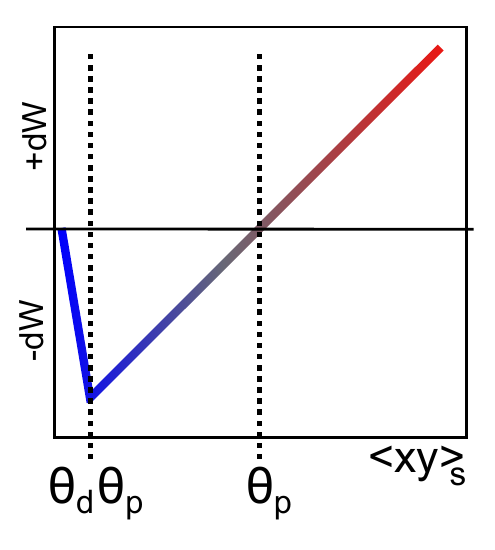}
 \caption{\small XCAL dWt function, shown with $\theta_p=0.5$, which determines
   the cross-over point between negative and positive weight changes, and
   $\theta_p \theta_d$ determines the inflection point at the left where the
   curve goes from a negative slope to a positive slope. This function fits
   the results of the highly detailed Urakubo et al \protect
   \cite{UrakuboHondaFroemkeEtAl08} model, with a correlation value of
   $r=0.89$.}
 \label{fig.xcal_dwt}
\end{figure}

The Leabra XCAL learning rule is based on a contrast between a sender-receiver activation product term (shown initially as just $xy$ -- relevant time scales of averaging for this term are elaborated below) and a dynamic plasticity threshold $\theta_p$ (also elaborated below), which are integrated in the XCAL learning function (Figure~\ref{fig.xcal_dwt}):
\begin{equation}
 \Delta_{xcal} \wij = f_{xcal} ( xy, \theta_p)
 \label{eq.xcal_simp}
\end{equation}
where the XCAL learning function was derived by fitting a piecewise-linear function to the Urakubo et al \cite{UrakuboHondaFroemkeEtAl08} simulation results based on synaptic drive levels (sender and receiver firing rates; the resulting fit was very good, with a correlation of $r=0.89$):
\begin{equation}
f_{xcal}(xy, \theta_p) = \left\{ \begin{array}{ll}
(xy - \theta_p) & \mbox{if} \; xy > \theta_p \theta_d \\
-xy (1 - \theta_d) / \theta_d & \mbox{otherwise} \end{array} \right.
\end{equation}
($\theta_d = .1$ is a constant that determines the point where the function reverses back toward zero within the weight decrease regime -- this reversalpoint occurs at $\theta_p \theta_d$, so that it adapts according to the dynamic $\theta_p$ value).

The BCM equation produces a curved quadratic function that has the same qualitative shape as the XCAL function (Figure~\ref{fig.xcal_dwt}). A critical feature of these functions is that they go to 0 as the synaptic activity goes to 0, which is in accord with available data, and that they exhibit a crossover point from LTD to LTP as a function of synaptic drive (which is represented biologically by intracellular Calcium levels). A nice advantage of the linear XCAL function is that, to first approximation, it is just computing the subtraction $xy - \theta_p$.

To achieve full error-driven learning within this XCAL framework, we just need to ensure that the core subtraction represents an error-driven learning term.  In the original Leabra, error-driven learning via the Contrastive Hebbian Learning algorithm (CHL) was computed as:
\begin{equation}
 \Delta_{chl} = x^+ y^+ - x^- y^-
 \label{eq:chl}
\end{equation}
where the superscripts represent the plus ($+$) and minus ($-$) phases. This equation was shown to compute the same error gradient as the backpropagation algorithm, subject to symmetry and a 2nd-order numerical integration technique known as the midpoint method, based the generalized recirculation algorithm (GeneRec; \cite{OReilly96}). In XCAL, we replace these values with time-averaged activations computed over different time scales:
\begin{itemize}
\item {\bf s} = short time scale, reflecting the most recent state of neural  activity (e.g., past 100-200 msec). This is considered the ``plus phase'' -- it represents the {\em outcome} information on the current trial, and in  general should be more correct than the medium time scale.
\item {\bf m} = medium time scale, which integrates over an entire  psychological ``trial'' of roughly a second or so -- this value contains a mixture of the ``minus phase'' and the ``plus phase'', but in contrasting it  with the short value, it plays the role of the minus phase value, or expectation about what the system thought should have happened on the  current trial.
\item {\bf l} = long time scale, which integrates over hours to days of processing -- this is the BCM-like threshold term.
\end{itemize}

Thus, the error-driven aspect of XCAL learning is driven essentially by the following term: 
\begin{equation}
 \Delta_{xcal-err} \wij = f_{xcal} ( x_s y_s, x_m y_m )
 \label{eq.xcal-err}
\end{equation}
However, consider the case where either of the short term values ($x_s$ or $y_s$) is 0, while both of the medium-term values are $>0$ -- from an error-driven learning perspective, this should result in a significant weight decrease, but because the XCAL function goes back to 0 when the input drive term is 0, the result is no weight change at all. To remedy this situation, we assume that the short-term value actually retains a small trace of the medium-term value:
\begin{equation}
 \Delta_{xcal-err} \wij = f_{xcal} ( \kappa x_s y_s + (1-\kappa) x_m y_m, x_m y_m)
 \label{eq.xcal-err2}
\end{equation}
(where $\kappa = .9$, such that only .1 of the medium-term averages are incorporated into the effective short-term average).

The self-organizing aspect of XCAL is driven by comparing this same synaptic drive term to a longer-term average, as in the BCM algorithm:
\begin{equation}
 \Delta_{xcal-so} \wij = f_{xcal} ( \kappa x_s y_s + (1-\kappa) x_m y_m, \gamma_l y_l)
 \label{eq.xcal-selforg}
\end{equation}
where $\gamma_l = 3$ is a constant that scales the long-term average threshold term (due to sparse activation levels, these long-term averages tend to be rather low, so the larger gain multiplier is necessary to make this term relevant whenever the units actually are active and adapting their weights).

Combining both of these forms of learning in the full XCAL learning rule amounts to computing an aggregate $\theta_p$ threshold that reflects a combination of both the self-organizing long-term average, and the medium-term minus-phase like average:
\begin{equation}
 \Delta_{xcal} \wij = f_{xcal} ( \kappa x_s y_s + (1-\kappa) x_m y_m, \lambda
 \gamma y_l + (1-\lambda) x_m y_m)
 \label{eq.xcal}
\end{equation}
where $\lambda = .01$ is a weighting factor determining the mixture of self-organizing and error-driven learning influences (as was the case with standard Leabra, the balance of error-driven and self-organizing is heavily weighted toward error driven, because error-gradients are often quite weak in comparison with local statistical information that the self-organizing system encodes).

The weight changes are subject to a soft-weight bounding to keep within the $0-1$ range:
\begin{equation}
 \Delta_{sb} \wij = [\Delta_{xcal}]_+ (1-\wij) + [\Delta_{xcal}]_- \wij
 \label{eq.err_soft_bound}
\end{equation}
where the $[]_+$ and $[]_-$ operators extract positive values or negative-values (respectively), otherwise 0.

Finally, as in the original Leabra model, the weights are subject to contrast enhancement, which magnifies the stronger weights and shrinks the smaller ones in a parametric, continuous fashion. This contrast enhancement is achieved by passing the linear weight values computed by the learning rule through a sigmoidal nonlinearity of the following form:
\begin{equation}
 \hat{w}_{ij} = \oneo{1 + \left(\frac{\wij}{\theta (1-\wij)}\right)^{-\gamma}}
 \label{eq.wt_off}
\end{equation}
where $\hat{w}_{ij}$ is the contrast-enhanced weight value, and the sigmoidal function is parameterized by an offset $\theta$ and a gain $\gamma$ (standard defaults of 1 and 6, respectively, used here). 


\end{document}